\RequirePackage{mathtools} 
\RequirePackage{empheq}
\documentclass[twocolumn,twocolappendix,numberedappendix]{openjournal}

\usepackage{amssymb}
\usepackage{amsfonts,amstext,amscd}   
\DeclareSymbolFont{bbold}{U}{bbold}{m}{n}%
\DeclareSymbolFontAlphabet{\mathbbold}{bbold}
\usepackage{esvect} 
\makeatletter
\let\@@array\relax  % Neutralize revtex4 redefinition
\makeatother 
\usepackage{array}   % load always after resetting @@array
\usepackage[detect-all]{siunitx}%need it to use the column type S in table

\usepackage{booktabs}
\usepackage{multirow}
\usepackage{tabstackengine}%\TABstackMath
\usepackage{booktabs}
\usepackage{makecell}
\usepackage{graphicx,xcolor,color}
\definecolor{azulelectrico}{rgb}{0., 0.3, 1.} 
\usepackage{float}%to use option [H] in figures
\usepackage{subfigure}%,lipsum,multicol}%placeins

\usepackage{enumitem}%for controlling lists of items
\usepackage{txfonts}  % Alternativa (puede causar conflictos)

\defcitealias{LL04}{LL04}
\defcitealias{van-Ballegooijen:1985wi}{vB85}
\defcitealias{Landi-Deglinnocenti:1985a}{LD85}
\defcitealias{Semel:1999aa}{SLA99}
\defcitealias{Lopez-Ariste:1999aa}{LAS99}
\usepackage[breaklinks,colorlinks=true,linkcolor=red,citecolor=azulelectrico,urlcolor=blue ]{hyperref}

\usepackage{orcidlink}

\usepackage{tikz}
\usetikzlibrary{positioning,shapes,matrix,calc}

\newcommand\mytri[3]{%
  \tikz[baseline=(char.base)]\node[draw,minimum width=0.207,inner sep=0.6pt,isosceles
  triangle,isosceles triangle apex angle=60,shape border uses
  incircle,rounded corners, shape border rotate=#1,scale=#2] (char) {#3};}

\newcommand\mytrin[3]{%
  \tikz[baseline=(char.base)]\node[draw,inner sep=-1.3pt,isosceles triangle, isosceles triangle apex angle=60,minimum width=0.20cm,rounded corners, shape border uses incircle,shape border rotate=#1,scale=#2](char) {#3};}

\newcommand\myrect[2]{%
  \tikz[baseline=(char.base)]\node[draw,inner sep=2pt,shape border uses incircle,minimum width=0.5cm,minimum height=0.52cm,yshift=0.2ex,scale=#1](char) {#2};}

\newcommand\myrectr[2]{%
  \tikz[baseline=(char.base)]\node[draw,inner sep=2pt,shape border uses incircle,minimum width=0.5cm,rounded corners, minimum height=0.52cm,yshift=0.2ex,scale=#1](char) {#2};}

\newcommand\mycircle[2]{%
  \tikz[baseline=(char.base)]\node[draw,circle,fill=gray!50,inner sep=3pt,shape border uses incircle,minimum height=0.45cm,scale=#1] (char) {#2};}

\newcommand\mydiam[2]{%
  \tikz[baseline= (char.base)]\node[diamond,draw=black,inner sep=1.8pt,fill=white!10,scale=1,yshift=-0.3ex,scale=#1](char) {#2};}

\graphicspath{{figures}} %figures/ 
\begin{document}

\title{\Large Reformulating polarized radiative transfer for astrophysical applications (I).\\ A formalism allowing non-local Magnus solutions}

%\title{Reformulating polarized radiative transfer for astrophysical applications}
%\secondtitle{I. A formalism allowing non-local Magnus solutions}

\shorttitle{Reformulating polarized radiative transfer (I)}
\shortauthors{Carlin, Blanes, \& Casas}
%short title to appear in every odd page

\author{\vspace{-0.17cm} {\large Edgar S. Carlin\,$^{1,*}$\orcidlink{0000-0002-0012-6581}}}
\author{ {\large Sergio Blanes\,$^{2}$\orcidlink{0000-0001-5819-8898}}}
\author{{\large Fernando Casas\,$^{3}$\orcidlink{0000-0002-6445-279X}}}

\affiliation{$^1$Instituto de Astrof\'isica de Canarias (IAC), E-38205, La Laguna, Tenerife, Spain}
\affiliation{$^2$Instituto de Matem\'atica Multidisciplinar, Universitat Polit\`ecnica de Valencia, E-46022 Valencia, Spain}
\affiliation{$^3$Departament de Matem\`atiques, Universitat Jaume I, E-12071 Castell\'on, Spain\vspace{0.05cm}}

%\date{Submitted: Dec 30, 2024; Accepted: March 17, 2025; Published: March 21, 2025}
%\submitted{Submitted: Jan 13, 2025; Accepted: March 17, 2025; Published: \today }
\submitted{Accepted: March 17, 2025. Published Version  March 27, 2025 }

%%\email{ecarlin@iac.es, e.carlin.mail@gmail.com}
\thanks{\vspace{-0.1cm}$^*$E-mails: ecarlin@iac.es, e.carlin.mail@gmail.com}
%\blfootnote{$^{\star}$ Corresponding author. Email: \href{mailto:s.wilkins@sussex.ac.uk}{s.wilkins@sussex.ac.uk}}

\begin{abstract}
  The solar atmosphere is diagnosed by solving the polarized radiative 
  transfer problem for plasmas in Non-Local Thermodynamic Equilibrium (NLTE). A key challenge in multidimensional NLTE diagnosis is to integrate efficiently the radiative transfer equation (RTE), but current methods are local, i.e. limited to constant propagation matrices. This paper introduces a formalism for 
  non-local integration of the RTE using the Magnus expansion. We begin by framing 
  the problem in terms of rotations within the Lorentz / Poincar\'e group 
  (Stokes formalism), motivating the use of the Magnus expansion. By 
  combining the latter with a highly detailed algebraic characterization 
  of the propagation matrix, we derive a compact analytical evolution 
  operator that supports arbitrary variations of the propagation matrix 
  and allows to increasingly consider any order in the Magnus expansion. 
  Additionally, we also reformulate the inhomogeneous part of the RTE, again using the Magnus expansion, and introducing the new concept of \textit{inhomogeneous} 
  evolution operator. This provides the first consistent, general, and 
  non-local formal solution to the RTE that is furthermore efficient and 
  separates integration from the algebraic formal solution. Our 
  framework is verified analytically and computationally, leading to a new 
  family of numerical radiative transfer methods and potential 
  applications such as accelerating NLTE 
  calculations. With minor adjustments, our results apply to other universal physical problems sharing the Lorentz / Poincar\'e algebra in special relativity and electromagnetism.
\end{abstract}


\maketitle

\section{Introduction}\label{sec:intro}
As astrophysics studies distant plasmas through scattered light, it is crucial to model and measure precisely how polarized light behaves. Currently, the measurement and interpretation of Stokes
spectropolarimetry in spectral lines is the best known way of diagnosing astrophysical plasmas \citep[][LL04 hereafter]{Stenflo:1994, LL04}. 
Thus, the Stokes 4-vector describes any partially polarized light beam, quantifying both the number of photons and their oscillation direction within the beam's reference frame \citep{Born:1980}. Once quantified, the next step is
 to describe its transference through the emitter plasma using the polarized radiative transfer equation (RTE). 

This key astrophysical problem is best studied in the nearby and well-resolved solar atmosphere, where most diagnostic techniques and methods have been developed. In such a context, radiative transfer is a fundamental part of the Non-Local Thermodynamic Equilibrium (NLTE) problem, in 
which radiation emitted at a given point in the solar atmosphere modifies, via radiative transfer, the physical state of other distant points. In Stokes formalism, the polarized RTE contains a $4\times 4$ propagation matrix 
quantifying the microphysics through optical coefficients that depend on angle, 
wavelength, and distance along the ray. This makes 
the integration of the RTE the most frequent and critical operation carried 
out in NLTE iterative schemes, especially when considering multidimensional solar models or inversions of optically thick lines. One
motivation for the present paper is to solve the RTE with more robust and accurate methods whose properties are preserved when dealing with large 
velocity gradients as those above photospheric 
layers. Also, boosting efficiency is a must for 
NLTE diagnosis in the era of big-data solar telescopes like DKIST. However, all this implies pursuing a more fundamental goal, which  
we consider the next level in polarized 
radiative transfer. Specifically, we aim at solving the RTE consistently in spatially resolved solar models \textit{without assuming constant local properties}. 

The use of constant propagation matrices is a valid and universally adopted technique for solving the RTE. It 
has been used since ~\cite
{unno1956} and~\cite{rachkovsky1967} first 
derived the full polarized RTE including magnetic fields and 
magneto-optical effects in the propagation matrix, 
together with its first analytical solution, valid for Milne-Eddington (i.e., constant, spatially homogeneous) atmospheres. Later, \cite{van-Ballegooijen:1985wi} and \cite{Landi-Deglinnocenti:1985a} \citep[hereafter vB85 and LD85; see 
also][]{Kalkofen:2009ty} provided analytical 
evolution operators solving the homogeneous RTE, again with constant propagation matrix in Jones and Stokes formalisms, respectively. 

However, there were soon evidences that spatial variations are not negligible in the solar atmosphere. On one hand, 
numerical solutions of the RTE were found to improve significantly when the atmosphere 
was subdivided into multiple layers \citep[e.g.,][]{rees+al1989, bellot_rubio+al1998}. On the other, numerical 
inversions of observed solar Stokes profiles confirmed the need for models with spatial variations \citep{Collados:1994, Del-Toro-Iniesta:1996tw}. Hence, many layers are needed both for models and numerical methods. %all seems to fit.

Initially, the inhomogeneous nature of the solar atmosphere was primarily attributed to  \textit{large-scale} stratification in density, temperature, and magnetic field. However, 
advances in both observations and simulations continue to demand ever-smaller resolved scales. Modern MHD solar models, for instance, still need finer resolution to address physical challenges like coronal and chromospheric heating \citep[e.g., ][]{Leenaarts:2009,Khomenko2018}. Early simulations also revealed large velocity-driven inhomogeneities, 
especially in the  chromosphere \citep[e.g.,][]{Carlsson:1997aa}. Indeed, plasma velocity gradients in sound/shock waves are primary sources of solar small-scale variations and 
discontinuities, with a large and specific 
impact in the polarization via radiative 
transfer effects \citep{Carlin:2013aa}, either due to their modulation of radiation field anisotropy \citep{Carlin:2015aa} and/or due to a \mbox{relatively} new phenomenon that we may call \textit{dynamic \mbox{dichroism}} \citep{Carlin:2019aa}.

To address spatial atmospheric variations in resolved atmospheres while using formalisms based on 
constant properties, numerical methods based on 
"ray characteristics" \cite[e.g.,][]{kunasz-auer} 
were developed. Some representative methods 
using constant properties are the piecewise constant 
Evolution Operator method (\citetalias{van-Ballegooijen:1985wi}, \citetalias{Landi-Deglinnocenti:1985a}), the “DELO” methods \citep{rees+al1989} 
of linear, semi- parabolic,
and parabolic kind \citep[see][]{Janett:2017aa};
or the order-3 DELO-Bezier \citep{delacruz_rodriguez+piskunov2013}.

Hence, we face a seemingly paradoxical situation. Despite increasingly smaller spatial variations continue to be observed in the sun and needed to 
model its atmosphere, for over $65$ years all mainstream numerical methods for solving the polarized RTE in (solar) astrophysics have remained limited to evolution operators with 
constant propagation matrices. These methods work  
very well as they asymptotically approach their regime 
of local constant properties but they have to solve the RTE 
at each numerical cell, i.e.~sequentially and many times along every ray. This process scales with the increasing 
resolution of the models, becoming a main source 
of computational cost. Clearly, the idealization of 
constant properties is never fully achieved 
because radiative transfer 
depends on several atmospheric quantities 
with different sources of gradients and scales of 
variation. Moreover, it cannot be achieved 
simultaneously and uniformly for all wavelengths (e.g.~around a spectral line) using a same discretization because the effective 
integration step of any numerical cell changes with 
wavelength (optical depth does it, and 
current methods integrate in optical depth scale). This fact is reinforced by the sensitivity of the 
optical depth step to Doppler shifts and opacity variations in the star. 

This situation remarks the importance of the evolution operator, which as the matrix that advances the solution of a differential equation, fully
characterizes the final solution physically 
and numerically \citep[][]{Gantmacher:1959aa}. \citetalias{van-Ballegooijen:1985wi} and \citetalias{Landi-Deglinnocenti:1985a} recognized this when they introduced in (solar) astrophysics 
the evolution operator method to solve the RTE. It is based on approximating the homogeneous 
evolution operator analytically by an exponential of a \textit
{constant} propagation matrix. To our knowledge, this method is currently used in solar physics only by the Hazel code \citep[][]{Asensio-Ramos:2008aa} mainly
to synthesize the He\textrm{I} $\lambda 10830$ line in optically thin conditions.  
Four reasons may explain this limited adoption: i) since it is piecewise constant, the method has low numerical 
order; ii) its evolution operator 
matrix is lengthy (spanning about a page) and lacks a clear physical
interpretation; iii) its numerical implementation is not efficient; and iv) it exhibits a limiting oscillatory behavior for strong (optically-thick) 
spectral lines with magneto-optical effects \citep[e.g.,][]{bellot_rubio+al1998}. In this paper we solve 
the first three issues. We believe 
that the fourth one may actually stem from the formulation with constant properties, and in this 
paper we set a basis that allows to investigate it.

In summary, currently adopted methods are limited by numerical approximations based on 
local constant properties, so it is unrealistic to hope for the large performance boosts demanded by the  general NLTE problem, which is non-local by definition. 
Here, we start exploring new general analytical solutions that also allow neat and fully consistent implementations of arbitrary spatial variations.
The only previous attempt to solve this general problem was by~\cite{Semel:1999aa} and ~\cite{Lopez-Ariste:1999aa}(hereafter SLA99 and LAS99). We think that \citetalias{Lopez-Ariste:1999aa}'s central intent, which is shared in the present paper, has a huge potential for solar diagnosis that was unfortunately overlooked by the solar community. \citetalias{Lopez-Ariste:1999aa} considered the group structure of the 
RTE and applied the \cite{Wei:1963aa} method to formally 
pose a non-local solution based on products of 
matrix exponentials. However, their scheme only provided a semi-closed formulation demanding to solve some coupled differential equations for the optical 
coefficients under poorly specified conditions, which anticipated a lack of economy and refrained a numerical materialization. %The authors stated that, although their approach is exact, it can hardly be used in practical computation because it would not be economic. 
Without discarding such results, we present a different approach based on the Magnus expansion \citep{Magnus:1954tj}. 

Enters the key problem of commutativity: when an atmosphere varies spatially, its propagation 
matrices do not commute between points,  
preventing exact solutions. \citetalias{Semel:1999aa} saw this as an intrinsic
limitation of the complex non-conmutative structure of the Magnus expansion. Thus, they avoided it explicitly, arguing (correctly) that without commuting layers the solution requires a constant 
propagation matrix that forces discretization. In our work, we shall assume spatial discretization and arbitrary spatial variations, fully accepting non-commutativity as an inherent
part of the problem, and exploring how to incorporate it 
in the theory with the Magnus expansion. Among its 
remarkable aspects, this expansion retains intrinsic properties of the exact solution by preserving its 
group structure after truncation in the Lie algebra \cite[e.g., ][]{Blanes:2009wr}. It is defined with a single exponential of integrals and commutators, which avoids 
products of matrix exponentials, differential formulations, and (intrinsically innaccurate) perturbative developments. 

This first paper presents the analytical foundational part of our formalism:
$\bullet$ Section 2 presents the problem in Stokes formalism, interpreting solutions to the RTE as rotations in a Lie group, and elaborating on the Magnus expansion.
$\bullet$Section 3 presents a detailed definition and characterization of the propagation vector and propagation matrix.
$\bullet$Section 4 derives a compact analytical evolution operator using Magnus to first order and delves into higher orders. 
$\bullet$Section 5 derives a novel formal solution based on the new inhomogeneous evolution operator.
$\bullet$Section 6 demonstrates our formalism in a model solar atmosphere and advance some applications.
%\bigskip

\section{Foundations: radiative transfer in terms of rotations with the Magnus expansion}\label{sec:evolop_sec}
\medskip
\subsection{Generic solution in Stokes formalism}
The polarized RTE is a system of four coupled first-order, ordinary differential equations whose homogeneous / inhomogeneous term contains the propagation / emissivity matrix / vector \citepalias[e.g.,][chap.~8]{LL04}:
\begin{equation}\label{eq:matrixrte}
\centering 
\frac{\rm{d}}{\rm{ds}} {\bf I} = \boldsymbol{\epsilon}- {\bf K I},
\end{equation}
with ${\boldsymbol \epsilon}$ the emissivity vector and ${\bf K}$ the 4x4 propagation matrix:
% \footnote{We use the
%   notation $\mathvec{\mathrm X}$ for physical vectors and $\mathbf{X}$ for formal
%   vectors. The former is reserved for vectorial
%   magnitudes with three components in the ordinary space, such as
%   the velocity or the magnetic field. The latter is for quantities
%   that are better described by collections of points involving dimensions different
%   than ordinary space, such as the Stokes vector}:
\begin{equation}\label{eq:basicrte}
\centering 
\frac{\rm{d}}{\rm{ds}}
\left( \begin{array}{c}
 I \\ 
 Q \\
 U\\
 V
\end{array} \right) 
=
\left(\begin{array}{c}
\epsilon_I \\ 
 \epsilon_Q \\
 \epsilon_U\\
 \epsilon_V
\end{array} \right) 
-
\left(\begin{array}{cccc}
 \eta_I & \eta_Q & \eta_U & \eta_V \\ 
 \eta_Q & \eta_I & \rho_V & -\rho_U \\
 \eta_U & -\rho_V & \eta_I & \rho_Q\\
 \eta_V & \rho_U & -\rho_Q & \eta_I
\end{array} \right) 
\left(\begin{array}{c}
 I \\ 
 Q \\
 U\\
 V
\end{array} \right).
\end{equation}
The homogenous part is then posed as the initial value problem:
\begin{equation}\label{eq:ivp_stokes}
  \frac{d\mathbf{I}(s)}{ds}=-\mathbf{K}(s)\mathbf{I}(s);\quad\quad \mathbf{I}(s_0)=\mathbf{I}_0,
\end{equation}
whose homogeneous formal solution at $s$ is given by an evolution 
operator $\mathbf{O}(s,s_0;\mathbf
{K})$ applied to the initial value at $\rm s_0$
\begin{equation}\label{eq:homog1}
\mathbf{I}(s) = \rm{\mathbf{O}} (s,s_0;{\bf K}) \cdot \mathbf{I}(s_0), 
\end{equation}
with $\rm{\mathbf{O}} (s,s_0;{\bf K})$ the solution to\footnote{To see this, insert Eq.(\ref
{eq:homog1}) into Eq.(\ref
{eq:ivp_stokes}) and do the derivative.}
\begin{equation}\label{eq:homog_oper}
  \begin{split}
%&\frac{d}{ds} {\mathbf{O}}^{s}_{s'} = -\rm{\mathbf{K}}(s) {\mathbf{O}}^{s}_{s'}\\
\frac{d}{ds} {\mathbf{O}}(s,s_0) = -\rm{\mathbf{K}}(s) {\mathbf{O}}(s,s_0), \quad\quad {\mathbf{O}}(s_0,s_0)
=\mathbbold{1}.%\\
%\left( \Rightarrow \frac{d}{ds_0} {\mathbf{O}}^{s}_{s_0} = {\mathbf{O}}^{s}_{s_0} \rm{\mathbf{K}}(s_0) \right),
\end{split}
\end{equation}
Knowing this operator, the inhomogenous solution to Eq. (\ref{eq:matrixrte}) is 
\begin{equation}\label{eq:inhomog1}
   \mathbf{I}(s) =  \rm{\mathbf{O}}(s,s_0)\mathbf{I}(s_0) + \int^s_{s_0}ds' \rm{\mathbf{O}}(s,s'){\boldsymbol \epsilon} (s').
\end{equation}
The suitability of this formal solution is physically and 
numerically determined by the expressions 
adopted for the evolution operator and for the inhomogeneous integral. Our goal is to formulate 
them in the best possible way.

\subsection{The currently adopted solution to the evolution operator is a series that cannot be truncated}\label{sec:evolop_series}
Applying the
fundamental theorem of calculus to Eq.(\ref{eq:homog_oper}), an evolution operator is obtained by~\citet{Volterra:1887aa}'s integral equation with $\mathbf{A}=-\mathbf{K}$:
\begin{equation}
\begin{split}
 & \mathbf{O}(s,s_0;\mathbf{A}) = \mathbbold{1} + \int^s_{s_0}dt \mathbf{A}(t)\mathbf{O}(t,s_0,\mathbf{A}),
 \end{split} 
\label{eq:volterraEO}
\end{equation} 
A Picard iteration on Eq. (\ref {eq:volterraEO}) can continue with $\mathbf{O}(t,s_0,\mathbf{A})=\mathbbold{1}$ and repeated infinitely. Thus,
a more accurate expression was obtained by~\cite{Peano:1888} (also~\cite{Peano:1890}, translated in~\cite{Peano1888_trans:2000}) and further studied by~\cite{Baker:1902,Baker:1905}. 
 It is the Peano-Baker series:
\begin{equation}
  \mathbf{O}(s,s_0;\mathbf{A}) = \mathbbold{1} +\int^s_{s_0}\mathbf{A}(t)dt+\int^s_{s_0}\mathbf{A}(t_1)\int^{t_1}_{s_0}\mathbf{A}(t_2)\,dt_2 \,dt_1+\cdots
\label{eq:peano_baker_int_1}
\end{equation}

%sec 16.5 Ince 1956, Sec 14 Gantmacher 1959
% Gantmacher Vol2 pag 127 and next chapters
This series is 
unique and converges absolutely and uniformly in every closed interval where $\mathbf{A}$ is continuous and bounded \citep[][]{ince1956,Gantmacher:1959aa}. 
It is applied in different 
contexts, most often known as the Neumann 
series or the Dyson 
perturbative solution \citep{Dyson:1949aa}
and simplified for constant $\mathbf{A}$. In our case, if $\mathbf{A}=-\mathbf{K}=ct.$, Eq. (\ref{eq:peano_baker_int_1}) becomes:
\begin{equation}\label{eq:dyson_taylor_2}  
    \mathbf{O}(s) = \mathbbold{1}+ \mathbf{A}(s-s_0)+\cdots+ \frac{\mathbf{A}^n{(s-s_0)}^n}{n!}+\cdots= e^{-\mathbf{K}(s-s_0)}
\end{equation}
We shall see that the problem with solutions (\ref{eq:peano_baker_int_1}) and (\ref{eq:dyson_taylor_2}) is that in general any 
truncation --also in numerical exponentiation-- makes them to leave the Lie group of the differential equation, no longer preserving the physical properties of the exact solution \citep[][]{Blanes:2009wr}.

In astrophysics, the Dyson solution for the Stokes RTE was specified for $\mathbf{K}=ct.$ by \citetalias{LL04} in two steps. Starting from
Eq.~(\ref{eq:peano_baker_int_1}):
\begin{equation*}
  \begin{split}
    \mathbf{O}(s,s_0) &= \mathbbold{1}+\sum^{\infty}_{n=1} (-1)^n\int^s_{s_0}ds_1\cdots%\\& \cdots
    \int^{s_{n-1}}_{s_0}ds_n\mathbf{K}(s_1)\cdots\mathbf{K}(s_n),
  \end{split}
  \end{equation*}
  and making the integration regions independent on integration variables, one obtains the "time-ordered" exponential in terms of the Dyson  chronological\footnote{For the RTE, it would be a space-ordered product instead.} product of operators $\mathcal{P}$ (\textbf{Step 1}):  
  \begin{equation*}
    \begin{split}
      \mathbf{O}(s,s_0) &= \mathbbold{1}+\sum^{\infty}_{n=1} \frac{(-1)^n}{n!}\int^s_{s_0}ds_1\cdots\int^s_{s_0}ds_n \mathcal{P} \{  \mathbf{K}(s_1)\cdots\mathbf{K}(s_n) \}=\\
      &=\mathcal{P}\big\{ e^{-\int^s_{s_0}dt \mathbf{K(t)} } \big\}
    \end{split}
    \end{equation*}
with $\mathcal{P}\{\mathbf{K}(s_1)\mathbf{K}(s_2)\cdots\mathbf{K}(s_n)\}=\mathbf{K}(s_{j_1})\mathbf{K}(s_{j_2})\cdots\mathbf{K}(s_{j_n})$ for $s_{j_1}\leq s_{j_2}\leq \cdots \leq s_{j_n}$.
The ordered exponential becomes an ordinary one 
when the ordered products commute among 
all of them ($[\mathbf{K}(s_i),\mathbf{K}(s_j)]=0$), such that can be reordered. 
Only then, they become ordinary products of a single integral, leading to the Taylor expansion of its exponential (\textbf{Step 2}): 
\begin{equation}\label{eq:Kexp_end}
  \begin{split}
    \mathbf{O}(s,s_0) &= \mathbbold{1}+\sum^{\infty}_{n=1} \frac{(-1)^n}{n!}\biggl[ \int^s_{s_0}dt \mathbf{K}(t) \biggr]^n=e^{-\int^s_{s_0}dt \mathbf{K}(t)} = e^{-\mathbf{K}\cdot \Delta s}
  \end{split}
  \end{equation}
We added the last equality to remark that, if commutation is assumed, the exponent is immediately forced to be calculated with $\mathbf{K}$ constant, a restrictive situation that is systematically adopted in  applications. Indeed, as it is imposible that 
propagation matrices commute simultaneously for all rays inside a realistic atmosphere, the only 
plausible assumption to achieve commutation is to 
restrict them to be constant along each integration step.
Apart from the obvious fact that atmospheres are 
not constant at small scales, this raises at least 
two sources of error. The first one comes from 
solutions assuming $\mathbf{K}$ constant 
but emissivity changing in the inhomogeneous 
integral. A fully consistent solution 
for $\mathbf{K}$ constant would then imply constant 
emissivity because their physical dependences are such that one cannot change if the other is constant (inconsistency).
A second issue is the computation of 
the exponential and of the terms mimicking the 
inhomogeneous integral in the numerical method. Such operations can 
imply indirect truncation of the Dyson series and damage to the matrix symmetry of 
the effective evolution operator (also inside the inhomogeneous solution), and 
therefore a systematic hidden error built in the numerical implementation (innacuracy). 
Other way to say this is that solutions do not 
evolve in the Lie group corresponding to the RTE. This 
restricts methods to be "local", i.e. to discrete 
numerical cells with constant properties. %, leading to the simplify Eq. (\ref{eq:inhomog1}) to 
%\begin{equation}\label{eq:inhomog1_kcte}
%  \begin{split}
%  \mathbf{I}(s) &=  e^{-\mathbf{K}(s-s_0)}\mathbf{I}_0 + e^{-\mathbf{K}s}\biggl[\int^s_{s_0}e^{\mathbf{K}t}dt\biggr] \,{\boldsymbol \epsilon}=\\
%  &=e^{-\mathbf{K}(s-s_0)}\mathbf{I}_0 +  (\mathbbold{1}-e^{-\mathbf{K}(s-s_0)})\mathbf{K}^{-1} \,{\boldsymbol \epsilon}
%\end{split}
%\end{equation}
 Hence, the price that methods pay to compensate that combination of errors is the need to be applied cell by cell (locally), whose cost roughly scales with the resolution of the numerical grid (inneficiency). 
 
 Our goal is to design an alternative accurate formalism that respects the Lie structure of the RTE and that is general enough to incorporate spatial inhomogeneity and non-locality.
%\medskip
%\clearpage
\begin{figure*}[t!]
  %\vspace{-3cm}
  \centering
 \begin{tabular}{ccc}
   \includegraphics[width=0.32\textwidth]{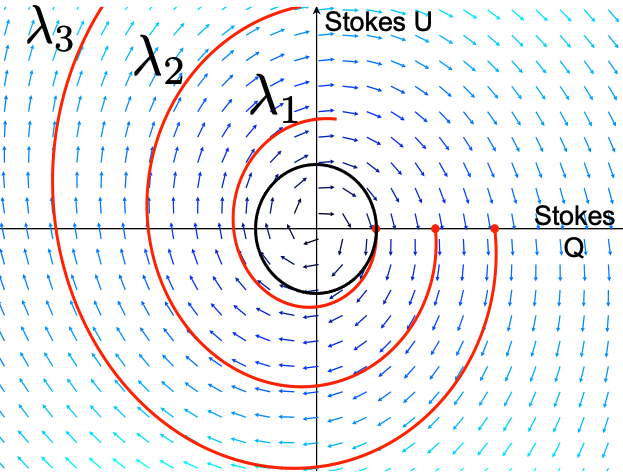} &
   \includegraphics[width=0.25\textwidth]{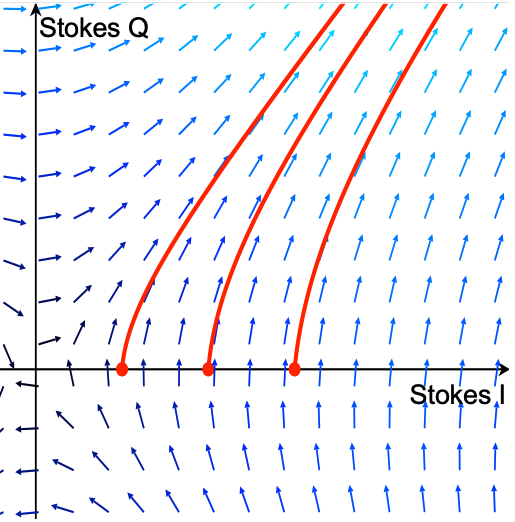} &
   \includegraphics[width=0.30\textwidth]{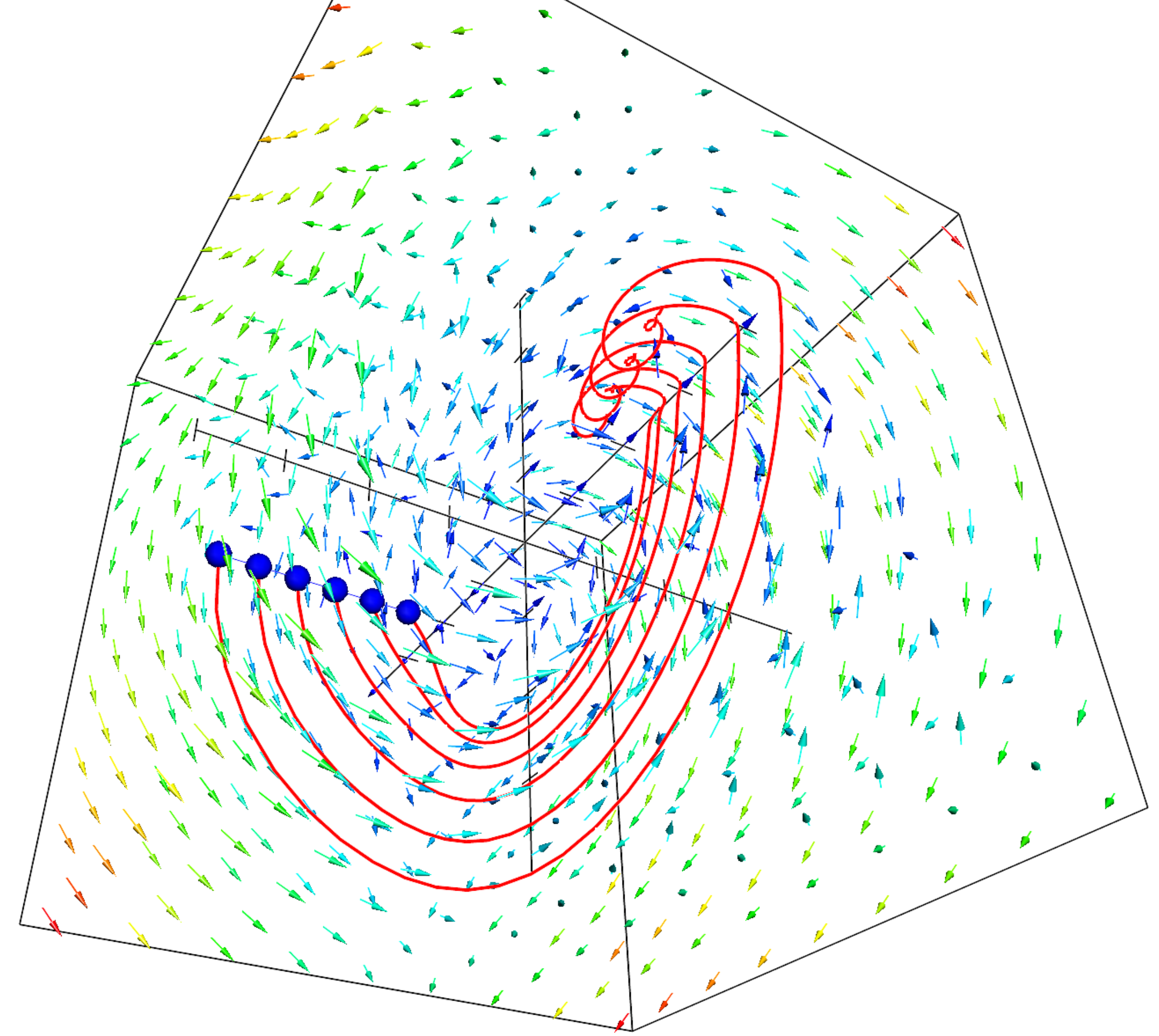} \\
 \end{tabular}
 \caption{Vector fields (arrows) and solutions (red lines) in different wavelengths for the homogeneous RTE.
 %for different initial conditions
 Left: rotational flow due to magneto-optical effects 
 for $\mathbf{K}$ constant in plane Q-U. The trajectory deviates from circularity to describe a spiral 
 due to the accumulated error in a  
 numerical method breaking Lie structure. Middle: idem for hyperbolic dichroic pseudo-rotation in I-V plane. Right: idem in QUV space for variable $\mathbf{K}$.}
 \label{fig:fig2}
\end{figure*}
\vspace{1cm}
 \subsection{Radiative transfer solutions as rotations in a Lie group}

\begin{figure}[H]
  \vspace{0.15cm}
  \includegraphics[scale=0.31]{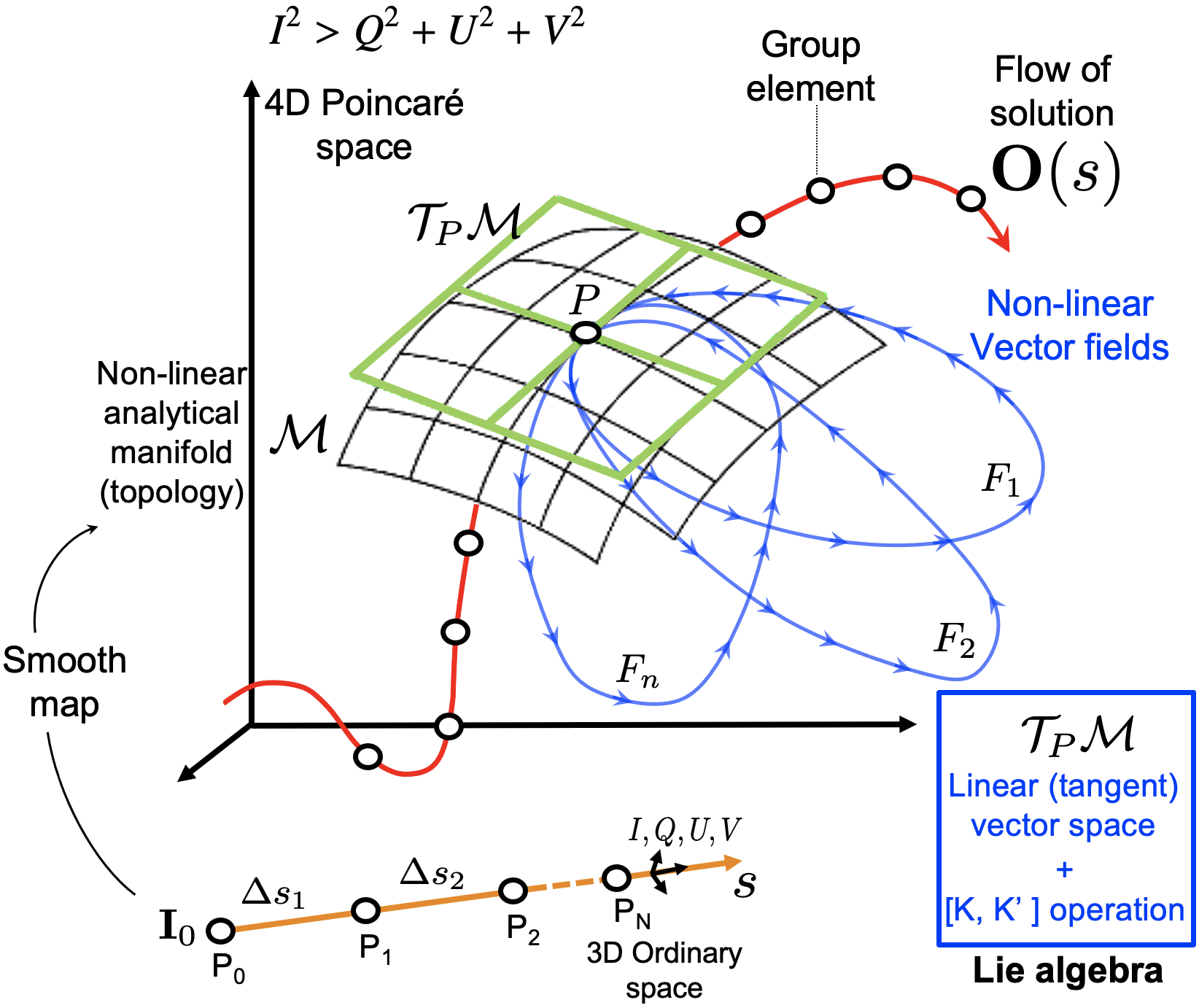}
  \caption{Local homogeneous solutions (white circles) to the RTE evolving in the ordinary 3D space of the ray (orange) and in the 4D space of 
  the Poincar\'e group (red). They result of locally applying 
  the exponential map to linear vector fields (blue) in the Lie algebra.}
  \label{fig:fig1}
\end{figure}

 Our first step is interpreting the evolution of the radiative transfer solution in terms of rotations in a Lie group.
 To start, consider that, for natural light, $I^2 > Q^2+U^2+V^2$,
  %\begin{equation}\label{eq:polconditions}
  % I^2 > Q^2+U^2+V^2  
 %\end{equation}
 which maps a Stokes vector to 
 a point inside the three-dimensional Poincar\'e sphere $Q/I, U/I, V/I$ \citep[e.g.,][]{Shurcliff:2013ve}. Similarly,  
 solutions at each location along the ray are now 
 seen as points forming a differentiable manifold $\mathcal{M}$ in a 4D Poincar\'e space mapping 
 each dimension to a Stokes parameter (Fig. \ref{fig:fig1}).
 In this scenario, the evolution operator $\mathbf{O}(s,s_0;\mathbf{A})
$ is a Lorentz-Poincar\'e group element called
flow of $\mathbf{A}$ because it follows $\mathbf{A}$ to advance the homogeneous solution from $\boldsymbol
{I}(s_0)$ to $\boldsymbol{I}(s)$ in $\mathcal{M}$ (see Eq. (\ref{eq:homog1})). Let us see now this advance as a rotation, $\mathbf{O}(s,s_0;\mathbf{A})
$ being a rotation matrix. Then, with Eq. (\ref{eq:homog_oper}) (or (\ref{eq:peano_baker_int_1})),   
$\mathbf{A}$ is seen as a vector field tangent 
to the s-parametric streamline $\boldsymbol
{I}(s_0) \rightarrow \boldsymbol{I}(s)$. I.e.,  $\mathbf{A}$ is\footnote{Also called \textit{infinitesimal 
generator} of rotations because, as of Eq.(\ref{eq:dyson_taylor_2}), it dominantes the rotation for infinitesimal $\Delta s$ : $\mathbf{O}(s) \approx \mathbbold{1}+ \mathbf{A}\cdot\Delta s$.} the derivative of the flow at $s=s_0$: 
\begin{equation}
\mathbf{A}=-\mathbf{K}=\left. \frac{d \mathbf{O}(s,s_0;\mathbf{A})}{ds}\right|_{s=s_0}
\label{eq:generator}
\end{equation}
This gives a local coordinate map linking $\mathcal{M}$ and its tangent linear space $\mathcal{T}_{\mathcal{P}}$, formed by all possible tangent vector fields $\mathbf{A}$ at the group identity 
$\mathcal{P}$, i.e., where $\mathbf{O}(s,s_0;\mathbf{A})=\mathbbold{1}$ for $s=s_0$. This condition is satisfied when $\mathbf{O}$ is an exponential map. More precisely, the link is between $\mathcal{M}$ 
and its Lie algebra ($\mathfrak{g}$), which is defined as $\mathcal{T}_{\mathcal{P}}$ with the operation of commutation. Thus, at $\mathcal{P}$ the group structure is 
locally captured by its algebra, a simpler object (vector space). 
%When the vector field of a d-dimensional Lie algebra $\mathfrak{g}$ is expressed in terms of basis matrices $\mathbf{B}_i$ ($i=1,\ldots,d$), the algebra 
%is characterized by its structure constants $c_{ijk}$:
%\begin{equation*}\label{eq:structu_constants}
% [\mathbf{B}_i,\mathbf{B}_j]= \sum_kc_{ijk}\mathbf{B}_k      \quad(\in \mathfrak{g})
%\end{equation*}
The mapping also allows solving the problem in $\mathcal{T}_{\mathcal{P}}$ (the Magnus expansion being the full solution).  Indeed, 
commutators of $\mathbf{A}$ among points 
along the ray shall appear in the Magnus expansion to represent the total 
derivative of the vector field $\mathbf{A}$. Such non-linear combination of fields gears a non-local evolution from $\mathfrak
{g}$, as if it was an ocean of small eddies or currents under the large-scale surface of the group. 

Again, further exponentiation converts 
that advance into a resultant rotation connecting two points in the group. Figure~\ref{fig:fig2} illustrates a particular example of this for the RTE in 
Stokes formalism. It also shows the key idea introduced before: to preserve the qualitative properties of the exact 
solution (e.g., the geometric shape of the rotation), we need analytical and numerical solutions respecting the group structure of 
the RTE. A detailed analysis of the propagation matrix in the Lorentz-Poincar\'e group associated to the RTE is presented in Sec. \ref{sec:structure}.
\vspace{-0.18cm}The framework based on rotations and groups 
helps to understand the Magnus expansion and to work with it. It also
suggests that more efficient rotation 
representations may exist for evaluating the expansion. Table \ref{tab:1} below summarizes some algebraic structures for representing rotations in three-dimensional and Minkowski (Poincar\'e) spaces. 

In general, the Jones formalism — which uses tensors and spinorial matrices to quantify polarization and its propagation — offers an alternative to the Stokes formalism. But currently, astrophysical radiative transfer is based on Stokes formalism since the Jones formalism, though equivalent, was shown not to be specially advantageous \citep{Sanchez-Almeida:1992ab}. However, it is possible that expressing the Magnus expansion in 
other \mbox{representations} could yield an advantage (e.g., eliminating 
redundant operations with $4\times 4$ matrices), which suggests a revision of this topic. While rotations are often represented by groups mapping onto (propagation) matrices that act on a vector space, matrices are not even the only 
possible representation. This should be discussed once 
the applicability of the Magnus expansion is demonstrated in Stokes 
formalism.   
\\
%\vspace{6.1cm}  
\begin{table*}\label{tab:1}
 \centering
 \caption{Group representations in ordinary and Minkowsky space. The matrix $\mathbf{J}=\mathrm{diag}\{1,-1,-1,-1\}$ induces reflection at origin in QUV subspace.
% The group elements $\mathbf{U}$ are rotational matrices associated to an axis of rotation obtained by exponentiation of algebra matrices.The axis asociated to the imaginary Pauli matrix $\sigma_3$ has been chosen along the third dimension, which is the direction of the ray.
}
%ROW C-----------------------
\newcommand{\prtCB}[1]{$\mathbb{R}^3$}
\newcommand{\prtCC}[1]{$\vv{r} \in \mathbb{R}^3$}
\newcommand{\prtCD}[1]{$\vv{n} \in \mathbb{R}^3$}
\newcommand{\prtCE}[1]{$\vv{n}=(n_1,n_2,n_3)$}
\newcommand{\prtCF}[1]{$\{(1,0,0),(0,1,0),(0,0,1)\}$}
\newcommand{\prtCG}[1]{$\vv{n}\times \vv{n}'$}
\newcommand{\prtCH}[1]{$\vv{n}$}
%ROW D-----------------------
\newcommand{\prtDB}[1]{SO$(3)$}
\newcommand{\prtDC}[1]{\makecell[cc]{$\mathbf{O}_{3\times 3}(\mathbb{R})$ \\ $\mathbf{O}^T\mathbf{O}=\mathbbold{1}$}}
\newcommand{\prtDD}[1]{\makecell[cc]{$\mathbf{N}\in\mathfrak{so}(3)$\\skew-symmetric}}
\newcommand{\prtDE}[1]{\makecell[cc]{\small$\setstacktabbedgap{1pt} \mathbf{N}  = \parenMatrixstack{0 & -n_3 & n_2 \\ n_3 & 0 & -n_1 \\ -n_2 & n_1 & 0}$\normalsize}}
\newcommand{\prtDF}[1]{\makecell[cc]{\small$\setstacktabbedgap{2pt}
\parenMatrixstack{0 & 0 & 0 \\ 0 & 0 & -1 \\ 0 & 1 & 0},\setstacktabbedgap{2pt}
\parenMatrixstack{0 & 0 & 1 \\ 0 & 0 & 0 \\ -1 & 0 & 0},\setstacktabbedgap{2pt}
\parenMatrixstack{0 & -1 & 0 \\ 1 & 0 & 0 \\ 0 & 0 & 0}$\normalsize}}
\newcommand{\prtDG}[1]{$[\mathbf{N},\mathbf{N}']$}
\newcommand{\prtDH}[1]{$\vv{n}$}
%ROW E-----------------------
\newcommand{\prtEB}[1]{SU$(2)$}
\newcommand{\prtEC}[1]{\makecell[cc]{$\mathbf{O}_{2\times 2}(\mathbb{C})$ \\ $\mathbf{O}^T\mathbf{O}=\mathbbold{1}$ \\  $|\mathbf{O}|=1$} }
\newcommand{\prtED}[1]{\makecell[cc]{$\mathbb{S}\in\mathfrak{su}(2)$\\ $\mathbb{S}=\mathbb{S}^{\dagger},n_k\in\mathbb{R}$\\traceless}}
\newcommand{\prtEE}[1]{\small$\setstacktabbedgap{1pt} \mathbb{S}  = \parenMatrixstack{n_1 & n_2-i n_3  \\ n_2+i n_3 & -n_1 }$\normalsize}
\newcommand{\prtEF}[1]{\makecell[cc]{$\sigma_i=$\small $\left\{ \setstacktabbedgap{2pt}
\parenMatrixstack{ 1 & 0  \\  0 & -1},\setstacktabbedgap{2pt}\parenMatrixstack{ 0 & 1  \\  1 & 0},
\setstacktabbedgap{2pt} \parenMatrixstack{0 & -i  \\ i & 0} \right\} $\normalsize}}
\newcommand{\prtEG}[1]{$[\mathbb{S},\mathbb{S}']$}
\newcommand{\prtEH}[1]{$\vv{n}$}
%ROW F-----------------------
%None
%None
\newcommand{\prtFD}[1]{$\vv{a} \in \mathbb{C}^3$}
\newcommand{\prtFE}[1]{\small$\vv{a}=(a_k)=(\hat{a}_k+i\tilde{a}_k)$ \normalsize}
\newcommand{\prtFF}[1]{\makecell[cc]{$\{(1,0,0),(0,1,0),(0,0,1),$\\$(i,0,0),(0,i,0),(0,0,i)\}$}}
\newcommand{\prtFG}[1]{$\vv{a}\times \vv{a}'$}
\newcommand{\prtFH}[1]{$\vv{a}$}
%ROW G-----------------------
\newcommand{\prtGB}[1]{\makecell[cc]{SO$(1,3)$ \\ (\small Lorentz) \\ (\small Stokes)\normalsize}}
\newcommand{\prtGC}[1]{\makecell[cc]{$\mathbf{O}_{4\times 4}(\mathbb{R})$ \\ $\mathbf{O}^T \mathbf{J} \mathbf{O}= \mathbf{J}$}}
\newcommand{\prtGD}[1]{\makecell[cc]{$\hat{\mathbb{L}}\in\mathfrak{so}(1,3)$\\ $\hat{\mathbb{L}}^{T}\mathbf{J}+\mathbf{J}\hat{\mathbb{L}}=0$}}
\newcommand{\prtGE}[1]{\makecell[cc]{\small$\setstacktabbedgap{1pt}
\hat{\mathbb{L}}  = \parenMatrixstack{0 & \hat{a}_1 & \hat{a}_2 & \hat{a}_3 \\ \hat{a}_1 & 0 & \tilde{a}_3 & -\tilde{a}_2 \\ \hat{a}_2 & -\tilde{a}_3 & 0 & \tilde{a}_1 \\ \hat{a}_3 & \tilde{a}_2 & -\tilde{a}_1 & 0
}$\normalsize}}
\newcommand{\prtGF}[1]{\makecell[cc]{\small Lorentz generators: $\mathbf{\hat{G}}_{i},\mathbf{\tilde{G}}_{i}$\\
Custom base matrices: $\mathbf{B}_i$\\(see text)\normalsize}}
\newcommand{\prtGG}[1]{$[\hat{\mathbb{L}},\hat{\mathbb{L}}']$}
\newcommand{\prtGH}[1]{\makecell[cc]{$\vv{a},\vv{a}^*$\\($\vv{a}_{\pm}$)}}
%ROW H-----------------------
\newcommand{\prtHB}[1]{\makecell[cc]{SL$(2,\mathbb{C})$ \\ (\small Spinorial) \\ (\small Jones)\normalsize}}
\newcommand{\prtHC}[1]{\makecell[cc]{$\mathbf{O}_{2\times 2}(\mathbb{C})$ \\ $\mathbf{O}^T\mathbf{O}=\mathbbold{1}$ \\ $|\mathbf{O}|=1$}}
\newcommand{\prtHD}[1]{\makecell[cc]{$\mathbb{S}\in\mathfrak{sl}(2,\mathbb{C})$\\$\mathbb{S}=\mathbb{S}^{\dagger},a_k\in\mathbb{C}$\\traceless}}
\newcommand{\prtHE}[1]{\small$\setstacktabbedgap{1pt}\mathbb{S}  = \parenMatrixstack{ a_1 & a_2-i a_3 \\ a_2+i a_3 & -a_1 }$\normalsize}
\newcommand{\prtHF}[1]{\makecell[cc]{$\sigma_i=$\small $\left\{ \setstacktabbedgap{2pt}
\parenMatrixstack{ 1 & 0  \\  0 & -1},\setstacktabbedgap{2pt}
\parenMatrixstack{0 & 1  \\  1 & 0},\setstacktabbedgap{2pt}
\parenMatrixstack{0 & -i  \\ i & 0} \right\}  $\normalsize}}
\newcommand{\prtHG}[1]{$[\mathbb{S},\mathbb{S}']$}
\newcommand{\prtHH}[1]{$\vv{a}$}

\resizebox{0.93\textwidth}{!}{
  \begin{tabular}{@{}c|cc|ccccc@{}}
  \toprule
  \multirow{2}{*}{ND} & \multicolumn{2}{c|}{\textbf{Lie Group}} & \multicolumn{5}{c}{\textbf{Lie Algebra}} \\
                      &  Name    &  Elements   & Representation & {Elements} & {Basis} & {Bracket}  &{Axis Rot.}  \\ 
  \midrule
  \multirow{3}{*}{\textbf{3D}} & \prtCB{} &  \prtCC{} & \prtCD{} & \prtCE{} & \prtCF{} & \prtCG{}  & \prtCH{}  \\[2mm]
                      & \prtDB{} &  \prtDC{} & \prtDD{} & \prtDE{} & \prtDF{} & \prtDG{}  & \prtDH{}  \\[2mm]      
                      & \prtEB{} &  \prtEC{} & \prtED{} & \prtEE{} & \prtEF{} & \prtEG{}  & \prtEH{}  \\%[1mm]
  \midrule
  \multirow{3}{*}{\textbf{4D}} &   {}       &   {}        & \prtFD{} & \prtFE{} & \prtFF{} & \prtFG{}  & \prtFH{}  \\[2mm]
                      & \prtGB{} &  \prtGC{} & \prtGD{} & \prtGE{} & \prtGF{} & \prtGG{}  & \prtGH{}  \\[2mm]
                      & \prtHB{} &  \prtHC{} & \prtHD{} & \prtHE{} & \prtHF{} & \prtHG{}  & \prtHH{}  \\ %[1mm]
  \bottomrule
  \end{tabular}
  }
\vspace{4mm} 
\end{table*}

\subsection{Magnus solution: an exact Lie-friendly evolution operator}\label{magnussect}
Instead of solving Eq. (\ref{eq:homog_oper}) in its Lie group with (\ref
{eq:peano_baker_int_1}), \cite{Hausdorff:1906aa} did it in its algebra ($\mathfrak{g}$) by assuming a single exponential for\footnote{As $\mathbf{\Omega}(s_0)=0 \Rightarrow \mathbf{O}(s_0)=\mathbbold{1}$, 
and $\mathbf{O}(s)$ is continuous and invertible 
around $s=s_0$. Then, its inverse is $\mathbf{O}
^{-1}=e^{-\mathbf{\Omega(s)}}$.}
$\mathbf{O}=e^{\mathbf{\Omega}(s)}$.
%$\mathbf{\Omega}(s)\in \mathfrak{g}$. 
From ($\mathbf{A}=-\mathbf{K}$ in our case)
\begin{equation}
  \frac{d\mathbf{O}}{ds}=\mathbf{A} \mathbf{O} 
  \quad\quad \Rightarrow \quad\quad \biggl ( \frac
  {d e^{\mathbf{\Omega}(s)}}{ds} \biggr ) e^{-\mathbf{\Omega}(s)}
  =\mathbf{A},
\end{equation}
he derived the non-linear differential Hausdorff equation for $\mathbf{\Omega}$:
\begin{equation}\label{eq:hausdorff}
\frac{d\mathbf{\Omega}}{ds}= \sum^{\infty}_{n=0}\frac{B_n}{n!}[\mathbf{A},\mathbf{\Omega}^{[n]}],
\end{equation}
where $ [\mathbf{A},\mathbf{\Omega}^{[n]}]=[ \cdots [[\mathbf{A},\overbrace{ \mathbf{\Omega}],\mathbf{\Omega}]\cdots ,\mathbf{\Omega} ]}^{n}$ is an n-times right-nested 
commutator with $[\mathbf{A},\mathbf{\Omega}^{[0]}]=\mathbf{A}$; and the $B_n$ are the Bernoulli numbers\footnote{The first values are $B_n= 1, -1/2, 1/6, 0, -1/30, 0, 1/42, 0, -1/30,\ldots$ with $B_n=0$ for any odd n different than $1$ \citep{Abramowitz:1972}.}.
Remarkably, the series in Eq. (\ref{eq:hausdorff}) is the matricial form of the generating scalar function (see Sec. \ref{subsec:constantprop}):
\begin{equation}\label{eq:generatrix}
  \frac{x}{e^{x}-1}= \sum^{\infty}_{n=0}\frac{B_n}{n!}x^n.
  \end{equation}
The infinite recursive series in Eq. (\ref{eq:hausdorff}) is analytic in all $\mathbb{C}$ except in the points $\mathbb{P} =  \{2\pi n i, n\in \mathbb{Z} >0\}$. 
Iteration on \mbox{Eq.~(\ref{eq:hausdorff})} 
led \cite{Magnus:1954tj} to his expansion $\mathbf{\Omega}(s)=\sum^
{\infty}_{n=1}\mathbf{\Omega}_n(s)$:
%\vspace{-0.4cm}
\begin{equation}  
  \vspace{-0.4cm}
  \begin{split}
   \vspace{-0.4cm} 
   & \mathbf{\Omega}_{1}(s) = \int^s_{s_0}d_1 \,\mathbf{A}_1,\\
   & \nonumber\mathbf{\Omega}_{2}(s) = -\frac{1}{2}\int^s_{s_0} \biggl[ 
    \int^{s_1}_{s_0}d_2 \,\mathbf{A}_2, \mathbf{A}_1 
    \biggr]d_1,\\
    & \mathbf{\Omega}_{3}(s) = \frac{1}{4}\int^s_{s_0} \biggl[\int^
    {s_1}_{s_0} \biggl[ \int^{s_2}_{s_0} \mathbf{A}_3d_3, 
    \mathbf{A}_2 \biggr]d_2 , \mathbf{A}_1 \biggr]d_1,\\
    & \mathbf{\Omega}_{4}(s) = \frac{1}{12}\int^s_{s_0} \biggl[\int^
    {s_1}_{s_0} \mathbf{A}_2 d_2, \biggl[ \int^{s_1}_
    {s_0} \mathbf{A}_2d_2, \mathbf{A}_1 \biggr]\biggr] 
    d_1,\\
    &\cdots
  \end{split}   
  \end{equation}
  \vspace{-0.1cm}
  \begin{equation}\label{eq:Magnus0} 
    \begin{split}
      &\cdots\\ 
      & \mathbf{\Omega}_{5}(s) = -\frac{1}{8}\int^s_{s_0} \biggl[\int^
      {s_1}_{s_0} \biggl[ \int^
      {s_2}_{s_0} \biggl[ \int^{s_3}_{s_0} \,\mathbf{A}
      _4d_4, \mathbf{A}_3 \biggr] d_3, \mathbf{A}_2 \biggr]
      d_2 , \mathbf{A}
        _1 \biggr]d_1,\\      
     & \mathbf{\Omega}_{6}(s) = -\frac{1}{24}\int^s_{s_0} \biggl[\int^
     {s_1}_{s_0} \mathbf{A}_2 d_2, \biggl[ \int^{s_1}_
     {s_0} \biggl[ \int^{s_2}_{s_0} \mathbf{A}_3d_3, 
     \mathbf{A}_2 \biggr]d_2, \mathbf{A}_1 \biggr]\biggr] d_1,\\
     & \mathbf{\Omega}_{7}(s) = -\frac{1}{24}\int^s_{s_0} \biggl[\int^
     {s_1}_{s_0} \biggl[ \int^{s_2}_{s_0} \mathbf{A}_3d_3, 
     \mathbf{A}_2 \biggr]d_2 , \biggl[ 
      \int^{s_1}_{s_0}d_2 \,\mathbf{A}_2, \mathbf{A}_1 \biggr] \biggr]d_1,\\
      & \mathbf{\Omega}_{8}(s) = -\frac{1}{24}\int^s_{s_0} \biggl
     [\int^{s_1}_{s_0} \biggl[ \int^
     {s_2}_{s_0} \mathbf{A}_3d_3, \biggl[ \int^
     {s_2}_{s_0} \mathbf{A}_3d_3, \mathbf{A}_2 \biggr] 
     \biggr]d_2 , \mathbf{A}
     _1 \biggr]d_1,\\
     &\cdots
    \end{split}     
\end{equation}
with $\mathbf{\Omega}(0)=\mathbf{0}$, $d_i=ds_i$ and $\mathbf{A}_i=\mathbf{A}(s_i)$. 
The series converges in the neighbourhood of 
$\exp{(\mathbf{\Omega}(s_0))}$ if the differences between any two eigenvalues of $\mathbf{\Omega}(s_0)$ is not in 
$\mathbb{P}$. We will come back to this in the numerical sequel of this 
paper. Now the first thing to do is to identify the 
level of importance of each term and their characteristics. Let us note that for $n>2$, more than one n-term $\mathbf{\Omega}_n$ is necessary to fully account for a certain convergence order $m$ of 
the expansion\footnote{There are three kinds of order: the iteration order $n$, which just label iteration terms arising 
during the obtention of the expansion; the Magnus convergence order $m$, formed by the $n$ terms sharing the highest power of the matrix in them; and the order of the numerical methods used to solve the integrals.}. Any such n-term belonging to $m$ order contains: $m$ 
propagation matrices, a multivariate integral with $m$ 
nested integrals, and $m-1$ nested commutators. Namely, grouping $n$-terms in $m$-terms: $\mathbf{\Omega}(s)=\sum^
{\infty}_{m=1}\mathbf{\Omega}^{[m]}(s)$, with  $\mathbf{\Omega}^{[1]}=\mathbf{\Omega}_1$,  $\mathbf{\Omega}^{[2]}=\mathbf{\Omega}_2$, $\mathbf{\Omega}^{[3]}=\mathbf{\Omega}_3+\mathbf{\Omega}_4$, $\mathbf{\Omega}^{[4]}=\mathbf{\Omega}_5+\mathbf{\Omega}_6+\mathbf{\Omega}_7+\mathbf{\Omega}_8$. 
Hence, 
$\mathbf{\Omega}_{1}$ and $\mathbf{\Omega}_{2}$ are special, not only being the simplest terms, but also the only ones requiring only one term to fully quantify their respectives orders, a property that remains invariant irrespective of any formulation of the expansion.

The properties of the elements of the Magnus expansion allow for alternative formulations. For instance, setting $s_0=0$ for simplicity, we find that all the eight terms shown in Eqs.~(\ref{eq:Magnus0}) can be substituted and 
combined by order of convergence in this neat recursive way:
% \begin{equation}\label{eq:Magnus0a}
%   \begin{split}
%     \mathbf{\Omega}^{[1]}(s) &= \int^s_{0}dt \,\mathbf{A}(t),\\
%     \mathbf{\Omega}^{[2]}(s) &=  \frac{1}{2}\int^s_{0}\biggl[\mathbf{A}(t),\mathbf{\Omega}^{[1]}(t) \biggr]dt ,\\
%     \mathbf{\Omega}^{[3]}(s) &=  \frac{1}{2}\int^s_{0}\biggr( \bigl[\mathbf{A}(t),\mathbf{\Omega}^{[2]}(t) \bigr] +\frac{1}{6}\bigl[\bigl[\mathbf{A}(t),\mathbf{\Omega}^{[1]}(t) \bigr],\mathbf{\Omega}^{[1]}(t) \bigr]\biggr)dt,\\
%     \mathbf{\Omega}^{[4]}(s) &= \frac{1}{2}\int^s_{0} \biggr( \biggl[\mathbf{A}(t),\mathbf{\Omega}^{[3]}(t) \biggr]+\bigl[\bigl[\mathbf{A}(t),\mathbf{\Omega}^{[1]}(t) \bigr],\mathbf{\Omega}^{[2]}(t) \bigr] +\\
%    &\quad\quad\quad\quad\quad\quad\quad\quad\quad+\bigl[\bigl[\mathbf{A}(t),\mathbf{\Omega}^{[2]}(t) \bigr],\mathbf{\Omega}^{[1]}(t) \bigr] \biggr)dt.
%    \end{split} 
%   \end{equation}
\begin{equation}\label{eq:Magnus0a}
  \begin{split}
    \mathbf{\Omega}^{[1]}(s) &= \int^s_{0}dt \,\mathbf{A}(t),\\
    \mathbf{\Omega}^{[2]}(s) &= \int^s_{0} \frac{1}{2}\biggl[\mathbf{A}(t),\mathbf{\Omega}^{[1]}(t) \biggr]dt ,\\
    \mathbf{\Omega}^{[3]}(s) &= \int^s_{0} \biggl(\frac{1}{2}\biggl[\mathbf{A}(t),\mathbf{\Omega}^{[2]}(t) \biggr] +\frac{1}{6}\biggl[\frac{d\mathbf{\Omega}^{[2]}}{dt},\mathbf{\Omega}^{[1]}(t) \biggr] \biggr)dt,\\
    \mathbf{\Omega}^{[4]}(s) &= \int^s_{0} \biggl( \frac{1}{2} \biggl[\mathbf{A}(t),\mathbf{\Omega}^{[3]}(t) \biggr]+\frac{1}{6}\biggl[\frac{d\mathbf{\Omega}^{[2]}}{dt},\mathbf{\Omega}^{[2]}(t) \biggr] +\\
    &\quad\quad+\frac{1}{2}\bigl[\bigl[\mathbf{A}(t),\mathbf{\Omega}^{[2]}(t) \bigr],\mathbf{\Omega}^{[1]}(t) \bigr] \biggr)\,dt,
%   &+\frac{1}{6}\biggl[\frac{d\mathbf{\Omega}^{[3]}}{dt}-\biggl[\frac{d\mathbf{\Omega}^{[2]}}{dt},\mathbf{\Omega}^{[1]}(t) \biggr],\mathbf{\Omega}^{[1]}(t) \biggr] \biggr)dt,\\
   \end{split} 
  \end{equation}

    % \begin{equation}\label{eq:Magnus0b}
    %   \begin{split}
    %    & \Omega_{1}(s) = \int^s_{s_0}d_1 \,\mathbf{A}(s_1),\\
    %    & \Omega_{2}(s) = -\frac{1}{2}\int^s_{s_0} \biggl[ \Omega_{1}(s_1), \mathbf{A}(s_1) \biggr]ds_1 ,\\
    %    & \Omega_{3}(s) = -\frac{1}{2}\int^s_{s_0} \biggl[\Omega_{2}(s_1), \mathbf{A}(s_1) \biggr]ds_1,\\
    %    & \Omega_{4}(s) = \frac{1}{12}\int^s_{s_0} \biggl[\Omega_{1}(s_1), \biggl[ \Omega_{1}(s_1), \mathbf{K}'(s_1) \biggr]\biggr] ds_1,\\
    %    & \Omega_{5}(s) = -\frac{1}{2}\int^s_{s_0} \biggl[\Omega_{3}(s_1) , \mathbf{A}(s_1) \biggr]ds_1,\\
    %  & \Omega_{6}(s) = -\frac{1}{2}\int^s_{s_0} \biggl[\Omega_{4}(s_1) , \mathbf{A}(s_1) \biggr]ds_1,\\
    %   & \hdots\\
    %    \end{split} 
    %   \end{equation}
  Here, one could also use $d\mathbf{\Omega}^{[2]}/dt=\frac{1}{2}\bigl[\mathbf{A}(t),\mathbf{\Omega}^{[1]}(t) \bigr]$. 
    Both Eq.(\ref{eq:Magnus0a}) and (\ref{eq:Magnus0}) point out that the Magnus expansion is a peculiar object that, as a result of the Hausdorff iteration, exhibits a formal fractal character. Namely, its terms can be written as 
    functions of lower-order terms changing at smaller integration scales\footnote{The fractal nature of this Magnus expansion can be seen in Eq. (\ref{eq:Magnus0}), where every new term can be obtained by substituting the 
    $\mathbf{A}$ of previous terms by the simplest elemental 
    commutator in $\Omega_{2}$.}. Despite being formal, it arises our interest in investigating the Magnus expansion as a fractal object.

By the properties of integrals, the Magnus expansion 
accomplishes $\mathbf{\Omega}
(s_2,s_1)+\mathbf{\Omega}(s_1,s_0)=\mathbf{\Omega}(s_2,s_0)$ and $\mathbf{\Omega}
(s,s_0)=-\mathbf{\Omega}(s_0,s)$. Hence, the Magnus evolution operator allows serialization and a symmetric homogeneous evolution:
\begin{subequations}
  \begin{align}
    &{\mathbf{O}}(s_2,s_0)= {\mathbf{O}}(s_2,s_1) {\mathbf{O}}(s_1,s_0),\label{eq:op2}\\
    &{\mathbf{O}}(s_0, s_0)=\mathbbold{1}={\mathbf{O}}(s,s_0)
  {\mathbf{O}^ {-1}}(s,s_0)\\
   &{\mathbf{O}^{-1}}(s,s_0)={\mathbf{O}}(s_0,s).\label{eq:op1}
  \end{align}
  \end{subequations}
  Contrary to truncating in the group with  Dyson series Eq.(\ref {eq:peano_baker_int_1}),
  Magnus' always stays in the algebra after truncation \citep[][]{Blanes:2009wr}. This allows to approximate $\mathbf{\Omega}$  while
  keeping the physical properties of the true solution. In this sense, the Magnus expansion is intrinsically accurate. However, 
  for this to fully apply, the solution must also stay in the group when exponentiating, hence the exponential of the Magnus expansion
  must be exact. For the RTE, the exponent contains Eqs. (\ref{eq:Magnus0}) with $\mathbf{A}(s)=-\mathbf{K}(s)=-\eta_0(s)\mathbbold{1}-\hat{\mathbb{L}}(s)$, and $\hat{\mathbb{L}}(s)$ the Lorentz matrix (see Table \ref{tab:1}, details in Sec. \ref{sec:structure}). And as $\mathbbold{1}$ commutes with all, the term in
$\eta_0$ cancels out in all commutators and survives in $\mathbf{\Omega}_1$ as the optical depth (accumulated opacity):
\begin{equation}\label{eq:tau}
  \tau=\int^{s}_{s_0}\eta_0(t)dt,
\end{equation}  
which being scalar can always be factored out as:
\begin{equation}\label{eq:high_magnus1}
\rm{\mathbf{O}}(s;\mathbf{K})= e^{-\tau}e^{\mathbf{\Omega}(\hat{\mathbb{L}})},
\end{equation}
where now $\mathbf{\Omega}$ contains Eqs.~(\ref{eq:Magnus0}) for $\mathbf{A}=-\hat{\mathbb{L}}$. Thus, an exact exponential in terms of Lorentz matrices is derived in Sec.~\ref{sec:homog}.
%Comment on the sign of the second coefficient, the handedness/sign of the commutators, write the first Bernoulli numbers, and check signs with MAgnus INcredible

\subsection{The Basic Evolution (BE) operator}\label{sec:bet}
The simplest evolution operator results of truncating\footnote{This is not the same 
as assuming constant propagation matrices, which 
would also cancel higher-order terms but reducing 
the first term to the common oversimplification given by Eq.(\ref{eq:Kexp_end}) } 
 the Magnus expansion to keep only its first 
term, which only contains the integral of $\mathbf{A}$. This is better than using Eq.(\ref{eq:Kexp_end}) as evolution operator, because the 
integral consistently preserves 
memory of the variation of 
the propagation matrix along the ray. Also note that all higher-order terms start by the same 
integral, so the whole expansion can be understood as a single integral of $\mathbf{A}$ plus corrections to $\mathbf{A}$. Then, let us now define a basic evolution operator as 
the matrix exponential \textit{of the integral of a 
propagation matrix}. 

Consider 
%the exponential of the generic integral $\int ds  \vv{n}\cdot\vv{\sigma}$, with 
a matrix $\mathbf{M}=f \vv{a}\cdot\vv{\mathbf{B}}$ that can 
be decomposed in terms of a constant $f$, a vector $\vv{a}=(a_1,a_2,a_3)$ with module $a$, and a vector $\vv{\mathbf{B}}=(\mathbf{B}_1,\mathbf{B}_2,\mathbf{B}_3)$ of 
basis matrices in algebra $\mathfrak{g}$:
\begin{equation}\label{eq:condition_theorem}
    \sigma_i\cdot\sigma_j = \delta_{ij}\mathbbold{1} + c\cdot\epsilon_{ijk}\cdot\sigma_k, \quad\quad(c= \mathrm{constant})
\end{equation}
with $\epsilon_{ijk}$ the Levi-Civita permutation symbol\footnote{It is $+1/-1$ if the number of permutations of $(i,j,k)$ to obtain $(1,2,3)$ is odd/even, or 0 if two indices are equal.}.
 Then, the exact exponential of the integral of $\mathbf{M}$ is (Appendix~\ref{sec:oursolution}):  
\begin{equation}\label{eq:evoltheorem_1}
  e^{\pm\int ds \,\vv{a}(s)\cdot\vv{\mathbf{B}}}=\mathrm{ch}{(b)}\mathbbold{1}\pm \mathrm{sh}{(b)}\vv{u}\cdot\vv{\mathbf{B}},
\end{equation}  
with $\vv{u}$ a unitary vector resulting of integrating $\vv{a}$:
\begin{subequations}\label{eq:evoltheorem_2}
  \begin{align}
  \vv{u}&=\frac{\vv{b}}{b}=\frac{(b_1,b_2,b_3)}{[b^2_1+b^2_2+b^2_3]^{1/2}}\\
  b_k&=\int ds \, a_k(s)
  \end{align}
\end{subequations}
Interpreting this in terms of rotations, we identify Eq. (\ref{eq:evoltheorem_1}) as a rotor, an element of Clifford 
geometric algebra \citep{Hestenes:2003aa} representing here a three-dimensional rotation around the axis $\vv{u}$. 

The easiest example is to particularize to SU$(2)$ representation, whose algebra basis are $2\times 2$ Pauli matrices (table~\ref{tab:1}): 
\begin{subequations}\label{exponentiation_1}
  \begin{align}
  & e^{\int ds \vv{a}\cdot\vv{\sigma}}= \mathrm{ch} (\theta) \mathbbold{1} + \mathrm{sh} (\theta)\vv{u}\cdot\vv{\sigma}=\\
  =& \left(\begin{array}{cc}
    \mathrm{ch}(\theta)+\mathrm{sh}(\theta)\,u_1 &  \mathrm{sh}(\theta)\,(u_2-i \,u_3) \\ 
    \mathrm{sh}(\theta)\,(u_2+i \,u_3) & \mathrm{ch}(\theta)-\mathrm{sh}(\theta)\,u_1
  \end{array} \right) \quad \in \mathrm{SU}(2).
  \end{align}
\end{subequations}
This matrix describes an evolution as hyperbolic rotations by an angle $\theta\equiv b$ in a plane 
 with normal unit vector $\vv{u} \in \mathbb{R}
 ^3$. Particularizing to $\mathbf{M}$ constant, $b_k=s\cdot a_k$, $b=s\cdot a$, and $\vv{u}=\vv{a}/a$, which gives the rotation used in quantum mechanics to quantify the spin operator. It is composed of three independent two-dimensional rotations, one per plane perpendicular to a cartesian basis axes\footnote{Pure rotations around such axes would be ($\theta_k=s\cdot n_k$):
\begin{subequations}
  \begin{align*}      
  e^{\theta_1}= \left(
  \begin{array}{cc}
    e^{\theta_1} &  0 \\
    0 & e^{-\theta_1}
  \end{array} \right);
  e^{\theta_2}= \left(
  \begin{array}{cc}
    \mathrm{ch}\theta_2 &  \mathrm{sh}\theta_2 \\ 
    \mathrm{sh}\theta_2 & \mathrm{ch}\theta_2
  \end{array} \right);
  e^{\theta_3}=\left(
  \begin{array}{cc}
   \mathrm{ch}\theta_3 &  -i \mathrm{sh}\theta_3 \\ 
    i\mathrm{sh}\theta_3 & \mathrm{ch}\theta_3
  \end{array} \right)      
\end{align*}
\end{subequations}
 }.
In the case of the 
polarization, the role of the axis of rotation is played by  the propagation vector, which is defined in the next section and contains all the physical information of the 
propagation matrix that is necessary to rotate a vector in the QUV Poincar\'e space. 
\begin{figure*}[t!]
  \centering$
  \begin{array}{cc} %[scale=0.3] [width=2.0in] --> options to control size
  \includegraphics[scale=0.28]{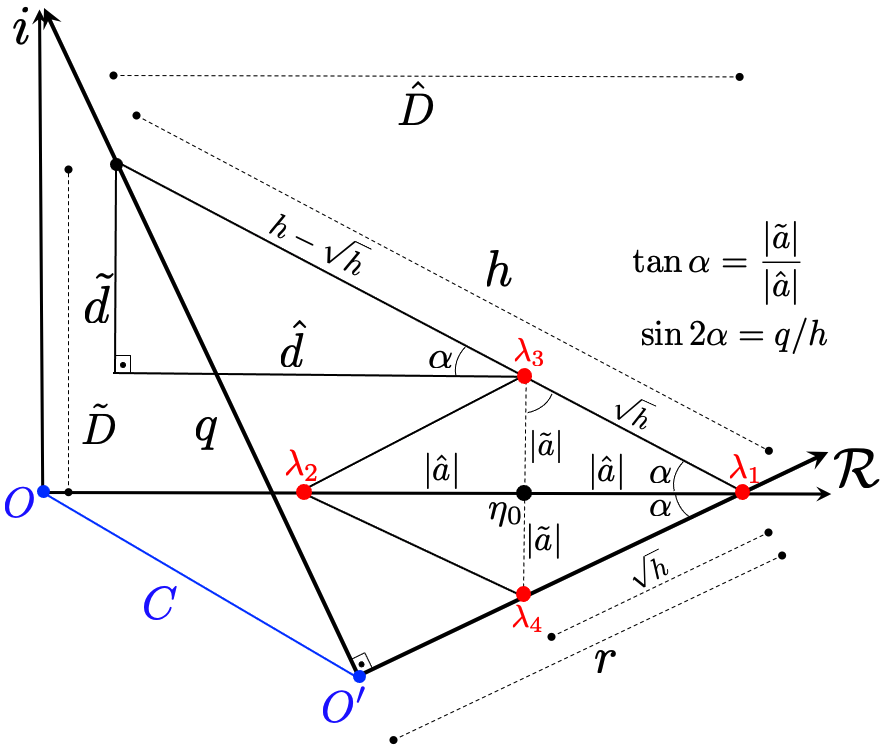} & \includegraphics[scale=0.28]{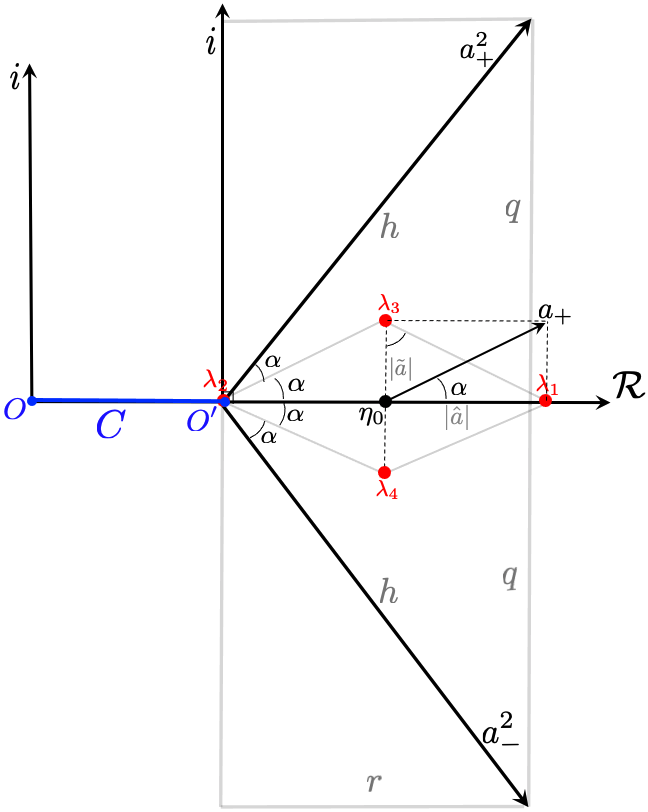} \\
  \end{array}$
  \caption{
    Geometry associated to $a_{\pm}$ and $a^2_{\pm}$. \textbf{Left panel}: steps deducing triangle r-q-h with a corner at $\lambda_1$. (1) As $|a_+|=\sqrt{h}$ is the side the romboid, $\vec{a}_{+}$ can be set from $\eta_0$ to the point $\lambda_1 +i\lambda_3$ (as if its origin  were at $\eta_0$); (2)
    Obtain $\alpha$ through the area of the romboid $A_{\diamondsuit}=2|\hat{a}||\tilde{a}|=|q|=h\sin(2\alpha)$; (3) As $h^2=q^2+r^2$ and $\sin(2\alpha)=q/h$, then h, q, and r can be associated with sides of a rectangular triangle with a corner at $\lambda_1$; (4) Identify auxiliary triangles with sides $\hat{d},\tilde{d}$ and $\hat{D},\tilde{D}$, finding: $\hat{d}+i\tilde{d}=(\sqrt{h}-1)\cdot(|\hat{a}|+i|\tilde{a}|)$ and $\hat{D}+i\tilde{D} =\sqrt{h}\cdot(|\hat{a}|+i|\tilde{a}|)$. 5.- Finally, we identify a vector $a^2_{\pm}=r+iq$ enclosed in the diagonal of the rectangle of sides r-q-h with origin in $O'=\hat{O}'+i\tilde{O}'=\lambda_1-re^{i\alpha}$. The distance $C=OO'$ is then: $C = [(\hat{O}')^2+(\tilde{O}')^2]^
    {1/2}=[\lambda^2_1-2r\cos(\alpha)\lambda_1+r^2]^{1/2}$.
    \textbf{Right panel}: $a^2_+$ and $a^2_-$ when their origin $O'$ is chosen at $\lambda_2$.}\label{fig:fig3}
  \end{figure*}  
\section{Algebraic analysis of the propagation vector and the propagation matrix}

\subsection{The propagation vector}\label{sec:prop_vector}
For convenience, we associate $Q,U,V$ with subindices $1,2,3$; and we define the propagation vector $\vv{a}$ and its complex conjugate with $+$ and $-$ symbols, respectively, as follows:
\begin{equation}\label{eq:algebra_aes_1}
  \vv{a}_{\pm} = \vv{\eta}\pm i\vv{\rho}=(\eta_1\pm i\rho_1,\eta_2\pm i\rho_2,\eta_3\pm i\rho_3).
\end{equation}
The module $a\in \mathbb{R}$ of both complex vectors $\vv{a}_+$ and $\vv{a}_-$ has:
\begin{subequations}
  \begin{align}\label{eq:algebra_aes_power}
    a^2 &=\vv{a}_{+}\cdot \vv{a}_{-}=\sum_k a_k a^*_k=\sum_k |a_k|^2=\eta^2+\rho^2,\\
    a^4&=(\eta^2-\rho^2)^2+(2\eta\rho)^2,
    \end{align}
\end{subequations}
where $\eta^2=\vv{\eta}^2$ and $\rho^2=\vv{\rho}^2$.
However, the vector modules of $\vv{a}_{+}$,$\vv{a}_{-}$ are complex numbers $a_+$ and $a_-$, fulfilling (see App. \ref{sec:acalc})
\begin{equation}\label{eq:algebra_aes_2}
 a^2_{\pm}=\vv{a}_{\pm}\cdot\vv{a}_{\pm}=r\pm i q \quad\Rightarrow \quad a_{\pm} =\hat{a}\pm i\tilde{a},
\end{equation}
whose solutions are
\begin{subequations}\label{eq:algebra_mns_all}
    \begin{align}\label{eq:algebra_mns}
      \hat{a}&=\hat{\sigma}\left(\frac{h + r}{2} \right) ^{1/2},\quad\tilde{a}=\tilde{\sigma}\left(\frac{h-r}{2}\right)^{1/2}\\
      h &=[r^2+q^2]^{1/2}= a_- \cdot a_+ = \hat{a}^2 +\tilde{a}^2 \quad(\in \mathbb{R})\label{eq:algebra_mns_b}\\
      r &= \hat{a}^2-\tilde{a}^2 = \eta^2-\rho^2\\
      q&=2\hat{a}\tilde{a}= 2(\vv{\eta}\cdot\vv{\rho})=2\eta\rho\cos(\theta)\\ 
      \sigma&=\hat{\sigma}\tilde{\sigma}=\text{sign}(\vv{\eta}\cdot \vv{\rho})\label{eq:algebra_mns_d}
      \end{align}
\end{subequations}
% \begin{subequations}
  %  \begin{align}\label{eq:algebra_aes_2}
  %     \theta_+&=[m^2+n^2]^{1/2} \quad \quad \theta_-= m\\
  %     \hat{a}&=\frac{1}{2}(\theta_+ + \theta_-)\quad\quad \tilde{a}=\frac{\sigma}{2}(\theta_+ - \theta_-)\\
  %     a_{\pm} &=\hat{a}\pm i \tilde{a}=\frac{\theta_+}{2}(1+i\sigma)+\frac{\theta_-}{2}(1-i\sigma)\\
  %     a^2_{\pm}&=\frac{m\pm i n}{2}
  %     \end{align}
  %   \end{subequations}
\begin{figure}[t!]
  \centering%$
    %\begin{array}{cc} %[scale=0.3] [width=2.0in] --> options to control size
      \includegraphics[scale=0.85]{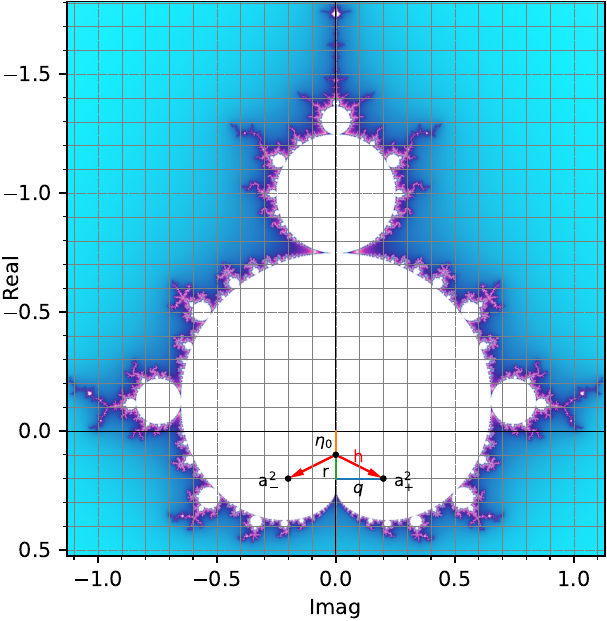} 
      \caption{Mandelbrot fractal arising when iterating the square of the propagation vector $C+a^2_{\pm}$ for different values of C. The iteration only converges in the white region. Two vectors for $C=\eta_0$ are represented.}
  \label{fig:fig4}
\end{figure}    
Here the signs $\hat{\sigma}$ and $\tilde{\sigma}$ are only constrained to Eq. (\ref{eq:algebra_mns_d}) (see Appendix \ref{sec:acalc}). An additional relation is $\hat{a}^4+\tilde{a}^4=(h^2+r^2)/2$.
Note that the real and imaginary parts of $a_{\pm}$ define the four eigenvalues of the propagation matrix $\lambda_{1,2,3,4}= \{\,\,\eta_0\pm |\hat{a}|\,\,,\,\,\eta_0\pm i|\tilde{a}| \,\, \}$,
such that they form an equilateral romboid of side $\sqrt{h}$ in the complex plane. This and the above algebraic relationships can be translated into the geometry of Fig. (\ref{fig:fig3}). As explained there, the distances between the 
eigenvalues of $\mathbf{K}$ can be related with the propagation vector and its square, given by $r$ and $q$ in Eq. (\ref
{eq:algebra_aes_2}).
By choosing a reference 
system defined by the distance $C=\overline{OO'}$ and inclined with respect to the original real axis 
by an angle $\alpha$, the projections $r$ and $q$ of $a^2_{+}$ can be made coincident with $\lambda_1$ and be related to the romboid sides. 
Once the relative size and inclination between $r$, $q$, and 
$h$ has been obtained, we could arbitrarily move and rotate their 
triangle to a different origin $O'$, as shown in the 
second panel of Fig.(\ref{fig:fig3}).  Interestingly, as $r$ and $q$ compose the squared complex number $a^2_{\pm}$, they form the Mandelbrot fractal \citep{Mandelbrot:1982}
when $(a^2_{\pm})_{n+1}=(a^2_{\pm})_n+C$ is iterated for all 
values of $C$ in the complex plane (see Fig. \ref{fig:fig4}). We suspect a possible application of the Mandelbrot fractal to analyze some aspects of the 
Magnus expansion (e.g., its convergence properties, or  relations among its terms, rotations, and the powers of $a_{\pm}$).

\subsection{Algebraic characterization of the propagation matrix}\label{sec:structure}
This section characterizes the Lie group and algebra of the RTE, which is useful to operate with the Magnus expansion. Readers uninterested in these technical details may skip to the next section; relevant results will be referenced as needed.

%,which corresponds to a Poincar\'e group plus dilatations. 
~\citetalias{Semel:1999aa} and~\citetalias
{Lopez-Ariste:1999aa} showed that the Stokes 4-vector can be seen as an element of a Minkowski-like $\mathbb{R}^{1,3}$
space, with $I$ as the temporal dimension.
%with a metric having the norm\footnote{In the solar atmosphere this norm is close to one,hence positive and timelike, because intensity and all its
%related quantities ($I$,$\eta_I$,$\rho_I$) are normally much
%larger than the polarization counterparts when the light is only partially polarized.} $\parallel \mathbf
%I\parallel \leq I^2-Q^2-U^2-V^2$.
Thus, the RTE describes infinitesimal transformations within
the ``Stokes cone of light'', categorized into 11 types forming a Poincar\'e group (the inhomogeneous generalization of a Lorentz group) allowing dilatations. The inhomogeneous term ($\boldsymbol{\epsilon}$) in the RTE accounts for four-dimensional translations, while dilatations appear from the diagonal of the propagation matrix in Eq. (\ref{eq:basicrte}):
 \begin{equation}\label{eq:lorentz1}
  \mathbf{K}(s)=\eta_0(s)\mathbbold{1}+\hat{\mathbb{L}}(s),
\end{equation} 
 which reduces/amplifies the number of reemitted photons through indiscriminate positive/negative absorption of photons in any polarization states as of $\rm
 \eta_0=\eta^{(A)}_0-\eta^{(S)}_0$ is dominated by ordinary absorption
 ($\eta_0^{(A)}$) as in the sun, or by stimulated emission ($\eta_0^{(S)}$). The remaining (homogeneous) part of the Poincar\'e group is the Lorentz group\footnote{As $I\geq0$, the group is SO$^+(1,3)$, only the "future cone of light" matters.} SO($1,3$), whose 
elements $\mathbf{O}$ can be expressed as exponentials of the Lorentz matrix $\hat{\mathbb{L}}$, accomplishing (Table \ref{tab:1}; $\boldsymbol{J}=\mathrm{diag}(1,-1,-1,-1)$):
  \begin{subequations}\label{eq:lorentz3}
    \begin{align}
    &\mathbf{O}^{T}\mathbf{JO}=\boldsymbol{J},\\
    &\hat{\mathbb{L}}^{T}\boldsymbol{J}+\boldsymbol{J}\hat{\mathbb{L}}=0.
  \end{align}
  \end{subequations}
In turn, $\hat{\mathbb{L}}$
%  The latter
%  is relevant when the intensity of the radiation is large enough for
%  producing an inversion of population between the lower and upper
%  levels of a given transition, which is typically assumed as unlikely
%  in ordinary solar conditions. 
is composed of the 6 infinitesimal
generators of rotations of the Lorentz group, i.e. the three $4\times4$
matrices $\tilde{\mathbf{G}}_1, \tilde{\mathbf{G}}_2, \tilde{\mathbf{G}}_3$ generating ordinary 3D rotations (anomalous dispersion) in the QUV space, 
plus the three $4\times4$
matrices $\hat{\mathbf{G}}_1, \hat{\mathbf{G}}_2, \hat{\mathbf{G}}_3$ producing
hyperbolic rotations (i.e., involving the first dimension) by dichroism (Lorentz boosts in relativity). 

Our characterization of the 
propagation matrix is based on analyzing $\hat{\mathbb{L}}$ in four algebraic levels decomposing it
in: 1) individual Lorentz generators; 2) 
base matrices; 3) hyperbolic and circular 
subspaces; and 4) complex Lorentz matrices.
In the first level, the Lorentz generators are 
encoded in $\hat{\mathbb{L}}$ as\footnote{
  %López Ariste and Semel called these matrices $\rm \mathbf{H}_Q,\mathbf{H}_U,\mathbf{H}_V$ and $\rm \mathbf{R}_Q,\mathbf{R}_U,\mathbf{R}_V$, respectively. 
  Our notation suggests they are real and imaginary parts of a same matrix.}:
\begin{equation}\label{eq:generators}
\begin{split}
% =\mathbf{H}_1^{\myrect{0.7}{$\phantom{1}$}}+\mathbf{H}_2^{\mydiam{0.7}{$\phantom{1}$}}+&\mathbf{H}_3^{\myrectr{0.8}{$\phantom{1}$}}+ \mathbbold{1}+\mathbf{R}_1^{\mytri{0}{0.7}{$\phantom{1}$}}
% +\mathbf{R}_2^{\mytri{45}{0.7}{$\phantom{1}$}}+\mathbf{R}_3^{\mytri{90}{0.7}{$\phantom{1}$}} 
\hat{\mathbf{G}}_{1}{\myrect{0.5}{$\phantom{1}$}}&+\hat{\mathbf{G}}_{2}{\mydiam{0.5}{$\phantom{1}$}}+\hat{\mathbf{G}}_{3}{\myrectr{0.7}{$\phantom{1}$}}+\tilde{\mathbf{G}}_{1}{\mytri{0}{0.6}{$\phantom{1}$}}
+\tilde{\mathbf{G}}_{2}{\mytri{45}{0.6}{$\phantom{1}$}}+\tilde{\mathbf{G}}_{3}{\mytri{90}{0.6}{$\phantom{1}$}} =\\
&=\footnotesize \left(
\begin{array}{cccc}
\mycircle{1}{$0$} & \myrect{1}{$1$} & \mydiam {1}{$1$} & \myrectr{1}{$1$}\\
 \myrect{1}{$1$} & \mycircle {1}{$0$} & \mytri{90}{1.1}{$1$} & \mytrin{45}{1.1}{$-1$}   \\
 \mydiam{1}{$1$} & \mytrin{90}{1.1}{$-1$} & \mycircle{1}{$0$} & \mytri{0}{1.1}{$1$}   \\
\myrectr{1}{$1$} & \mytri{45}{1.1}{$1$} & \mytrin{0}{1.1}{$-1$} & \mycircle{1}{$0$}  
\end{array}
\right) \normalsize.%=\\
%&\mathbf{H}_Q^{\myrect{0.7}{$\phantom{1}$}}+\mathbf{H}_U^{\mydiam{0.7}{$\phantom{1}$}}+\mathbf{H}_V^{\myrectr{0.8}{$\phantom{1}$}}+ \mathbf{D}_I^{\mycircle{0.7}{$\phantom{1}$}}+\mathbf{R}_Q^{\mytri{0}{0.7}{$\phantom{1}$}}
%+\mathbf{R}_U^{\mytri{45}{0.7}{$\phantom{1}$}}+\mathbf{R}_V^{\mytri{90}{0.7}{$\phantom{1}$}}
\end{split}
\end{equation}
We also introduce a set of \textit{dual} generators:
\begin{equation}\label{eq:dual_generators}
  \begin{split} 
  \hat{\mathbf{D}}_{1}{\myrect{0.5}{$\phantom{1}$}}&+\hat{\mathbf{D}}_{2}{\mydiam{0.5}{$\phantom{1}$}}+\hat{\mathbf{D}}_{3}{\myrectr{0.7}{$\phantom{1}$}}+\tilde{\mathbf{D}}_{1}{\mytri{0}{0.6}{$\phantom{1}$}}
  +\tilde{\mathbf{D}}_{2}{\mytri{45}{0.6}{$\phantom{1}$}}+\tilde{\mathbf{D}}_{3}{\mytri{90}{0.6}{$\phantom{1}$}} =\\
  &=\footnotesize 
  \renewcommand{\arraystretch}{0.2} \left(
  \begin{array}{cccc}
  \mycircle{1}{$0$} & \myrect{1}{$1$} & \mydiam{0.9}{$-1$} & \myrectr{1}{$1$}\\
   \myrect{1}{$-1$} & \mycircle {1}{$0$} & \mytri{90}{1.1}{$1$} & \mytrin{45}{1.1}{$\,\, 1\,\,$}   \\
   \mydiam{1}{$1$} & \mytrin{90}{1.1}{$\,\, 1\,\,$} & \mycircle{1}{$0$} & \mytri{0}{1.1}{$1$}   \\
  \myrectr{1}{$-1$} & \mytri{45}{1.1}{$1$} & \mytrin{0}{1.1}{$\,\, 1\,\,$} & \mycircle{1}{$0$}  
  \end{array}
  \right) \normalsize.%=\\
  \end{split}
  \end{equation}
Every generator $\mathbf{\hat{G}}_k$, $\mathbf{\tilde{G}}_k$, $\mathbf{\hat{D}}_k$, $\mathbf{\tilde{D}}_k$ is a $4\times4$ matrix describing rotations in a subspace whose only entries different than zero correspond to the pictoric symbols. For all $k$, we find:
\begin{subequations}\label{eq:algebra_orto}
  \begin{align}
    \mathbf{\hat{G}}_k^2-\mathbf{\tilde{G}}_k^2  &=\mathbbold{1}\quad\quad;\quad\quad
    \mathbf{\hat{G}}_k\cdot\mathbf{\tilde{G}}_k=0\\
    \mathbf{\hat{G}}_k^2 &= \mathbbold{1}_k \quad\Rightarrow \quad \mathbf{\tilde{G}}_k^2  =\mathbbold{1}_k-\mathbbold{1},
  \end{align}
  \end{subequations}
%\begin{equation}\label{eq:unity_submatrix}
%  \mathbbold{1}_k=  
%  \left(\begin{array}{cccc}
%    1 &  0 & 0 &  0\\ 
%    0 &  \delta_{1,k}  & 0 &  0\\ 
%    0 &  0 & \delta_{2,k} &  0\\ 
%    0 &  0 & 0 &  \delta_{3,k}\\ 
%  \end{array} \right).
%\end{equation} 
where $\mathbbold{1}_k=diag\{1,\delta_{1,k},\delta_{2,k},\delta_{3,k}\}$. Equivalently, all the above implies $(\mathbf{\hat{G}}_k\pm i \mathbf{\tilde{G}}_k)^2 =\mathbbold{1}$.
%\begin{equation}\label{eq:agen}
%  (\mathbf{\hat{G}}_k\pm i \mathbf{\tilde{G}}_k)^2 =\mathbbold{1}
%\end{equation}
 These relations facilitate obtaining the commutators and anticommutators of the generators as:
\begin{subequations}\label{eq:algebra_2a}
  \begin{align}
    [\mathbf{\hat{G}}_i,\mathbf{\tilde{G}}_j]  &= -\epsilon_{ijk} \mathbf{\hat{G}}_k \quad(\forall i, j)\\
    [\mathbf{\hat{G}}_i,\mathbf{\hat{G}}_j]  &= \epsilon_{ijk} \mathbf{\tilde{G}}_k \quad(\forall i, j)\\
    [\mathbf{\tilde{G}}_i,\mathbf{\tilde{G}}_j]  &= -\epsilon_{ijk} \mathbf{\tilde{G}}_k \quad(\forall i, j)\\
    [\mathbf{\hat{G}}_i,\mathbf{\hat{G}}_j]_+  &= 2\cdot\delta_{ij}\mathbbold
    {1}_i+|\epsilon_{ijk}|\mathbf{\tilde{D}}_k \quad (\forall i,j)\\
    [\mathbf{\tilde{G}}_i,\mathbf{\tilde{G}}_j]_+  &= 2(\mathbbold{1}_i-\mathbbold{1}) +|\epsilon_{ijk}|\mathbf{\tilde{D}}_k\quad (\forall i,j)\\
    [\mathbf{\hat{G}}_i,\mathbf{\tilde{G}}_j]_+  &= \epsilon_{ijk}(-1)^k\mathbf{\hat{D}}_k\quad (\forall i,j)
    %[\mathbf{\tilde{G}}_i,\mathbf{\tilde{G}}_j]_+  &= [\mathbf{\hat
    %{G}}_i,\mathbf{\hat{G}}_j]_+= |\mathbf{\tilde{G}}_k| 
    %\quad (i\neq j \neq k)\\
  \end{align}
  \end{subequations}  
  Instead of infinitesimal generators, sometimes it is convenient to use denser matrices.  
Thus, a second algebraic level considers a 
decomposition in larger subspaces. Namely, we define vectors of $4\times4$ hyperbolic and ordinary 
rotation matrices:
\begin{equation}\label{eq:Gvectors}
    \vv{\mathbf{\hat{G}}}=(\mathbf{\hat{G}}_1,\mathbf{\hat{G}}_2 , \mathbf{\hat{G}}_3 )\quad\quad;\quad\quad
      \vv{\mathbf{\tilde{G}}}=(\mathbf{\tilde{G}}_1,\mathbf{\tilde{G}}_2 , \mathbf{\tilde{G}}_3 ),
\end{equation}
such that any Lorentz matrix $\hat{\mathbb{L}}(s_i)\equiv \hat{\mathbb{L}}_i=\mathbf{H}_i+\mathbf{R}_i$  
at a point $i$ can be neatly decomposed into a 
hyperbolic rotation $\mathbf{H}_i=\vv
{\eta}_i\cdot \vv{\mathbf{\hat{G}}}$ 
and an ordinary rotation $\mathbf{R}
_i=\vv{\rho}_i\cdot \vv{\mathbf{\tilde{G}}}$:
\begin{equation}\label{decomposing_M_N}
    \hat{\mathbb{L}}_i = 
  \left(\begin{array}{cccc}
    0 &  \eta_{1,i} & \eta_{2,i} &  \eta_{3,i}\\ 
    \eta_{1,i} & 0  & 0 &  0\\ 
    \eta_{2,i} &  0 & 0 &  0\\ 
    \eta_{3,i} &  0 & 0 &  0\\ 
  \end{array} \right) + 
  \left(\begin{array}{cccc}
    0 &  0 & 0 &  0\\ 
    0 & 0  & \rho_{3,i} &  -\rho_{2,i} \\ 
    0 &  -\rho_{3,i} & 0 &  \rho_{1,i}\\ 
    0 &  \rho_{2,i} & -\rho_{1,i} &  0\\ 
  \end{array} \right).
\end{equation}
Remarkably, any power $n\in\mathbb{Z}$ of $\mathbf{H}$ and $\mathbf{R}$ can be reduced to
\begin{subequations}
  \begin{align}\label{eq:powers_HR}
    \mathbf{H}^{2n}  &= \eta^{2(n-1)}\mathbf{H}^2\quad\quad;\quad\quad\quad
    \mathbf{H}^{2n+1}  = \eta^{2n}\mathbf{H}\\
    \mathbf{R}^{2n}  &= (-1)^{n-1}\rho^{2(n-1)}\mathbf{R}^2\quad;\quad
    \mathbf{R}^{2n+1}  = (-1)^{n}\rho^{2n}\mathbf{R},
  \end{align}
  \end{subequations}
which allows to use power series to calculate exactly any function of $\mathbf{H}$ or $\mathbf{R}$ in terms of only  $\mathbf{H},\mathbf{R},\mathbf{H}^2,\mathbf{R}^2$, where $\mathbf{H}^2,\mathbf{R}^2$ are easily obtained with
\begin{equation}\label{eq:H2R2}
    \mathbf{H}^{2} = \mathbf{C}(\vv{\eta})\quad\quad;\quad\quad
    \mathbf{R}^{2}  = \mathbf{C}(\vv{\rho})-\rho^2\mathbbold{1},
  \end{equation}
and $\mathbf{C}$ is the generic correlation matrix defined as:
\begin{equation}\label{eq:corr_matrix}
  \mathbf{C}(\vv{u})=  
  \left(\begin{array}{cccc}
    u^2 &  0 & 0 &  0\\ 
    0 &  u^2_1  & u_1u_2 &  u_1u_3\\ 
    0 &  u_1u_2 & u^2_2 &  u_2u_3\\ 
    0 &  u_1u_3 & u_2u_3 &  u^2_3\\ 
  \end{array} \right).
\end{equation}
We also find interesting that
\begin{subequations}
  \begin{align}
    \mathbf{HRH}  &= \mathbf{RHR} = (\mathbf{HR})^n= (\mathbf{RH})^n= \mathbf{0}\quad (n\in\mathbb{Z}),\\\label{eq:algebra_HR}
    &-\mathbf{Q}^T=\mathbf{R}\mathbf{H}= 
    \left(\begin{array}{cccc}
      0 & 0 & 0 &  0\\ 
      S_1 & 0  & 0 &  0\\ 
       S_2 &  0 & 0 &  0\\ 
      S_3 &  0 & 0 &  0\\ 
    \end{array} \right),    
  \end{align}
  \end{subequations}
%\begin{equation}\label{eq:matrixQ}
%  -\mathbf{Q}^T=\mathbf{R}\mathbf{H}= 
%\left(\begin{array}{cccc}
%  0 & 0 & 0 &  0\\ 
%  (\vv{\eta}\times\vv{\rho})_1 & 0  & 0 &  0\\ 
%   (\vv{\eta}\times\vv{\rho})_2 &  0 & 0 &  0\\ 
%  (\vv{\eta}\times\vv{\rho})_3 &  0 & 0 &  0\\ 
%\end{array} \right),
%\end{equation}
for $\mathbf{Q}=\mathbf{H}\mathbf{R}$ and $\vv{S}=\vv{\eta}\times\vv{\rho}$.
% the oriented surface vector with module the area of the paralellogram enclosed by $\vv{\eta}$ and $\vv{\rho}$. 
And as $\mathbf{H}$ and $\mathbf{R}$ do not 
commute: 
\vspace{-0.3cm}
\begin{subequations}
  \begin{align}\label{eq:algebra_M_N}
    &\hspace{-0.23cm}[\mathbf{H}_i,\mathbf{R}_j]  = \mathbf{Q}+\mathbf{Q}^T=-(\vv{\eta_i}\times \vv{\rho_j})\cdot\vv{\mathbf{\hat{G}}}\,\hspace{1.2cm} (\mathbf{H}-\mathrm{like})\\
    &\hspace{-0.23cm}[\mathbf{H}_i,\mathbf{H}_j] = (\vv{\eta_i}\times \vv{\eta_j})\cdot\vv{\mathbf{\tilde{G}}}\,\hspace{3cm} (\mathbf{R}-\mathrm{like})\\
    &\hspace{-0.23cm}[\mathbf{R}_i,\mathbf{R}_j] = -(\vv{\rho_i}\times \vv{\rho_j})\cdot\vv{\mathbf{\tilde{G}}}%\quad\Rightarrow 
    \hspace{2.9cm} (\mathbf{R}-\mathrm{like})\\
    \nonumber
    &\hspace{-0.23cm}[\ldots[[\mathbf{H}_1,\mathbf{R}_1],\mathbf{R}_2]\ldots,\mathbf{R}_n] =\phantom{\vv{\mathbf{\tilde{G}}}}\\
    &\hspace{-0.23cm} \,=(-1)^{n}(\ldots((\vv{\eta_1}\times \vv{\rho_1})\times \vv{\rho_2})\times\ldots \times \vv{\rho_n})\cdot\vv{\mathbf{\hat{G}}}\quad (\mathbf{H}-\mathrm{like})\\\nonumber
    &\hspace{-0.23cm}[\ldots[[\mathbf{H}_0,\mathbf{H}_1],\mathbf{H}_2]\ldots,\mathbf{H}_n] =\phantom{\vv{\mathbf{\tilde{G}}}}\\
    &\hspace{-0.23cm} \,=(\ldots((\vv{\eta_0}\times \vv{\eta_1})\times \vv{\eta_2})\times\ldots \times \vv{\eta_n})\cdot\vv{\mathbf{\tilde{G}}}\quad\quad\quad\, (\mathbf{R}-\mathrm{like})\\\nonumber
    &\hspace{-0.23cm}[\ldots[[\mathbf{R}_0,\mathbf{R}_1],\mathbf{R}_2]\ldots,\mathbf{R}_n] =\phantom{\vv{\mathbf{\tilde{G}}}}\\
    &\hspace{-0.23cm}=(-1)^{n}(\ldots((\vv{\rho_0}\times \vv{\rho_1})\times \vv{\rho_2})\times\ldots \times \vv{\rho_n})\cdot\vv{\mathbf{\tilde{G}}}\quad\,\, (\mathbf{R}-\mathrm{like}) 
  \end{align}
  \end{subequations}
These commutators may be useful to operate with the terms of the Magnus expansion.
The third algebraic level arises decomposing in basis matrices. In terms of generators, 
 we define a basis with $\mathbf{B}_0=\mathbbold{1}$ and:
 \begin{equation}\label{eq:Bmats}
      \mathbf{B}_k = \hat{\mathbf{G}}_k-i\tilde{\mathbf{G}}_k \quad \Rightarrow\quad  
      \vv{\mathbf{B}}=(\mathbf{B}_1,\mathbf{B}_2 , \mathbf{B}_3 ),
\end{equation}
for $k=1,2,3$. Then,we find for any $i,j>0$:
\begin{subequations}
  \begin{align}\label{eq:algebra_0}
    %  \mathbf{L}^{\pm}_k &= \frac{\mathbf{B}_k \pm \mathbf{B}^{*}_k}{2}\\
    &\mathbf{B}_i\mathbf{B}_j = \delta_{ij}\mathbbold{1} 
    +i\epsilon_{ijk} \mathbf{B}_k=\frac{[\mathbf{B}_i,\mathbf{B}_j]_++[\mathbf{B}_i,\mathbf{B}_j]}{2}\\%\quad (\forall i,j)\\
    &\mathbf{B}^*_i\mathbf{B}^*_j= \delta_{ij}\mathbbold{1} 
    -i\epsilon_{ijk} \mathbf{B}^*_k=\frac{[\mathbf{B}^*_i,\mathbf{B}^*_j]_++[\mathbf{B}^*_i,\mathbf{B}^*_j]}{2} \\%\quad (\forall i,j)\\
    &\mathbf{B}_k\cdot\mathbf{B}^*_k = \mathbbold{1} +2\mathbf{\tilde{G}}^2_k=2\mathbbold{1}_k - \mathbbold{1} \quad (\forall k)\\
    &[\mathbf{B}_i,\mathbf{B}^*_j]= [\mathbf{B}^*_i,\mathbf{B}_j]
    =0\label{eq:algebra_0b}\\%\quad (\forall i,j)
    &[\mathbf{B}_i,\mathbf{B}_j] = 2 i\epsilon_{ijk} \mathbf{B}_k \quad \Rightarrow \quad 
    [\mathbf{B}^*_i,\mathbf{B}^*_j] = -2 i\epsilon_{ijk} \mathbf{B}
    ^*_k\\%  \quad (\forall i, j)\\
    &[\mathbf{B}_i,\mathbf{B}_j]_+ = 2 \delta_{ij}\mathbbold{1}\\% \quad (\forall i, j)\\
    &[\mathbf{B}_i,\mathbf{B}^*_j]_+ = 2 |\epsilon_{ijk}|\mathbf{\tilde{D}}_k+2i\epsilon_{ijk}(-1)^k\mathbf{\hat{D}}_k% \quad (\forall i, j)          
    \end{align}
  \end{subequations}
The vector of matrices in (\ref{eq:Bmats}) leads to a fourth algebraic level involving the propagation vector. Here, we propose the existence of a generalized (complex) Lorentz matrix:
\begin{equation}\label{eq:lorentz_generalized}
  \mathbb{L}=\vv{a}\vv{\mathbf{B}}=\hat{\mathbb{L}}+ i \tilde{\mathbb{L}}.
\end{equation}
Its real part is the ordinary Lorentz matrix $\hat{\mathbb{L}}$ in Eq. (\ref{eq:lorentz1}), while the imaginary part is a new \textit{dual Lorentz matrix} containing the same physical information as $\hat{\mathbb{L}}$ but reorganized such that %fulfilling $\tilde{\mathbb{L}}(\rho_k,-\eta_k)=\hat{\mathbb{L}}(\eta_k,\rho_k)$
\begin{subequations}\label{eq:K_decomp_1}
  \begin{align}
    \label{eq:inverse_lorentz}
    \tilde{\mathbb{L}}&=(\vv{\eta}\cdot\vv{\rho} )\, \hat{\mathbb{L}}^{-1}\\
    \label{eq:KdecompA}
  \hat{\mathbb{L}} &= \frac{\mathbb{L} + \mathbb{L}^*}{2}
  =\vv{\eta}\cdot \vv{\hat{G}} +\vv{\rho}\cdot \vv{\tilde{G}}\\
  %\sum_k (\eta_k\hat{\mathbf{G}}_k +\rho_k \tilde{\mathbf{G}}_k)\\
  \label{eq:niceLtilde}
  i \,\tilde{\mathbb{L}} &= \frac{\mathbb{L} - \mathbb{L}^*}{2}=i 
  (\vv{\rho}\cdot \vv{\hat{G}} -\vv{\eta}\cdot \vv{\tilde{G}})\\
  %\sum_k(\rho_k\hat{\mathbf{G}}_k -\eta_k \tilde{\mathbf{G}}_k),\\
  \label{eq:KdecompC}
  &[\mathbb{L},\mathbb{L}^*]=[\tilde{\mathbb{L}},\hat{\mathbb{L}}]=0.
  \end{align}
\end{subequations}
These expressions are verified with a direct calculation. They show that $\hat{\mathbb{L}}$ and $\tilde{\mathbb{L}}$
can be decomposed either with non-commuting Lorentz generators or with $\mathbb{L}$ and $\mathbb{L}^*$, which commute locally. Note also that $\hat{\mathbb{L}}^{-1}$ (Eq.~\ref{eq:inverse_lorentz}) can be easily calculated 
by rearranging elements in $\hat{\mathbb{L}}$ (Eq.~\ref{eq:KdecompA}) to obtain $\tilde{\mathbb{L}}$ (Eq. \ref{eq:niceLtilde}) and dividing by a scalar $\vv{\eta}\cdot\vv{\rho}$. Let us remind that the stardard formula 
for $\hat{\mathbb{L}}^{-1}$ diverges when $|\hat{\mathbb{L}}|= 0$. The determinant of a matrix is 
the product of all its eigenvalues, which for $\hat{\mathbb{L}}$ and $\mathbf{K}$ are given by the sets 
 \begin{subequations}\label{eqs:eigenvalues_L}
   \begin{align}
     \lambda'_{1,2,3,4}&= \{\,\,\pm|\hat{a}|\,\,,\,\, \pm i|\tilde{a}|\,\, \}\\
     \lambda_{1,2,3,4}&= \{\,\,\eta_0\pm |\hat{a}|\,\,,\,\,\eta_0\pm i|\tilde{a}| \,\, \},
   \end{align}
 \end{subequations}
containing the components of $a_{\pm}$ (see Eq. \ref{eq:algebra_aes_2}). 
% The properties of $\eta_0$ imply a spectrum of eigenvalues
% \begin{equation}\label{sigma_eta0}
%  \sigma(-\mathbf{K}(s))\in \mathbb{C}^-. 
%\end{equation}
 %The propagation matrix can also be characterized by its eigenvalues and eigevectors. DO IT: define the metric and eigenvalues an eigenvectors, plot them, define them using the a vectors in terms of module ad angle too.
 Multiplying all values in each set, we obtain the determinants:
 \begin{subequations}\label{eqs:determinants}
  \begin{align}
    |\hat{\mathbb{L}}|&=|\tilde{\mathbb{L}}|=-p^2\label{eq:determinant_lorentz}\\
    |\mathbf{K}|&= \eta_0^2[\eta_0^2-r] + |\hat{\mathbb{L}}|.\label{eq:det2}
  \end{align}
\end{subequations}
with $p=\vv{\eta}\cdot\vv{\rho}$ and $r=\eta^2-\rho^2$ (Eqs. \ref{eq:algebra_mns_all}). Hence, standard $\hat{\mathbb{L}}^{-1}$ diverges for\footnote{The determinant $p=\vv{\eta}\cdot\vv{\rho}$ defines 
whether the algebraic systems represented by $\hat{\mathbb{L}}$ and $\hat{\mathbb{L}}^{-1}$ has a unique 
solution. $p=0$ implies 
 degenerated eigenvalues (matrix rows/columns not linearly independent), which reduces the dimensionality of the problem.} $p=0$, i.e. if $\eta$ 
or $\rho$ are zero\footnote{The wavelength symmetries of $\vv{\eta}$ or $\vv{\rho}$ makes them zero at a few discrete wavelengths (e.g. $\vv{\rho}=0$ at 
line center of the atomic transition).}, or if 
 \mbox{$\vv{\eta}\perp \vv{\rho}$}. This is not a problem in our formulation because, as said, Eq. (\ref{eq:inverse_lorentz}) gives a 
 stable fast way of obtaining $\hat{\mathbb{L}}^{-1}$ without divisions by zero, and because our solutions will not actually depend on $\hat{\mathbb{L}}^{-1}$. This is not the case in piecewise-constant numerical methods such as the standard evolution operator.
%, and therefore a standard diagonalization would not be possible
%\footnote{The only possibility would then be to reduce the propagation matrix to an almost-diagonal matrix with its Jordan normal and the use of generalized eigenvectors.}. 

Several key expressions arise from relating the basis and the Lorentz matrices. Using subscript $i$ for a quantity at point $s_i$:
\begin{subequations}\label{eq:from_comm_to_cross}
  \begin{align}
    &\mathbb{\hat{L}}_i= \frac{\mathbb{L} + \mathbb{L}^*}{2}=\frac{\vv{a}_i\vv{\mathbf{B}} + \vv{a}^*_i\vv{\mathbf{B}}^*}{2},\\
    &[\vv{a}_1\cdot\vv{\mathbf{B}},\vv{a}_2\cdot\vv{\mathbf{B}}]=2i(\vv{a}_1\times\vv{a}_2)\cdot\vv{\mathbf{B}},\\
    &[\vv{a}_1^*\cdot\vv{\mathbf{B}}^*,\vv{a}_2^*\cdot\vv{\mathbf{B}}^*]=-2i(\vv{a}_1^*\times\vv{a}_2^*)\cdot\vv{\mathbf{B}}^*.
  \end{align}
  \end{subequations} 
Besides, considering the normalized vector
\begin{equation}\label{eq:uuuu}
    \vv{u} = \frac{\vv{a}}{a_+}=\frac{a_{-}\cdot\vv{a}}{h}=\frac{(\hat{a}-i\tilde{a})\cdot\vv{a}}{h}
\end{equation}
%and its complex conjugate ($\vv{u}^*=\vv{a}^*/a_-$)
 it is direct to show that:
\begin{subequations}\label{eq:K_decomp_2}
  \begin{align}
    &\vv{u}\vv{B} = \frac{a_-\cdot \mathbb{L}}{h}\\
    &\vv{u}^*\vv{B}^* = \frac{a_+\cdot\mathbb{L}^*}{h}\\ 
    &\frac{\vv{u}\vv{B} + \vv{u}^*\vv{B}^*}{2} = \frac{\hat{a}\hat{\mathbb{L}} +\tilde{a}\tilde{\mathbb{L}}}{h}\\
   &\frac{\vv{u}\vv{B} - \vv{u}^*\vv{B}^*}{2} = i\frac{\hat{a}\tilde{\mathbb{L}} - \tilde{a}\hat{\mathbb{L}}}{h}\\
   &(\vv{u}\vv{B}) \cdot (\vv{u}^*\vv{B}^*)=\frac{\mathbb{L}\cdot\mathbb{L}^*}{h}.
  \end{align}
\end{subequations}
Finally, using the commutation rules obtained, we discover: 
\begin{subequations}\label{eq:algebra_3}
  \begin{align}
    \hat{\mathbb{L}}^2-\tilde{\mathbb{L}}^2&= r\cdot \mathbbold{1}\label{eq:aux1}\\
    \mathbb{L}^2=a_+^2\mathbbold{1}\quad&\,;\quad (\mathbb{L^*})^2=a_-^2\mathbbold{1}\\
    \frac{[\tilde{\mathbb{L}},\hat{\mathbb{L}}]_+}{2}= \hat{\mathbb{L}}\cdot \tilde{\mathbb{L}}&= i\frac{(\mathbb{L}^2-(\mathbb{L^*})^2)}{4}=p\cdot \mathbbold{1}\label{eq:guau}\\
    \hspace{-0.3cm}\frac{[\mathbb{L},\mathbb{L}^*]_+}{2} = \mathbb{L}\cdot \mathbb{L}^*=\hat{\mathbb{L}}^2+\tilde{\mathbb{L}}^2&=2\mathbf{K}^2-4\eta_0\mathbf{K}+(2\eta^2_0-r)\mathbbold{1}
  \end{align}
\end{subequations}
To obtain the last equality\footnote{Note that while $\mathbb{L}^2$ and $(\mathbb{L}^*)^2$ are diagonal, $\mathbb{L}$ and $\mathbb{L}^*$ are not: there can be an infinite set of nondiagonal square roots to a square matrix.}
in Eq.(\ref{eq:algebra_3}d), we applied Eq. (\ref{eq:aux1}) and $\hat{\mathbb{L}}^2=(\mathbf{K}-\eta_0\mathbbold{1})^2=\mathbf{K}^2-2\eta_0\mathbf{K}+\eta^2_0\mathbbold{1}$. 
Using (\ref{eq:aux1}) and (\ref{eq:guau}), any power of $\hat{\mathbb{L}}$ is recursively obtained from $\hat{\mathbb{L}},\hat{\mathbb{L}}^2$, and $\tilde{\mathbb{L}}$: 
\begin{subequations}\label{eq:powers_Lorentz}
  \begin{align}
    \hspace{0.5cm}\hat{\mathbb{L}}^3&=r\hat{\mathbb{L}}+p\tilde{\mathbb{L}}\\
    \hat{\mathbb{L}}^4&=r\hat{\mathbb{L}}^2+p^2\mathbbold{1}\\
    \hat{\mathbb{L}}^5&=(r^2+p^2)\hat{\mathbb{L}}+rp\tilde{\mathbb{L}}\\
    \hat{\mathbb{L}}^6&=(r^2+p^2)\hat{\mathbb{L}}^2+rp^2\mathbbold{1}\\
    \hat{\mathbb{L}}^7&=(r^2+2p^2)r\hat{\mathbb{L}}+(r^2+p^2)p\tilde{\mathbb{L}}\quad
  \end{align}
\end{subequations}
For $n\geq 2$ and starting from an odd power of the kind $\hat{\mathbb{L}}^{2n-1}=\alpha\hat{\mathbb{L}}+\beta\tilde{\mathbb{L}}$, the rules to obtain the next even and odd powers are:
\begin{subequations}%\label{eq:powers_Lor_gen}
  \begin{align}
    &\hat{\mathbb{L}}^{2n}=\alpha\hat{\mathbb{L}}^2+p\beta\mathbbold{1}\\
    &\hat{\mathbb{L}}^{2n+1}=(r\alpha+p\beta)\hat{\mathbb{L}}+p\alpha\tilde{\mathbb{L}}
 \end{align}
\end{subequations}
And similarly, we obtain:
\begin{subequations}\label{eq:powers_LL}
  \begin{align}
    \hat{\mathbb{L}}^2+\tilde{\mathbb{L}}^2&=2\hat{\mathbb{L}}^2-r\mathbbold{1}\label{eq:powers_LL_a}\\
    \hat{\mathbb{L}}^3+\tilde{\mathbb{L}}^3&=(p+r)\hat{\mathbb{L}}+(p-r)\tilde{\mathbb{L}}\\
    \hat{\mathbb{L}}^4+\tilde{\mathbb{L}}^4&=(2p^2+r^2)\mathbbold{1} \quad\ldots\nonumber
  \end{align}
\end{subequations}
We also analyzed the specific structure of $\hat{\mathbb{L}}^2$ and $\hat{\mathbb{L}}^3$. The latter (and any odd power of $\hat{\mathbb{L}}$) can be decomposed with $(\mathbf{H}+\mathbf{R})
^3=\mathbf{H}^3+\mathbf{R}^3+[\mathbf{H}^2,
\mathbf{R}]_+ +[\mathbf{R}^2,
\mathbf{H}]_+ +(\mathbf{H}\mathbf{R}\mathbf{H}+\mathbf{R}\mathbf{H}\mathbf{R})$, which using Eqs. (\ref{eq:powers_HR}) and (\ref{eq:algebra_HR}) gives same structure as $\hat{\mathbb{L}}$ :
\begin{equation}\label{eq:L3}
  \hat{\mathbb{L}}^3= [\eta^2\vv{\eta}+(\vv{S}\times \vv{\rho})]\cdot\vv{\mathbf{\hat{G}}}-[\rho^2\vv{\rho}+(\vv{S}\times \vv{\eta})]\cdot\vv{\mathbf{\tilde{G}}},
\end{equation}
%For $\hat{\mathbb{L}}^2$, we wrote that 
%$\hat{\mathbb{L}}^2=\mathbf{K}^2-2\eta_0\mathbf{K}%+\eta^2_0\mathbbold{1}$. 

%Despite 
%being skew-symmetric in the hyperbolic space, 
%it cannot be neatly expressed in terms of vector of 
%matrices with the generators that we have defined, due to an alterning sign:
%\begin{subequations}\label{eq:anticomm_HR}
%  \begin{align}
%    [\mathbf{H}_i,\mathbf{R}_j]_+ &= 
%    \sum_k(\vv{\eta}_i\times \vv{\rho}_j)_k(-1)^{k}\mathbf{\hat{D}}_k=\\
%    =&\left(\begin{array}{cccc}
%      0 & -S_1 & -S_2 &  -S_3\\ 
%      S_1 & 0  & 0 &  0\\ 
%      S_2 &  0 & 0 &  0\\ 
%      S_3 &  0 & 0 &  0\\ 
%    \end{array} \right).
%  \end{align}
%  \end{subequations}
$\hat{\mathbb{L}}^2$ cannot be so cleanly decomposed using $(\mathbf{H}+\mathbf{R})
^2=\mathbf{H}^2+\mathbf{R}^2+[\mathbf{H},
\mathbf{R}]_+$ because despite 
the element $[\mathbf{H}_i,\mathbf{R}_j]_+=\sum_k(\vv{\eta}_i\times \vv{\rho}_j)_k(-1)^{k}\mathbf{\hat{D}}_k$ is skew-symmetric in the hyperbolic space, it contains a $(-1)^{k}$.
%Eqs. (\ref{eq:H2R2}) and (\ref{eq:anticomm_HR}), 
%the structure of $\hat{\mathbb{L}}^2$ can be calculated as :
%\begin{equation}\label{eq:L2}
%      \hat{\mathbb{L}}^2= (\mathbf{U}(a)-\rho^2\mathbbold{1})+\vv{d}\vv{\mathbf{\tilde{D}}}+\sum_kS_k(-1)^{k}\mathbf{\hat{D}}_k,
%\end{equation}
%with $\mathbf{U}(a)=diag\{a^2,a_1^2,a_2^2,
%a^2_3\}$ ($a_k$ given in Eq.\ref
%{eq:algebra_aes_power}), $d_k=\eta_i\eta_j
%+\rho_i\rho_j\, (i\neq j\neq k)$ and $\vv{S}
%=\vv{\eta}\times\vv{\rho}$. 
To obtain a neater expression for $\hat{\mathbb{L}}^2$, 
we combine Eqs. (\ref{eq:powers_LL_a}) and (\ref{eq:algebra_3}d) into 
$\hat{\mathbb{L}}^2=\frac{1}{2}(\mathbb{L}\mathbb{L}^*+r\mathbbold{1})$. Developing the summations in $\mathbb{L}\cdot \mathbb{L}^*$, we reach:
\begin{equation}\label{eq:long_product}
  \nonumber
  \mathbb{L}\cdot \mathbb{L}^*=2\sum_k |a_k|^2\mathbbold{1}_k-a^2\mathbbold{1}+\sum_{k=1,2,3}(g_k \mathbf{D}_k+g_k^*\mathbf{D}^*_k),
\end{equation}
 where $\mathbf{D}_k=\mathbf{\tilde{D}}_k+i(-1)^k\mathbf{\hat{D}}_k$ are basis matrices obtained from dual generators and $g_k=a_ia^*_j\epsilon_{ijk}$ has coefficients cyclically ordered to give $\epsilon_{ijk}=1$. Writing these quantities as vectors, and  calling $\mathbf{U}(a)=diag\{a^2,a_1^2,a_2^2,
 a^2_3\}$ ($a_k$ given in Eq.\ref
 {eq:algebra_aes_power}), we get
 \begin{equation}\label{eq:L2_again}
  \hat{\mathbb{L}}^2= \biggl(\mathbf{U}(a)-\rho^2\mathbbold{1}\biggr)+\Re e \{ \vv{g} \cdot\vv{\mathbf{D}} \}.
\end{equation} 

\section{Reformulation of the homogeneous solution}\label{sec:homog}
\subsection{Derivation of the evolution 
operator for non-constant $\mathbf{K}$}\label{sec:evolop}
To derive an evolution operator for arbitrary variations of $\mathbf{K}$, start truncating the Magnus expansion to first order in Eq. (\ref{eq:high_magnus1})  
%The algebraic analysis performed in Section 
%(\ref{sec:structure}) allows to calculate it 
%explicitly in more than one way. 
and decompose $\hat{\mathbb{L}}(s)$ in 
generalized Lorentz matrices with Eq.(\ref{eq:from_comm_to_cross}a)
\begin{equation}\label{eq:eo4_step1}
  \begin{split}
\rm{\mathbf{O}}(s)= e^{-\int^{s}_{s_0} dt \mathbf{K}(t)} =e^{-\tau}\cdot e^{-\frac{1}{2}\int^{s}_{s_0}dt\vv{a}(t)\vv{\mathbf{B}}}\cdot
e^{-\frac{1}{2}\int^{s}_{s_0}dt\vv{a}^*(t)\vv{\mathbf{B}}^*}
  \end{split}
\end{equation}
The commutation among $\mathbb{L}=\vv{a}(t)\vv{\mathbf{B}}$ and $\mathbb{L}^*$  in Eq. (\ref{eq:KdecompC}) allows dividing the Magnus  
evolution operator in those two matrix exponentials.
Yet, the 
exponents are integrals combining 
all points along the ray and there was the problem of commutativity, so let us be more specific. Calling $\vv{b}=\int^{s}_{s_0}dt\vv{a}(t)$, the 
two matrix exponents can be 
joined as
\begin{equation*}\label{eq:specify1}
  \begin{split}
-\frac{1}{2}[\vv{\mathbf{B}}\vv{b} + \vv{\mathbf{B}}^*\vv{b}^*]
  \end{split}
\end{equation*}
because $\vv{\mathbf{B}}$ is a vector of basis 
(i.e., constant) matrices. To split the two exponentials, any component of $\vv{\mathbf{B}}\vv{b}$ must commute with any other of $\vv
{\mathbf{B}}^*\vv{b}^*$, which occurs due to Eq. (\ref{eq:algebra_0b}) because
\begin{equation}\label{eq:specify2}
  \begin{split}
[\mathbf{B}_ib_i,\mathbf{B}^*_jb^*_j]&=
(b_i\mathbf{B}_i)(b^*_j\mathbf{B}^*_j)-
(b^*_j\mathbf{B}^*_j)(b_i\mathbf{B}_i)=\\
&=b_ib^*_j\mathbf{B}_i\mathbf{B}
^*_j-b_ib^*_j\mathbf{B}^*_j\mathbf{B}
_i=b_ib^*_j[\mathbf{B}_i,\mathbf{B}^*_j]=0,
  \end{split}
\end{equation}
 Now we can apply the basic rotor Eq. (\ref{eq:evoltheorem_1}) to each matrix 
 exponential\footnote{The conditions for Eq. (\ref{eq:evoltheorem_1}) are fulfilled because the $\mathbf{B}
 _k$ matrices form a basis of the Lorentz 
 algebra that accomplishes with Eq. (\ref 
 {eq:condition_theorem}) in Eq. (\ref 
 {eq:algebra_0}).} in Eq. (\ref{eq:eo4_step1}), obtaining 
\begin{equation}\label{eq:eo4_step2}
  \begin{split}
\rm{\mathbf{O}}(s)= e^{-\tau}
\left[ch\left(\frac{b}{2}\right)\mathbbold{1}-sh\left(\frac{b}{2}\right)\vv{u}\vv{\mathbf{B}}\right]
\left[ch\left(\frac{b^{*}}{2}\right)\mathbbold{1}-sh\left(\frac{b^{*}}{2}\right)\vv{u}^*\vv{\mathbf{B}}^*\right],
  \end{split}
\end{equation}
where $b$ and $\vv{u}$ cast $a_+$ and $\vv{a}/a_+$ in Sec. \ref{sec:prop_vector}, but built from integrated coefficients $\eta'_k$ and $\rho'_k$ ($k=1,2,3$):
\begin{subequations}\label{eq:bks}
  \begin{align}
  \label{eq_bksok_U}
 % &\tau=\int^{s}_{s_0}\eta_0(t)dt\\
  &\vv{u}=\frac{\vv{b}}{b}=\frac{(b_1,b_2,b_3)}{[b^2_1+b^2_2+b^2_3]^{1/2}}=\frac{(\hat{b}-i\tilde{b})}{h}(b_1,b_2,b_3)\\\label{eq_bksok}
  &b_k(s)=\int^{s}_{s_0} a_k(t) dt = \int^{s}_{s_0}\eta_k(t)dt +i\int^{s}_{s_0}\rho_k(t)dt=\eta'_k +i\rho'_k\\
  &\hat{b}^2=(h+r)/2\\
  &\tilde{b}^2=(h-r)/2\\
  &h=[r^2+q^2]^{1/2}\quad(= \hat{b}^2 +\tilde{b}^2)\\
  &r= (\vv{\eta}')^2-(\vv{\rho}')^2\quad(= \hat{b}^2-\tilde{b}^2 )\\
  &q= 2(\vv{\eta}'\cdot\vv{\rho}')\quad(=2\hat{b}\tilde{b})
\end{align}
\end{subequations}
This step allows to express everything in terms of the 
seven basic integrals $b_k$ as elementary coefficients. Next, we perform the products in Eq. (\ref{eq:eo4_step2}), identify 
matrix terms with Eqs. (\ref
{eq:K_decomp_2}), and apply 
Appendix \ref{sec:trigo} to separate the real 
($\hat{b}$) and imaginary ($\tilde{b}$) parts of $b$ and $b^*$. This leads to:
\begin{equation}\label{eq:unimportant}
  \begin{split}
\mathbf{O}(s)= e^{-\tau}\cdot \Bigl\{
c_0\mathbbold{1}-\frac{1}{h}\left[ c_1\hat{\mathbb{L}}+ c_2\tilde{\mathbb{L}}
+c_3(\hat{\mathbb{L}}^2 +\tilde{\mathbb{L}}^2)\right] \Bigr\},
\end{split}
\end{equation}
with
\begin{equation}
    c_{0,3}=\frac{\mathrm{ch}(\hat{b})\pm\cos(\tilde{b})}{2} \quad;
    c_{1,2}=\frac{\mathrm{sh}(\hat{b})\mp i\sin(\tilde{b})}{2}.
\end{equation}
For simplicity we keep calling $\hat{\mathbb{L}}$ and $\tilde{\mathbb{L}}$ as in previous sections, but hereafter they  contain integrated coefficients.
Finally, substituting Eq. (\ref{eq:powers_LL_a}) for $\hat{\mathbb{L}}^2 +\tilde{\mathbb{L}}^2$ and operating, we find
\begin{equation}\label{eq:eo4_step3}
  \begin{split}
\mathbf{O}(s)= e^{-\tau}\cdot \Bigl\{
f_0\mathbbold{1}
+ f_{1a}\hat{\mathbb{L}}+ f_{1b}\tilde{\mathbb{L}}
+f_2\hat{\mathbb{L}}^2\Bigr\},
\end{split}
\end{equation}
with scalar functions
%$\tilde{\mathbb{L}}=p\hat{\mathbb{L}}^{-1}$ 
\begin{equation}\label{eq:eo4_efes}
  \begin{split}
f_0&= 
\frac{\tilde{b}^2\mathrm{ch}(\hat{b})+\hat{b}^2\cos(\tilde{b})}{\hat{b}^2+\tilde{b}^2}\quad;
f_{1a}= -\left[\frac{
  \hat{b}\,\mathrm{sh}(\hat{b})+\tilde{b}\sin(\tilde{b})}{\hat{b}^2+\tilde{b}^2}\right]\\
  f_2&= \frac{\mathrm{ch}(\hat{b})-\cos(\tilde{b})}{\hat{b}^2+\tilde{b}^2}\quad\quad\quad
;f_{1b}=\sigma\frac{\hat{b}\sin(\tilde{b})-
  \tilde{b}\,\mathrm{sh}(\hat{b})}{\hat{b}^2+\tilde{b}^2}.
\end{split}
\end{equation}
%same equations but with alternative parameters
%f_0&= 
%\frac{\tilde{b}^2\mathrm{ch}(2\hat{b})+\hat{b}^2\cos(2\tilde{b})}{\hat{b}^2+\tilde{b}^2}=(1-\varepsilon)\mathrm{ch}(2\hat{b})+\varepsilon\cos(2\tilde{b})\\
%f_{1a}&= -\left[\frac{
%\hat{b}\,\mathrm{sh}(2\hat{b})+\tilde{b}\sin(2\tilde{b})}{\hat{b}^2+\tilde{b}^2}\right]=-\varepsilon\frac{\mathrm{sh}(2\hat{b})}{\hat{b}}-(1-\varepsilon)\frac{\sin(2\tilde{b})}{\tilde{b}}\\
%f_{1b}&=-\frac{
%\tilde{b}\,\mathrm{sh}(2\hat{b})-\hat{b}\sin(2\tilde{b})}{\hat{b}^2+\tilde{b}^2}=-\delta \varepsilon\left[\frac{\mathrm{sh}(2\hat{b})}{\hat{b}}-\frac{\sin(2\tilde{b})}{\tilde{b}}\right]\\
%f_2&= \frac{\mathrm{ch}(2\hat{b})-\cos(2\tilde{b})}{\hat{b}^2+\tilde{b}^2}=\frac{2\delta \varepsilon}{q}[\mathrm{ch}(2\hat{b})-\cos(2\tilde{b})],
%containing well-behaving sync functions 
%\begin{equation}\label{eq:lims}
%  \lim_{\hat{b} \to 0} \frac{\mathrm{sh}(2\hat{b})}{\hat{b}} = \lim_{\tilde{b} \to 0} \frac{\sin(2\tilde{b})}{\tilde{b}}=1
%\end{equation}
We calculate $\hat{b}$ and $\tilde{b}$ from the positive root of their squares as in Eqs.(\ref{eq:bks}d,e). Thereby, $\sigma=\text{sign}(\vv{\eta}\cdot\vv{\rho})$ is 
added in $f_{1b}$ for sign correction\footnote{The specific 
sign of $\hat{b}$ and $\tilde{b}$ is irrelevant, only their 
product ($\sigma$), matters. To see this, propagate the signs from 
the argument of the trigonometric functions in Eqs. (\ref{eq:eo4_efes}) outwards, and apply Eqs.(\ref{eq:algebra_mns_all}a) and (\ref{eq:algebra_mns_all}e) to $\hat{b}$ and $\tilde{b}$.}.

Alternative expressions can be obtained using Eqs. (\ref{eq:powers_Lorentz}) to substitute
the $\tilde{\mathbb{L}}$ in Eq.(\ref
{eq:eo4_step3}) by a function of 
any odd power of $\hat{\mathbb{L}}$. For 
instance, in 
terms of $\hat{\mathbb{L}}$ and $\hat{\mathbb{L}}
^3$ with Eq. (\ref{eq:powers_Lorentz}a), we 
obtain:
\begin{equation}\label{eq:eo4_step4}
  \begin{split}
\mathbf{O}(s)= e^{-\tau}\cdot \Bigl\{
f_0\mathbbold{1}
+ f_1\hat{\mathbb{L}}+ f_2\hat{\mathbb{L}}^2
+f_3\hat{\mathbb{L}}^3\Bigr\},
\end{split}
\end{equation}
where the new functions are just
\begin{equation}\label{eq:eo4_efes2}
  \begin{split}
f_1&= -\left[\frac{
  \tilde{b}^2\,\mathrm{sh}(\hat{b})+\hat{b}^2\sin(\tilde{b})}{\hat{b}^2+\tilde{b}^2}\right];\,\, 
f_3= -\left[ \frac{\mathrm{sh}(\hat{b})}{\hat{b}}+\frac{\sin(\tilde{b})}{\tilde{b}}.\right]
\end{split}
\end{equation}
The evolution operator
can be made even more compact by dividing $\hat{\mathbb{L}}$ and $\tilde{\mathbb{L}}$ into hyperbolic ($\vv{\mathbf{\hat{G}}}$) 
and rotational ($\vv{\mathbf{\tilde{G}}}$) subspaces. Thus, 
operating, and regrouping, Eq. (\ref{eq:eo4_step3}) becomes: 
%\begin{equation}\label{eq:eo4_step5}
%  \begin{split}
%\mathbf{O}(s)= e^{-\tau}f_2\cdot \Big[\frac{f_0}{f_2}%\mathbbold{1}+\mathbf{\mathring{L}}+ \hat{\mathbb{L}}%^2\Big],
%\end{split}
%\end{equation}
\begin{equation}\label{eq:eo4_step5}
  \begin{split}
\mathbf{O}(s)= e^{-\tau}\cdot (\mathbf{\mathring{K}}+ f_2\hat{\mathbb{L}}^2),
\end{split}
\end{equation}
whose new matrix $\mathbf{\mathring{K}}=f_0\mathbbold{1}+\vv{\hat{\alpha}}\cdot\vv{\mathbf{\hat{G}}}+\vv{\tilde{\alpha}}\cdot\vv{\mathbf{\tilde{G}}}$ has same structure as $\mathbf{K}$ but with the new propagation vector
%\begin{equation}\label{eq:finalkprime}
%  \begin{split}
%\vv{\alpha}= \vv{\hat{\alpha}}+i\vv{\tilde{\alpha}}=&-%[x+iy]\vv{u}  
%\end{split}
%\end{equation}
\begin{equation}\label{eq:finalkprime}
  \begin{split}
\vv{\alpha}&= \vv{\hat{\alpha}}+i\vv{\tilde{\alpha}}=-[\mathrm{sh}(\hat{b})+i\sin(\tilde{b})]\vv{u},
\end{split}
\end{equation}
being $\vv{u}$ still given by Eq.(\ref{eq_bksok_U}).
Our three expressions for the evolution 
operator, Eqs. (\ref{eq:eo4_step3}), (\ref
{eq:eo4_step4}), and (\ref{eq:eo4_step5}), 
are fully equivalent. They all are 
more general and significantly
simpler than the one given by \citetalias{Landi-Deglinnocenti:1985a} for $\mathbf{K}$ constant. They are more
general because, being exact exponentials of any propagation (Lorentz) matrix, they can still preserve memory of arbitrary spatial variations within the 
ray path integrals of optical 
coefficients, as defined
in Eq.(\ref{eq:bks}c), while also keeping their analytical form when higher order terms of the Magnus expansion are 
considered. They express everything in terms of 
\textit{seven basic scalar integrals} ($\tau$ and $b_k$), which seems to be the minimal and most efficient integration 
possible to solve the RTE for non-constant 
properties. This innovative separation between 
integration and algebraic calculation will also be present in the inhomogeneous solution. 

The relative simplicity of our solution 
comes from expressing it in terms of the 
(integrated) Lorentz matrix $\hat{\mathbb{L}}=  \mathbf{K}'
(s) - \eta'_0(s)\mathbbold{1} =\vv{\eta}'\cdot \vv{\hat{G}} +\vv{\rho}'\cdot \vv{\tilde{G}}$ and its dual $\tilde{\mathbb{L}}$, which furthermore has a similar structure. Namely, from Eqs.(\ref{eq:K_decomp_1}), $\tilde{\mathbb{L}}$
 is both $\propto\hat{\mathbb{L}}^{-1}$ and $\tilde{\mathbb{L}}=  \vv{\rho}'\cdot \vv{\hat{G}} -\vv{\eta}'\cdot \vv{\tilde{G}}$, hence $\tilde{\mathbb{L}}$ and $\hat{\mathbb{L}}$ are effortlessly 
built from each other by exchanging $\eta'_k \leftrightarrow 
\rho'_k$ and $\rho'_k \leftrightarrow -\eta'_k$.

Considering also that $\hat{\mathbb{L}}^2$ is directly calculated from $\hat{\mathbb{L}}$ or by Eq.(\ref{eq:L2_again}), we see that our homogenous solution is 
efficient. E.g., in Eq. (\ref{eq:eo4_step5}) we just need to sum 
two composed matrices ($\hat{\mathbb{L}}^2$ and $\mathbf{\mathring{K}}$), half of them containing redundant 
(symmetric) information.
Hence, our solution is analytically compact, efficient, and suitable for considering spatial variations of $\mathbf{K}$.    

Eq. (\ref{eq:eo4_step5}) reveals explicitly other insight: $\hat{\mathbb{L}}^2$ is the only term in the homogeneous solution whose structure differs from that of $\mathbf{K}$. In other words, it is the only algebraic difference between an element of
the Lorentz group (like the evolution operator) and one in
its Lie algebra ($\mathbf{K}$, or $\hat{\mathbb{L}}$). This shows why a numerical method calculating the exponential by truncating the 
Dyson series Eq. (\ref{eq:dyson_taylor_2}) at an 
arbitrary power would be unsuitable, in general 
breaking the group structure. 

Our result belongs to the Lorentz group because fulfills Eq. (\ref{eq:lorentz3}a). Particularizing Eq.(\ref{eq:eo4_step3}) to zero dichroism
($\eta'=0$), then $q=\hat{b}=0, r=-h=\tilde{b}^2=(\rho')^2$, $\hat{\mathbb{L}}=\mathbf{R}$ by 
Eq.(\ref{decomposing_M_N}), and substituting we obtain a Lorentz-Poincar\'e $4\times4$ generalization of the Rodrigues formula for rotations:
\begin{equation}\label{eq:rodriguesR}
\mathbf{O}(s)= e^{-\tau}\Biggl[
\mathbbold{1}+ \sin(\rho')\Biggl( \frac{\mathbf{R}}{\rho'} \Biggr) + \Big(1-\cos(\rho')\Big)\Biggl( \frac{\mathbf{R}}{\rho'} \Biggr)^2 \Biggr].
\end{equation}
Similarly, without magneto-optical effects $\rho'=q=\tilde{b}=0, r=h=\hat{b}^2=(\eta')^2$, $\hat{\mathbb{L}}=\mathbf{H}$, and we get the hyperbolic analogue:
\begin{equation}\label{eq:rodriguesH}
  \mathbf{O}(s)= e^{-\tau}\Biggl[
  \mathbbold{1}+ \mathrm{sh}(\eta')\Biggl( \frac{\mathbf{H}}{\eta'} \Biggr) - \Big(1-\mathrm{ch}(\eta')\Big)\Biggl( \frac{\mathbf{H}}{\eta'} \Biggr)^2 \Biggr].
  \end{equation}
Finally, particularizing our result in Eq.(\ref{eq:eo4_step3}) to constant propagation matrix, the evolution operator of \citetalias{Landi-Deglinnocenti:1985a} is recovered.

\subsection{Higher order terms of the Magnus expansion}
Now consider again the Magnus expansion in Eq. (\ref{eq:high_magnus1}) and let us explore what 
happens when including orders larger 
than one. To make things clearer, let us also rename the integration variables $s_1,s_2,\ldots,s_n,$ as $s_n,
s_{n-1},\ldots,s_1$ in Eq. (\ref{eq:Magnus0}). Thus, $\mathbf{\Omega}_2$ gives: 
%after decomposing with Eqs. (\ref{eq:KdecompA}) and 
%(\ref{eq:KdecompC}):
\begin{equation}\label{eq:magnus2}
  \begin{split}
   \mathbf{\Omega}_{2}(s) &= -\frac{1}{2}\int^s_{s_0} d_1 
    \int^{s_1}_{s_0}d_2 \biggl[\,\mathbf{K}(s_2)\mathbf{K}(s_1)-\mathbf{K}(s_1)\mathbf{K}(s_2)
    \biggr] =\\
  &= -\frac{1}{2}\int^s_{s_0} d_2 
  \int^{s_2}_{s_0}d_1 \biggl[\mathbb{\hat{L}}(s_1),\mathbb{\hat{L}}(s_2)\biggr].
  \end{split}
\end{equation}
Labelling the dependence on $s_i$ with a subindex $i$ and developing the commutator with Eqs. (\ref{eq:from_comm_to_cross}), involving
the vector of basis matrices $\vv{\mathbf{B}}$, we obtain
\begin{equation}\label{eq:magnus2b}
  \begin{split}
  & [\mathbb{\hat{L}}_1,\mathbb{\hat{L}}_2]=\frac{i}{2}\biggl[  (\vv{a}_1\times\vv{a}_2)\vv{\mathbf{B}}-(\vv{a}^*_1\times\vv{a}^*_2)\vv{\mathbf{B}^*}\biggr]=\Re e 
  \Biggl\{i\vv{a}_{1x2}\cdot\vv{\mathbf{B}}
  \Biggr\}=\\
  &=\biggl[(\vv{\eta}_1\times\vv{\eta}_2)-(\vv{\rho}_1\times\vv{\rho}_2)\biggr]\vv{\mathbf{\tilde{G}}}-\biggl[(\vv{\eta}_1\times\vv{\rho}_2)+(\vv{\rho}_1\times\vv{\eta}_2)\biggr]\vv{\mathbf{\hat{G}}},
  \end{split}
\end{equation}
with $\vv{a}_{1x2}=\vv{a}_1\times\vv{a}_2$.
Developing $\mathbf{\Omega}_3$ in a similar way:
\begin{equation}\label{eq:magnus3}
   \mathbf{\Omega}_{3}(s) = -\frac{1}{4}\int^s_{s_0} d_3 
    \int^{s_3}_{s_0}d_2\int^{s_2}_{s_0}d_1 \Bigl[\Bigl[\mathbb{\hat{L}}_1,\mathbb{\hat{L}}_2\Bigr],\mathbb{\hat{L}}_3\Bigr].
\end{equation}
we obtain $[[\mathbb{\hat{L}}_1,\mathbb{\hat{L}}_2],\mathbb{\hat{L}}_3]=-\Re e 
\Bigl\{\vv{a}_{(12)3}\cdot\vv{\mathbf{B}}
\Bigr\}$ with $\vv{a}_{(12)3}=(\vv{a}_{1}\times\vv{a}_{2})\times\vv{a}_{3}$. Moving each integral inward to the propagation vectors, and using Eq.(\ref{eq:Bmats}) and Eqs. (\ref{eq:from_comm_to_cross}) again, the Magnus expansion can then be written by subspaces as:
\begin{equation}\label{eq:magnus3b}
  \mathbf{\Omega}'(s) = \Re e \Bigl\{\vv{f}(s)\cdot\vv{\mathbf{B}}\Bigr\}= \vv{f}_w(s)\cdot \vv{\mathbf{\hat{G}}}+\vv{f}_y(s)\cdot \vv{\mathbf{\tilde{G}}}
\end{equation}
with
%where we have the three-component vector function:
%\begin{equation}\label{eq:magnus_expan1}
%   \vv{f}(s) = -\vv{b}_1(s)-\frac{i}{2}\vv{b}_{2}(s)+\frac{1}{4}\vv{b}_{3}(s)+\ldots 
%\end{equation}
%with
%\begin{equation}\label{eq:magnus_expan1_coeffs}
%  \vv{b}_n(s)=\int^s_{s_0} \vv{b}_{n-1}(t)\times\vv{a}(t) \,dt 
  %\vv{b}_0(t)=\vv{\eta}(t)+i\vv{\rho}(t)) 
%\end{equation}
%starting with $\vv{b}_1(s)=\int^s_{s_0} \vv{a}(t) \,dt$. The second way is more explicit, with:
\begin{subequations}\label{eq:magnus_expan2}
  \begin{align}
  \vv{f}_w(s) &= -\vv{w}_1(s)-\frac{1}{2}\vv{w}_{2}(s)+\frac{1}{4}\vv{w}_{3}(s)+\ldots\\
  \vv{f}_y(s) &= -\vv{y}_1(s)-\frac{1}{2}\vv{y}_{2}(s)+\frac{1}{4}\vv{y}_{3}(s)+\ldots 
  \end{align}
\end{subequations}
and
\begin{subequations}\label{eq:magnus2bcoeffs}
  \begin{align}
    \vv{w}_{1}(s)&=\int^s_{s_0} \vv{\eta}(t)\, dt\quad;\quad\vv{y}_{1}(s)=\int^s_{s_0} \vv{\rho}(t)\, dt\\
    \vv{w}_{2}(s)&=-\int^s_{s_0}\biggl[ \vv{y_1}(t)\times\vv{\eta}(t)+\vv{w}_{1}(t)\times\vv{\rho}(t)\biggr]dt\\
    \vv{y}_{2}(s)&=\int^s_{s_0}\biggl[ \vv{w_1}(t)\times\vv{\eta}(t)-\vv{y}_{1}(t)\times\vv{\rho}(t)\biggr]dt\\
    \vv{w}_{3}(s)&=\int^s_{s_0}\biggl[ \vv{y_2}(t)\times\vv{\eta}(t)+\vv{w}_{2}(t)\times\vv{\rho}(t)\biggr]dt\\
    \vv{y}_{3}(s)&=\int^s_{s_0}\biggl[-\vv{w_2}(t)\times\vv{\eta}(t)+\vv{y}_{2}(t)\times\vv{\rho}(t)\biggr]dt
  \end{align}
\end{subequations}
The complexity of $\vv{f}_w(s)$ and $\vv{f}_y(s)$ increases in a non-obvious way when more terms are added to the Magnus expansion. 
Comparison of Eq. (\ref{eq:magnus3b}) with Eq. (\ref{eq:KdecompA}) 
shows explicitly that, as expected, $\mathbf{\Omega}$ stays in the Lie 
algebra, having the same structure as 
the Lorentz matrix for any truncation of the Magnus expansion. As a result, the corresponding exponential evolution operator in Eq. (\ref{eq:high_magnus1}) has the same form as Eq.(\ref
{eq:eo4_step3}), but now the quantities in Eqs. (\ref{eq:bks}) are just obtained substituting $\vv{\eta}'$ and $\vv{\rho}'$ by $\vv{f}_w$ and $\vv{f}_y$, which contain integral corrections. Thus, the calculation Eq.(\ref
{eq:eo4_step3}) applied to higher orders is increasingly complicated by 
the presence of nested integrals of vector products, as those in Eqs. (\ref{eq:magnus2bcoeffs}).
% Eq. (\ref{eq:magnus2b}). 
We dont expect corrections beyond order 2 to be calculated efficiently by brute force. It is a 
matter of ongoing numerical investigation to 
find out how to calculate these integrals 
efficiently when 
extending our methods to higher orders of the expansion.  

\section{Reformulation of the inhomogeneous problem}
The homogeneous
evolution operator derived in the previous sections can be inserted into Eq. (\ref
{eq:inhomog1}) to obtain a new family 
of numerical methods based on the Magnus 
expansion. 
However, the following issue motivates 
an alternative formulation for the inhomogeneous 
problem too. The inhomogeneous RTE:
\begin{equation}\label{eq:inhomog_scalar}
  \begin{split}
    \mathbf{I}'(s) &= \mathbf{A}(s)\mathbf{I}(s)+\boldsymbol{\epsilon}(s) \quad\quad ; \mathbf{I}(0)=\mathbf{I}_0
  \end{split}
\end{equation}
with $\mathbf{A}(s)=-\mathbf{K}(s)$ can be solved with Eq. (\ref{eq:inhomog1}) using the full Magnus exponential as evolution operators in it. If we start considering Eq. (\ref
{eq:inhomog_scalar}) with a Magnus 
expansion to first order allowing for non-constant $\mathbf{A}(s)$, then:
%$\mathbf{I}'(s)=\mathbf{A}(s)\mathbf{I}(s)+\boldsymbol{\epsilon}(s)$, (with
%$\mathbf{A}=-\mathbf{K}$ and $\mathbf{I}(0)=\mathbf{I}_0$)
\begin{equation}\label{eq:inhomog_sol}
  \begin{split}
    \mathbf{I}(s) &= e^{\int^s_0\mathbf{A}(\tau)d\tau}\mathbf{I}_0 + \int^{s}_{0} e^{\int^{s}_{t}\mathbf{A}(\tau)d\tau}\boldsymbol{\epsilon}(t)\, dt.
  \end{split}
\end{equation}
The 
problem here is that the calculation of the 
inhomogeneous integral with the nested integral in the 
inner exponent is costly and difficult to 
evaluate because we need two different 
sets of quadrature points for nested integrals 
changing in different intervals. To solve this problem, we have reformulated it.
\subsection{Reformulating the case with constant properties}\label{subsec:constantprop}
 In order to introduce our method of solution for the general case, we first consider the simplest consistent system, given by Eq. (\ref{eq:inhomog_scalar}) 
with both $\mathbf{A}$ and $\boldsymbol{\epsilon}$ constant. It is easy to see that its
solution can be written in terms of a matrix function $\boldsymbol{\phi}(s\mathbf{A})$, both as
%\begin{equation}\label{eq:inhomog_5D_1}
%  \begin{split}
%\mathbf{I}' &= \mathbf{A} \cdot \mathbf{I} %+\mathbf{\epsilon} \quad\quad ; \mathbf{I}(0)%=\mathbf{I}_0
%\end{split}
%\end{equation}
%with $\mathbf{A}=-\mathbf{K}$. The solution is
\begin{equation}\label{eq:inhomog5Dsol1}
\mathbf{I}(s) = e^{s\mathbf{A}}\mathbf{I}_0 
+ \int^{s}_{0} e^{(s-t)\mathbf{A}}\boldsymbol{\epsilon}\, dt=e^{s\mathbf{A}}\mathbf{I}_0 + 
s\boldsymbol{\phi}(s\mathbf{A})\boldsymbol{\epsilon}
\end{equation}
or as
\begin{equation}\label{eq:inhomog5Dsol2}
  \begin{split}
\mathbf{I}(s) &= e^{s\mathbf{A}}\mathbf{I}_0 + (e^{s\mathbf{A}}-\mathbbold{1})\mathbf{A}^{-1}\boldsymbol{\epsilon}=e^{s\mathbf{A}}(\mathbf{I}_0+\mathbf
{A}^{-1}\mathbf{\epsilon})-\mathbf{A}^{-1}
\boldsymbol{\epsilon}%\Biggr\}
=\\
&=\mathbf{I}_0+s\boldsymbol{\phi}
(s\mathbf{A})(\mathbf{A}\mathbf{I}_0+
\boldsymbol{\epsilon}),
\end{split}
\end{equation}
where we define $\boldsymbol{\phi}(s\mathbf{A})$ in several equivalent ways:
\begin{subequations}\label{eq:phi1}
  \begin{align}
  \label{eq:phi1a}   
  \boldsymbol{\phi}(s\mathbf{A})&=\frac{1}{s}\int^s_0e^{(s-t)\mathbf{A}}\,dt=\frac{1}{s}\int^s_0e^{t\mathbf{A}}\,dt=\\\label{eq:phi1b}
 &=\sum^{\infty}_{n=0}\frac{(s\mathbf{A})^{n}}{(n+1)!}=\\\label{eq:phi1c}
 &=(e^{s\mathbf{A}}-\mathbbold{1})(s\mathbf{A})^{-1}=\quad\biggl( \equiv \frac{e^{x}-\mathbbold{1}}{x}\biggr)\\\label{eq:phi1d}
 &=\int^1_0e^{\xi s\mathbf{A}}\,d\xi.
\end{align}
\end{subequations}
Eq. (\ref{eq:phi1a}) shows the relation of $\boldsymbol{\phi}(s\mathbf{A})$ with Eq. (\ref{eq:inhomog5Dsol1}), while Eq. (\ref{eq:phi1b}) shows its relation with the exponential
\begin{equation}\label{eq:phifuncts}
  \begin{split}
 e^{s\mathbf{A}}&=\sum^{\infty}_{n=0}\frac{(s\mathbf{A})^{n}}{n!}.\\
\end{split}
\end{equation}
Alternatively, Eq. (\ref{eq:phi1c}) gives an \textit{inefficient} way of calculating $\boldsymbol{\phi}(s\mathbf{A})$ inverting a matrix, but also points out the relation with the generatrix function of the inverse of the Magnus expansion 
itself (see Eq. (\ref{eq:generatrix})). Finally, eq. (\ref{eq:phi1d}) gives an efficient integral definition in terms of a parameter $\xi$. 

The second key point of our method of solution is the realization that the Eq. (\ref{eq:inhomog_scalar}) with both $\mathbf{A}$ and $\boldsymbol{\epsilon}$ constant
is equivalent to a $5\times5$ homogeneous system
\begin{equation}\label{inhomog_5D_4_cte}
  \begin{split}
  \frac{d}{ds} 
  \left(\begin{array}{c}
    \mathbf{I}(s)  \\ 
    1
  \end{array} \right) &=
  \left(\begin{array}{cc}
    \mathbf{A} & \boldsymbol{\epsilon}  \\ 
    \mathbf{0}^{\intercal} & 0
  \end{array} \right) 
   \left(\begin{array}{c}
    \mathbf{I}(s)  \\ 
    1
  \end{array} \right)
\end{split}
\end{equation}
with $\mathbf{I}(0)=\mathbf{I}_0$, and whose new 
$5\times 5$ propagation matrix $\mathcal{A}_
{5}$ contains $\mathbf{A}\equiv\mathbf
{A}_{4\times4}$. Correspondingly, from Eq. (\ref{eq:homog1}), the  
solution in Eq. (\ref{eq:inhomog5Dsol1}) 
must be equivalent to the 5D solution:
\begin{equation}\label{eq:sol5Dcte}
  \mathbf{I}_5(s)=e^{s\mathcal{A}_{5}}\mathbf{I}_{5,0}=
  \left(\begin{array}{cc}
    e^{s\mathbf{A}} & s\boldsymbol{\phi}(s\mathbf{A})\boldsymbol{\epsilon}  \\ 
    \mathbf{0}^{\intercal} & 0
  \end{array} \right) 
  \mathbf{I}_{5,0}
\end{equation}
with $\mathbf{I}_{5,0}\equiv\mathbf{I}_{5}(0)=(\mathbf{I}_0,1)^
{\intercal}$. Eq. (\ref{eq:sol5Dcte}) tells us that 
the inhomogeneous solution for $\mathcal{A}_
{5}$ constant can be written as 
a ${5\times5}$ evolution operator containing both the corresponding ${4\times4}$ evolution operator and a special product involving $\boldsymbol{\phi}$. Let us now apply this to the general case.

\subsection{Reformulating the general case with arbitrary variations}\label{sec:5x5}
The method that we propose consists in extending by one the 
dimension of the inhomogeneous problem to convert it 
in a five-dimensional homogeneous one, and 
thus solve it with the Magnus expansion again. Namely, Eq. (\ref{eq:inhomog_sol}) 
is equivalent to a $5\times5$ homogeneous system
\begin{equation}\label{inhomog_5D_4}
  \begin{split}
  \frac{d}{ds} 
  \left(\begin{array}{c}
    \mathbf{I}(s)  \\ 
    1
  \end{array} \right) &=
  \left(\begin{array}{cc}
    \mathbf{A}(s) & \boldsymbol{\epsilon}(s)  \\ 
    \mathbf{0}^T & 0
  \end{array} \right) 
   \left(\begin{array}{c}
    \mathbf{I}(s)  \\ 
    1
  \end{array} \right)
%\\
%  \left(\begin{array}{c}
%    \mathbf{I}(0)  \\ 
%    I_t(0)
%  \end{array} \right)&=
%  \left(\begin{array}{c}
%    \mathbf{I}_0  \\ 
%    1
%  \end{array} \right)
\end{split}
\end{equation}
with $\mathbf{I}_{5,0}\equiv\mathbf{I}_{5}(0)=(\mathbf{I}_0,1)^
{\intercal}$, and
where again we call $\mathbf{I}_{5}$(s) to the solution vector of unknowns and $\mathcal{A}_{5}$(s) to the new propagation matrix containing the original $\mathbf{A}(s)=-\mathbf{K}(s)$. Being homogeneous, this system can be solved 
applying the Magnus expansion to 
$\mathcal{A}_{5}(s)$. We do this first considering only $\mathbf{\Omega}_1$ in the Magnus expansion, to keep consistency with the evolution operator that we derived in the previous section. Thus, we have to calculate:
\begin{equation}\label{eq:magnus_5x5}
  \begin{split}
    \rm{\mathbf{O}}_{5}(s)=& e^{\int^{s}_{s_0}  \mathcal{A}_{5}(t)dt}=e^{\bar{\mathcal{A}}_{5}}=\sum^{\infty}_{n=0}\frac{\bar{\mathcal{A}}_{5}^{n}}{n!}.
  \end{split}
\end{equation}
%    \rm{\mathbf{O}}(s)=& e^{\int^{s}_{s_0}  \mathbf{A}_{5\times5}(t)dt}=\sum^{\infty}_{n=0}\frac{s^n}{n!}\mathbf{A}_{5\times5}^{n}.
where the overbars mean integration hereafter:
\begin{subequations}\label{eq:A_5x5}
  \begin{align}
    \bar{\mathcal{A}}_{5}
    =\left( \begin{array}{cc}
      \bar{\mathbf{A}} & \bar{\boldsymbol{\epsilon}}  \\ 
      \mathbf{0}^{\intercal} & 0
    \end{array}\right) & \\
    \bar{\mathbf{A}}=-\int^{s}_{s_0} \mathbf{K}(t)dt\quad;\quad
    \bar{\boldsymbol{\epsilon}}=&\int^{s}_{s_0} \boldsymbol{\epsilon}(t)dt \hspace{0.5cm}
  \end{align}
\end{subequations}
These integrals contain the six ray-path 
scalar integrals of Eq.(\ref{eq_bksok}) $\eta'_k$ and $\rho'_k (k=1,2,3)$, and the four ones of 
$\epsilon'_k (k=0,1,2,3)$. Since the powers of $\bar{\mathcal{A}}_{5}$ show the simple general form
\begin{equation}\label{A5x5_general}
  \begin{split}
  \bar{\mathcal{A}}_{5}^{n}
  &=\left(\begin{array}{cc}
    \bar{\mathbf{A}}^n & \bar{\mathbf{A}}^{n-1}\bar{\boldsymbol{\epsilon}}  \\ 
    \mathbf{0}^{\intercal} & 0
  \end{array} \right), 
\end{split}
\end{equation}
all terms in Eq. (\ref{eq:magnus_5x5}) can be readily resummed to obtain
\begin{subequations}\label{eq:magnus_5x5sol1}
  \begin{align}
    \label{eq:magnus_5x5sol1_a}
    \rm{\mathbf{O}}_{5}(s)&=\left(\begin{array}{cc}
      e^{\bar{\mathbf{A}}} & \Biggl[\sum^
      {\infty}_{n=0}\frac{\bar{\mathbf{A}}^
      {n}}{(n+1)!}\Biggr]\bar{\boldsymbol{\epsilon}}  \\ 
      \mathbf{0}^{\intercal} & 1
    \end{array} \right)=
    \left(\begin{array}{cc}
      e^{\bar{\mathbf{A}}} & \boldsymbol{\phi}
      (\bar{\mathbf{A}})\bar{\boldsymbol{\epsilon}}\\ 
      \mathbf{0}^{\intercal} & 1
    \end{array} \right)\\
    &=\mathbbold{1}_5+
    \left(\begin{array}{cc}
      \boldsymbol{\phi}(\bar{\mathbf{A}})\bar{\mathbf{A}} & \boldsymbol{\phi}(\bar{\mathbf{A}})\bar{\boldsymbol{\epsilon}}\\ 
      \mathbf{0}^{\intercal} & 0
    \end{array} \right),
  \end{align}
\end{subequations}
where we have used the definition of $\boldsymbol{\phi}$ in Eq. (\ref{eq:phi1b}). Inserting the above equivalent expressions into the general solution $\mathbf{I}_5(s)=\rm{\mathbf{O}}_
{5}(s)\mathbf{I}_{5,0}$, and taking the 
$4\times4$ subspace, we find:
\begin{subequations}\label{eq:sol5Dfin1}
  \begin{align}
  \mathbf{I}(s)&=e^{\bar{\mathbf{A}}}\mathbf{I}_{0}+\boldsymbol{\phi}(\bar{\mathbf{A}})\bar{\boldsymbol{\epsilon}}=\\
  &=\mathbf{I}_{0}+\boldsymbol{\phi}(\bar{\mathbf{A}})(\bar{\mathbf{A}}\mathbf{I}_{0}+\bar{\boldsymbol{\epsilon}}),
\end{align}
\end{subequations}
which reduces to Eqs. (\ref{eq:inhomog5Dsol1}) and (\ref{eq:inhomog5Dsol2}) when $\mathbf{A}$ and $\boldsymbol{\epsilon}$ are constant. 

What has happened in Eqs. 
(\ref{eq:sol5Dfin1})? The inhomogeneous formal integral that has 
always characterized the  
radiative transfer solution (e.g., in Eq.~\ref{eq:inhomog1} or \ref{eq:inhomog_sol}) has been absorbed into an algebraic expression. Eq. 
(\ref{eq:sol5Dfin1}) replaces it with the product of an 
integrated emissivity vector and a special function of 
an integrated propagation matrix.
We explain this by noting that solving the $4\times 4$ inhomogeneous problem via the Magnus solution to the $5\times 5$ homogeneous problem is like treating the emissivity \textit{translations} in Poincar\'e space as \textit{higher-dimensional rotations} described by the $5\times 5$ Magnus algebra. Thus, Eqs. (\ref{eq:sol5Dfin1}) offers a novel approach to solving the radiative transfer problem.

This can be extended to higher orders. Eq. (\ref{eq:magnus_5x5sol1_a}) shows 
that the algebra of the 5D propagation matrix 
is such that the 4D subspace containing the 4D 
homogeneous evolution 
operator is always independent on the 
inhomogeneous part of the system containing 
 $\boldsymbol{\phi}$ and $\boldsymbol{\epsilon}$. This implies that if we extend the 
Magnus expansion to higher orders, we can 
always continue using the corresponding 
evolution operator because the only thing 
that changes is how the function $\boldsymbol{
\phi}$ is combined with $\boldsymbol{\epsilon}$. For 
instance, adding $\mathbf{\Omega}_2$ to 
the Magnus expansion we would be adding a 
term of the kind of Eq.(\ref{eq:magnus2b}), but with the commutator :
\begin{equation}\label{eq:comm_omega2}
  \begin{split}
  [\mathcal{A}_{5}(s_1), &\mathcal{A}_{5}(s_2)]=\mathcal{A}_{5}(s_1)\mathcal{A}_{5}(s_2)-\mathcal{A}_{5}(s_2),\mathcal{A}_{5}(s_1)=\\
  &=
  \left(\begin{array}{cc}
    [\mathbf{A}(s_1),\mathbf{A}(s_2)] & \mathbf{A}(s_1)\boldsymbol{\epsilon}(s_2) - \mathbf{A}(s_2)\boldsymbol{\epsilon}(s_1)\\ 
    \mathbf{0}^{\intercal} & 0
  \end{array} \right).
  \end{split}
\end{equation}
Here, the $4\times4$ subspace is preserved, containing a commutator similar to that in the l.h.s.
Thus, after integrating and exponentiating, the new solution will have the new evolution operator (now including $\Omega_2$) and a composition of matrix-vector products between $\boldsymbol{
\phi}$ and integrals of $\mathbf{A}$ and $\boldsymbol{\epsilon}$. We will have to find the right balance between more general theoretical description (higher order in Magnus) and more efficient computational representation (higher numerical order of integration and fastest calculation).

\subsection{Calculation of $\mathbf{\phi}$: introducing the inhomogeneous evolution operator}
An advantage of our approach is that $\boldsymbol{\phi}(\bar{\mathbf{A}})$ in Eq. (\ref{eq:sol5Dfin1}a) can be calculated analytically. Namely, applying Eq. (\ref{eq:phi1d})
\begin{equation}\label{eq:phi1_final1}
  \boldsymbol{\phi}(-\bar{\mathbf{K}})=\int^1_0e^{-\int^{s}_{s_0} \xi \mathbf{K}(t)d\xi}\,d\xi=\int^1_0 O(t,\xi)\,d\xi,
 \end{equation}
we see that $\boldsymbol{\phi}$ 
is obtained by integrating our 
evolution operator Eq. (\ref{eq:eo4_step3}) along a 
parameter $\xi$ assigned to every matrix component 
of $\mathbf{\bar{K}}$, i.e.  
to every 
$\eta'$ and $\rho'$ in Eqs. (\ref{eq_bksok}) and in $\hat{\mathbb{L}}$, $\hat{\mathbb{L}}^2$. However,
$\xi$ only survives 
inside the exponential and trigonometrical expressions 
because its dependence cancels out when multiplying 
every matrix (containing $\eta$'s and $\rho$'s to the 
same power of the matrix\footnote{See structure of $\hat{\mathbb{L}}^2$ in (\ref{eq:L2_again})}) by its accompanying factor in 
$f_0,f_{1a},f_{1b},f_2$. This is easy to see by writing 
(\ref{eq:eo4_step5}) as:
\begin{equation}\label{eq:pre_integration}
  \begin{split}
\mathbf{O}(s,\xi)&= e^{-\tau\xi}\mathrm{ch}(\hat{b}\xi)
\cdot \left(\frac{\tilde{b}^2\mathbbold{1}+\hat{\mathbb{L}}^2}{\hat{b}^2+\tilde{b}^2}\right) +e^{-\tau\xi}\cos
(\tilde{b}\xi)\cdot \left(\frac{\hat{b}^2\mathbbold{1}
-\hat{\mathbb{L}}^2}{\hat{b}^2+\tilde{b}^2}\right)
- \\
&-e^{-\tau\xi}\mathrm{sh}(\hat{b}\xi)\cdot \left(\frac{\hat{b}\hat{\mathbb{L}}+\tilde{b}\tilde{\mathbb{L}} }{\hat{b}^2+\tilde{b}^2}\right)+e^{-\tau\xi}\sin(\tilde{b}\xi)\cdot \left(\frac{\hat{b}\tilde{\mathbb{L}}- \tilde{b}\hat{\mathbb{L}}}{\hat{b}^2+\tilde{b}^2}\right).
\end{split}
\end{equation}
% Guided by those elements in  Eq. (\ref{eq:eo4_step5}) (or also by Eq. (\ref{eq:eo4_step2})), and propagating $\xi$ to all the subsequent quantities, it is easy to  
% see that $\xi$ only survives 
% inside the exponential and trigonometrical expressions, 
% because any other quantity depending on $\eta'$ and $\rho'$ (e.g., $b_k$, $\hat{b}$, or $\tilde{b}$) is always divided by something ($\hat{b}^2$,$\tilde{b}^2$) cancelling out that 
% dependence.
% Specifying all the subspaces in Eq. (\ref{eq:eo4_step5}) for $\mathbf{O}(s,z)$ and adding $\xi$ where it remains:
% \begin{equation}\label{eq:integrated_evolop}
%   \begin{split}
% \mathbf{O}(s,\xi)&= e^{-\tau\xi}\cdot \Biggl\{ \frac{\tilde{b}^2\mathrm{ch}(\hat{b}\xi)+\hat{b}^2\cos(\tilde{b}\xi)}{\hat{b}^2+\tilde{b}^2}\mathbbold{1}+\\
% &-
% \Re e \Bigl\{[\mathrm{sh}(\hat{b}\xi)+i\sin(\tilde{b}\xi)]\vv{u}\Bigr\}
% \cdot\vv{\mathbf{\hat{G}}}\,+\\
% &-\Im m\Bigl\{[\mathrm{sh}(\hat{b}\xi)+i\sin(\tilde{b}\xi)]\vv{u}\Bigr\}\cdot\vv{\mathbf{\tilde{G}}}\,+\\
% &+ \frac{\mathrm{ch}(\hat{b}\xi)-\cos(\tilde{b}\xi)}{\hat{b}^2+\tilde{b}^2} \biggl(\mathbf{U}(b)-\rho^2\mathbbold{1}+ \Re e \{ \vv{g} \cdot\vv{\mathbf{D}} \}
%  \biggr)\Biggr\},
% \end{split}
% \end{equation}
% with $\mathbf{U}(b)=diag\{b^2,b_1^2,b_2^2,
% b^2_3\}$, $g_k=b_ib^*_j\epsilon_{ijk}$, and $b_k$ given in Eq.(\ref
% {eq:algebra_aes_power}).
Then, in solving for Eq. (\ref{eq:phi1_final1}), only 
the following integrals appear. Integrating \citep[e.g., 
with][]{Gradshteyn:2015vy}, and defining the ratios $\hat{p}=\hat{b}/\tau$ and $\tilde{p}=\tilde{b}/\tau$, we obtain:  
\begin{equation}\label{eq:integrals}
  \begin{split}
    &\hat{c}_1=\int^1_0 e^{-\tau\xi}\mathrm{ch}(\hat{b}\xi)d\xi=\frac{(1-\hat{c})-\hat{p}\hat{s}}{\tau(1-\hat{p}^2)}\quad\,\,\,\Leftarrow\quad\hat{c}=e^{-\tau}\mathrm{ch}(\hat{b})\\
    &\tilde{c}_1=\int^1_0 e^{-\tau\xi}\cos(\tilde{b}\xi)d\xi=\frac{(1-\tilde{c})+\tilde{p}\tilde{s}}{\tau(1+\tilde{p}^2)}\quad\Leftarrow\quad\tilde{c}=e^{-\tau}\cos(\tilde{b})\\
    &\hat{s}_1=\int^1_0 e^{-\tau\xi}\mathrm{sh}(\hat{b}\xi)d\xi=\frac{\hat{p}(1-\hat{c})-\hat{s}}{\tau(1-\hat{p}^2)}\quad\,\,\,\Leftarrow\quad\hat{s}=e^{-\tau}\mathrm{sh}(\hat{b})\\
    &\tilde{s}_1=\int^1_0 e^{-\tau\xi}\sin(\tilde{b}\xi)d\xi=\frac{\tilde{p}(1-\tilde{c})-\tilde{s}}{\tau(1+\tilde{p}^2)}\quad\Leftarrow\quad\tilde{s}=e^{-\tau}\sin(\tilde{b})
\end{split}
\end{equation}
where we have signified that $\hat{c},\tilde{c},\hat{s},\tilde{s}$ (on the right) has been transformed to $\hat{c}_1,\tilde{c}_1,\hat{s}_1,\tilde{s}_1$ (on the left). 
Since the rest of the homogeneous evolution operator (\ref{eq:eo4_step5}) remains unchanged, the result to Eq. (\ref{eq:phi1_final1}) is a strictly analogous 
operator with the same algebraic structure: it depends on the same matrices and is multiplied by similar 
scalar functions $f_0,f_{1a},f_{1b},f_2$ from (\ref{eq:eo4_efes}), but now with the  
substitution shown above. Hence, in a boast of creativity, we refer to $\boldsymbol{\phi}$ as the \textit{inhomogeneous evolution operator}. Since the same Lorentz matrices built 
from integrated optical profiles can be reused for both operators, and as the formal solution (\ref{eq:sol5Dfin1}a) 
involves only two matrix-vector products, the computational cost of our analytical solution is 
quiet contained. Note also that the full development always remains mathematically exact up to the order chosen for truncating the Magnus expansion.  

\begin{figure*}[t!]
  \hspace{-0.2cm}%\centering
  $\begin{array}{c} %[scale=0.3] [width=2.0in] --> options to control size
  \includegraphics[width=\textwidth]{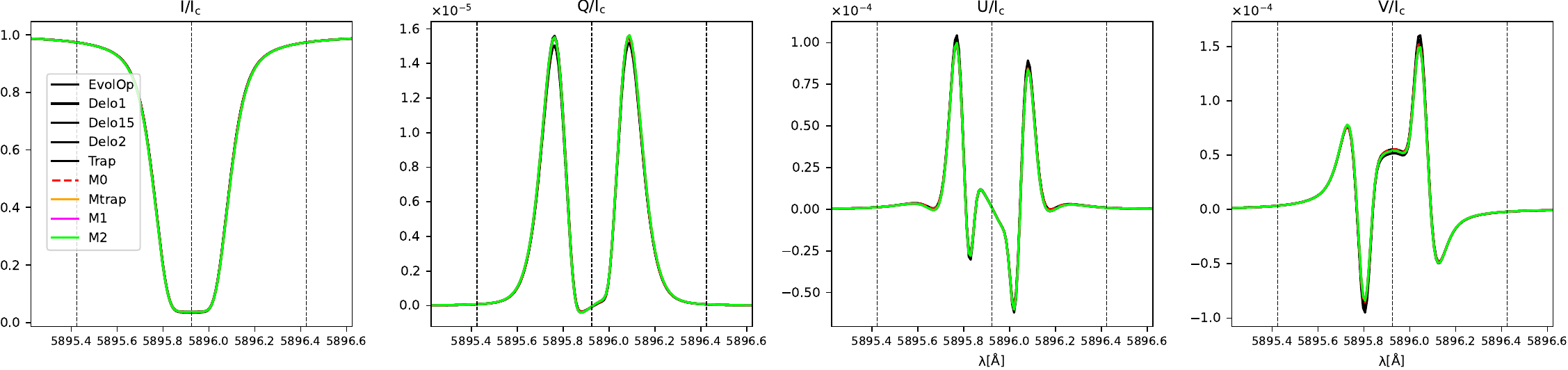} \\ \includegraphics[width=\textwidth]{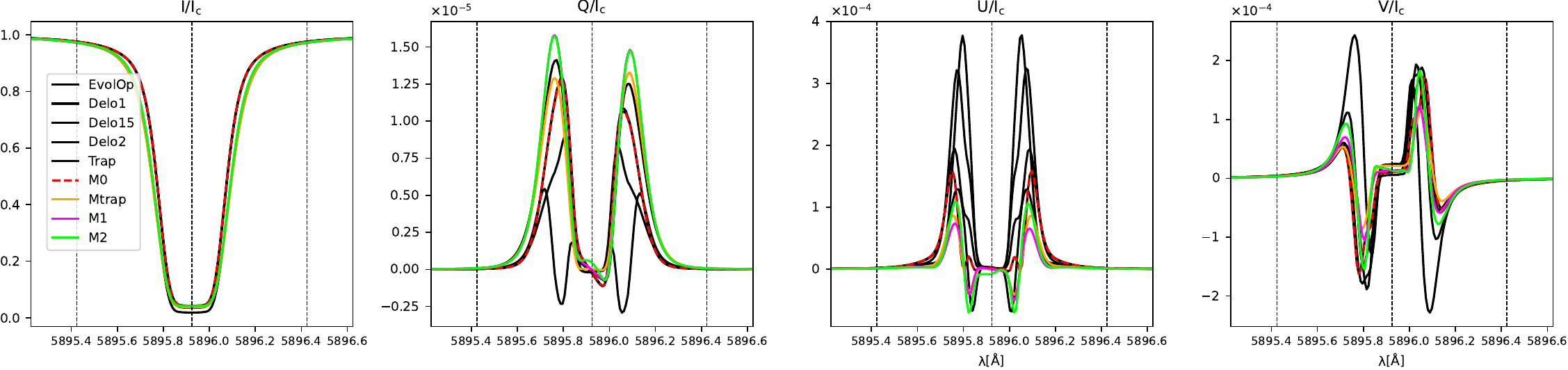} 
  \end{array}$
  \caption{Stokes profiles in the Na${\rm I}$ D1 line for an arbitrary parametric solar atmosphere (Appendix \ref{sec:atm_hazel_exp}) at solar disk center ($\mu=1$) with $7$ points (bottom panels) and $97$ points (top panels). The methods are (see Sec. \ref{sec:intro}): Trapezoidal, DELO order 1, DELOpar order 1.5, DELOparabolic order 2, Evolution Operator, Magnus degraded to piecewise constant (M0), Magnus-order 1 with order-3 integrals (M1), Magnus-order 1 with trapezoidal integrals (Mtrap), Magnus-order 2 with order-3 integrals (M2).}
  \label{fig:figlast}
\end{figure*}  

\section{Applications}\label{sec:applications}
\subsection{Numerical demonstration}\label{subsec:numerical}
Let us sum up and sketch an implementation of our formalism. First, calculate the optical depth $\tau = \int^s_{s_0} \eta_0(t)\, dt$ and the unsigned Magnus-wise 
integrals\footnote{Again, the subscripts refer to the integration variable (i.e., $d_k\equiv ds_{k}$). Our 
methods do not require the common use 
of $\tau$ as integration scale. 
Note also that we 
slightly change to a cleaner notation in 
this section, representing vectors with 
boldface ($\boldsymbol{\eta}$, 
$\boldsymbol{\mathring{\rho}}$, $\boldsymbol{\hat{G}}$,$\ldots$) 
instead of arrows on top.} of $\boldsymbol{\eta},\boldsymbol{\rho},\boldsymbol{\epsilon}$ (e.g, up to m-order 2):
\begin{subequations}\label{eq:magnus_expan_end}
  \begin{align}
    \boldsymbol{\mathring{\eta}} &= \int^s_{s_0} \boldsymbol{\eta}_1\, d_1+\frac{1}{2}\int^s_{s_0} d_2 
    \int^{s_2}_{s_0}d_1\biggl[\boldsymbol{\eta}_1\times\boldsymbol{\rho}_2-\boldsymbol{\rho}_1\times\boldsymbol{\eta}_2\biggr]\\
    \boldsymbol{\mathring{\rho}}&= \int^s_{s_0} \boldsymbol{\rho}_1\, d_1+\frac{1}{2}\int^s_{s_0} d_2 
    \int^{s_2}_{s_0}d_1\biggl[\boldsymbol{\eta}_1\times\boldsymbol{\eta}_2-\boldsymbol{\rho}_1\times\boldsymbol{\rho}_2\biggr]\\
    \boldsymbol{\mathring{\epsilon}}&= \int^s_{s_0} \boldsymbol{\epsilon}_1\, d_1+\frac{1}{2}\int^s_{s_0} d_2 
    \int^{s_2}_{s_0}d_1\biggl[\hat{\mathbf{\Omega}}_1\boldsymbol{\epsilon}_{2}-\hat{\mathbf{\Omega}}_2\boldsymbol{\epsilon}_{1}\biggr]
  \end{align}
\end{subequations}
Now build the Magnus-wise Lorentz matrix $\hat{\mathbf{\Omega}}(\boldsymbol{\mathring{\eta}},\boldsymbol{\mathring{\rho}})=\boldsymbol{\mathring{\eta}}\cdot \boldsymbol{\hat{G}} +\boldsymbol{\mathring{\rho}}\cdot \boldsymbol{\widetilde{G}}$, its 
dual $\widetilde{\mathbf{\Omega}}(\boldsymbol{\mathring{\eta}},\boldsymbol{\mathring{\rho}})= \hat{\mathbf{\Omega}}(\boldsymbol{\mathring{\rho}},-\boldsymbol{\mathring{\eta}})$, and the squared propagation vector components $q= 2(\boldsymbol{\mathring{\eta}}\cdot\boldsymbol{\mathring{\rho}})$ and $r= \mathring{\boldsymbol{\eta}}^2-\mathring{\boldsymbol{\rho}}^2$.
Then, build the two formally-identical evolution operators (from Eq. \ref{eq:eo4_step3}):
\begin{equation}\label{eq:last1}
  \begin{split}
    \mathbf{O}_{h,n}(s;\hat{\mathbf{\Omega}})= e^{-\tau}&e^{-\mathbf{\hat{\Omega}}}=h^{-1}\Big(     g_0\mathbbold{1}
    - g_{1a}\hat{\mathbf{\Omega}}- g_{1b}\widetilde{\mathbf{\Omega}}
    +g_2\hat{\mathbf{\Omega}}^2\Big),\\
g_0&=\tilde{b}^2\hat{c}+\hat{b}^2\tilde{c}\quad;\quad     g_{1a}=-(\hat{b}\hat{s}+\tilde{b}\tilde{s})\\
g_2&=\hat{c}-\tilde{c}\quad\quad;\quad    
g_{1b}=\sigma (\hat{b}\tilde{s}-\tilde{b}\hat{s}),
\end{split}
\end{equation}
% \begin{equation}\label{eq:last1}
%   \begin{split}
%     \mathbf{O}_{h,n}(s;\hat{\mathbf{\Omega}})= e^{-\tau}&e^{-\mathbf{\hat{\Omega}}}=
%      g_0\mathbbold{1}
%     - g_{1a}\hat{\mathbf{\Omega}}- g_{1b}\widetilde{\mathbf{\Omega}}
%     +g_2\hat{\mathbf{\Omega}}^2,\\
% g_0&=\frac{\tilde{b}^2\hat{c}+\hat{b}^2\tilde{c}}{\hat{b}^2+\tilde{b}^2}\quad;\quad     g_{1a}=\frac{-(\hat{b}\hat{s}+\tilde{b}\tilde{s})}{\hat{b}^2+\tilde{b}^2}\\
% g_2&=\frac{\hat{c}-\tilde{c}}{\hat{b}^2+\tilde{b}^2}\quad\quad;\quad    
% g_{1b}=\sigma \frac{\hat{b}\tilde{s}-\tilde{b}\hat{s}}{\hat{b}^2+\tilde{b}^2},
% \end{split}
% \end{equation}
with $\hat{c},\tilde{c},\hat{s},\tilde{s}$ given by Eq. (\ref{eq:integrals})-right or -left, for 
$\mathbf{O}_{h}$ or $\mathbf{O}_{n}$ respectively, and $\hat{b}^2=(h+r)/2$, $\tilde{b}^2=(h-r)/2$, $h=[r^2+q^2]^{1/2}$.
Finally, solve the 
RTE $\rm{d}\mathbf{I}(s)/\rm{ds}= \boldsymbol{\epsilon}(s)- \mathbf{K}(s)\mathbf{I}(s)$ with $\mathbf{I}(s_0)=\mathbf{I}_0$ by applying the exact general formal solution (from Eq. \ref{eq:sol5Dfin1}a):
\begin{equation}\label{eq:last0}
  \begin{split}
  \mathbf{I}(s)&=\mathbf{O}_h\cdot\mathbf{I}_{0}+\mathbf{O}_n\cdot\boldsymbol{\mathring{\epsilon}}.\\
  \end{split}
\end{equation}

The new family of methods arising from this first version of our formalism comes from solving Eqs. (\ref{eq:magnus_expan_end}) with different strategies and orders of approximation. To make them easily
reproducible and accessible, we have modified the community code 
HAZEL-2 \citep[][]{Asensio-Ramos:2008aa} to create HAZEL-Exp. This new online\footnote{\url{https://github.com/edgecarlin/hazel2\_experimental}} 
code intents to offer a community tool for numerical and analytical benchmarking and experimentation with 
spectral-line polarization. The code can define and 
process optically thick atmospheres and spectral lines with versatility and minimal syntaxis. Although yet in development, first results in 
Fig.~(\ref{fig:figlast}) demonstrate the suitability of our formulation with just two facts. First, when the grid 
resolution is increased (i.e. going from bottom to top panels), all methods converge to a common asymptotic 
solution, which validates our solution and implementation. And 
second, when our simplest method is degraded to 
advance point by point with constant propagation 
matrix (M0 method, red line), the result 
fits perfectly \citetalias{Landi-Deglinnocenti:1985a}'s method, which is bounded to be piecewise constant. The lines for both 
methods perfectly overlap each other: note that 
the red discontinuous line (M0 method) exposes a 
black line just behind, which corresponds to the 
EvolOp method, hence the coincidence. All this 
demonstrates a suitable numerical behavior and 
that our analytical equations are fully correct.

Regarding the computing times for different methods, we anticipate that comparing 
them at this point might not be the best way of 
evaluating them. Yet, as a 
representative fact, we observe that our 
best Magnus method (M2) is roughly $1.8$ faster 
than Delopar and $2.3$ faster than the Evolution 
Operator method for a ray with 97 points. 
These improvements grow with the 
number of points along the ray and the quadrature 
order. The proper numerical description,  quantification, and comparison with other methods, shall be presented in our numerical paper (Paper II).
\subsection{What can be done with this new formalism?}\label{subsec:comments}
\begin{itemize}[leftmargin=0cm, rightmargin=0cm, itemindent=0.25cm]
\item 1) Ensure consistent solutions that respect the Lie group structure, accurately solving the physical problem. A natural step now is to extend our formulation to include higher-order terms of the Magnus 
expansion and alternative group rotation representations like Jones', or even
abstracting out the physical problem from the  
representation to simplify calculations. 
\item 2) Explore the new family of numerical methods arising from Eq.(\ref{eq:last0}). These methods are advantageous due to their accurate, compact, and efficient  
representation of the physical problem, and because part of the calculation has been already solved analytically (the evolutions operators). 
However, the non-locality of our theory must be investigated 
for every Magnus expansion truncation order, as its convergence may set a maximum radius. A sufficiently large radius could enable scale-independent integration methods.
\item 3) Understand polarized radiative transfer effects in depth. As our formal 
solution is analytical and not restricted to regions of constant properties, the general 
non-local solution to the RTE can be studied 
analytically as a 
function of its parameters, thus transcending previous  models. For instance, now it would be possible to explain the joint action of magneto-optical and dichroic effects in certain 
wavelengths of the polarization profiles, which before could only be done approximately for separated mechanisms \citep[as done for dichroism in][]{Carlin:2019aa}. 

\item 4) Build 
analytical models of complex astrophysical objects (e.g., a whole star or a prominence) that can be 
however taylored to the amount of physical information 
available. One could analytically set a 
parametrical geometry in the radiative transfer 
problem and insightfully study the physics of these 
objects. This is possible because the input for our equations, 
the eleven ray-path scalar integrals of the optical 
coefficients, cannot only be provided by numerical atmosphere models but also by exact prescribed parametrical variations.    
\item 5) Derive non-local (multiscale) methods of solution for the solar NLTE 
problem with polarization. By considering variations of $\mathbf{K}$, our homogeneous and inhomogeneous solutions are 
\textit{non-markovian in space}, i.e. they preserve memory of the 
atmospheric physics along the 
ray\footnote{This problem is equivalent to considering partial 
redistribution in the 
equations of the NLTE problem. However, in that case a true non-markovian character arises from integrating along time the Schrödinger 
equation that describes the evolution of matter-radiation interaction.}. This memory can in 
principle be used, perhaps in tandem with multigrid methods, to accelerate the NLTE iteration between the RTE and the rate equations for the atomic populations. Our solution separates the integration from 
the formal solution, which 
permits to compute the latter only directly at the end of a (hopefully large and dynamic) portion of the ray, and reduces the numerical 
task to the simplest integration possible (i.e. 
directly to the eleven scalar optical 
coefficients of the problem).    
\item 6) Solve general physical problems with a similar algebraic structure to that of the RTE. With small cosmetic changes (renaming 
physical quantities), our solutions are also applicable to other homogeneous and inhomogeneous 
problems of universal interest that share the 
Lorentz group structure, again allowing non-local variations of their system matrices. Possible applications are the motion of masses at relativistic speeds, and the calculation of Lorentz forces on charged particles moving in an electromagnetic field. 
\end{itemize}

\section{Conclusions}
We have presented the first accurate and consistent formalism for solving polarized radiative transfer in Stokes representation using the Magnus expansion with exact exponential operators. This approach inherently preserves the Lorentz-Poincar\'e group structure and fully allows arbitrary non-local variations of the propagation matrix along the ray of light.

The homogeneous solution was interpreted in Poincar\'e space as continuous rotations whose directions and angles are defined by the direction and module of a varying \textit{propagation vector} (Sec. \ref{sec:prop_vector}). Using the Magnus expansion, we derived compact analytical 
homogeneous evolution operators (Eqs. \ref{eq:eo4_step3}-\ref{eq:eo4_step5}) that quantify arbitrary spatial variations of the propagation matrix via the Lorentz matrix (i.e., the propagation matrix without the diagonal). These operators 
are expressed in terms of Magnus-wise integrated Lorentz 
matrices and remain formally valid for any truncation order 
of the Magnus expansion. This result stems from a detailed 
algebraic characterization of the Lorentz matrix, after 
introducing a \textit{generalized} complex Lorentz matrix 
with a \textit{dual} matrix that is both a rearrangement of the Lorentz matrix and proportional to its inverse (Sec. \ref{sec:structure}).

For the inhomogeneous solution, we realized the 
inefficiency of nested quadratures that are intrinsic to the 
general inhomogeneous integral and we addressed it by 
reformulating the $4\times4$ inhomogeneous problem as a $5\times5$ homogeneous one (Sec.~\ref{sec:5x5}). This approach, solved using the 
Magnus expansion to first order and extended to higher 
orders (made here explictly up to second order), transforms the standard inhomogeneous integral into an analytically solvable integral of the homogeneous evolution operator (Eq.~\ref{eq:phi1_final1}). Thus, we obtained the novel 
\textit{inhomogeneous} evolution operator, an analytical
object formally identical to its homogeneous counterpart but 
with new trigonometrical coefficients increasing recursively in complexity with the Magnus truncation order (Eqs. \ref{eq:last1} and \ref{eq:integrals}). This advance should imply 
improved numerical methods, since part of the solution is now exactly precalculated analytically.

Combining these results, we formulated an explicit and elegant full radiative transfer solution (Eq. \ref{eq:last0}), whose inputs are Magnus-wise integrals of the scalar 
optical coefficients along the ray path (Eqs. \ref{eq:magnus_expan_end}). The 
formalism’s suitability is demonstrated analytically and numerically, supported by the new online code HAZEL-Exp (Sec. \ref{subsec:numerical}). In a forthcoming paper we are 
quantifying in full detail the numerical aspects and limitations, considering the 
algebraic complexity and convergence radius of the Magnus expansion.

Our approach opens new possibilities for solar spectropolarimetry (Sec. \ref{subsec:comments}). It decouples non-local integration of optical coefficients from the algebraic composition of a semi-formal solution, 
enabling strategies such as along-LOS parallelization, multi-scale integration, and possibly step-size-independent 
computational costs. Additionally, it allows for studying 
the parametric dependence of the solution Stokes vector and of NLTE response functions. These advances 
suggest reformulating the polarized NLTE problem with intrinsic non-local methods, potentially accelerating Stokes profile inversions and opening new avenues for enhancing solar diagnostics.

Furthermore, our 
analysis suggests 
reformulating the Magnus solution in alternative group representations (e.g., Jones') or even abstracting it out from a representation altogether. 
Finally, given the formal equivalence with other physical universal problems with Lorentz/Poincar\'e algebra, these can be directly benefited from our novel solutions after a mere 
renaming of the physical quantities, which reinforces interesting connections between solar physics and other fields.

Twenty-five years ago, \citetalias{Semel:1999aa} concluded that solar physics was not yet mature enough to solve the RTE using the Magnus expansion. 
Since then, the topic has remained unexplored, 
eclipsed by global research trends in observations, instrumentation, and simulations. Our findings mark a departure from this trend, 
offering a robust analytical framework to investigate radiative transfer and related physical problems through the Magnus expansion. This advance challenges 
the limitations of conventional formalisms, showing that analytical solutions are very valuable to improve the description of complex physical processes and guide their technological implementations. 

\begin{acknowledgements}
  E.S.C. dedicates this work to the memory of Vasile Ploscar. We thank the anonymous referee for their valuable revision. E.S.C. acknowledges financial support from the Spanish Ministry of Science and Innovation (MICINN) through the Spanish State Research Agency, under Severo Ochoa Centres of Excellence Programme 2020-2023 (CEX2019-000920-S). Part of his work has been funded by Ministerio de Ciencia e Innovación (Spain) through project PID2022-136585NB-C21, MCIN/AEI/10.13039/501100011033/FEDER, UE, and also by Generalitat Valenciana (Spain) through project CIAICO/2021/180. 
\end{acknowledgements}

\bibliographystyle{mnras} 
\bibliography{cbc2025a}
%\bibliography{mybibdesk_1}

\begin{appendix} %First appendix 

  \section{Some terminology regarding Lie groups}\label{app:Liegroup}
  In general, a group $\mathcal{G}$ is a set of elements 
  $A_i \in \mathcal{G}$ that together with a binary 
  operation fulfills the axioms of closure ($A_1\cdot 
  A_2=A_3 \in \mathcal{G}$), associativity ($A\cdot(B\cdot 
  C)=(A\cdot B)\cdot C$), and existence of neutral, identity, and inverse ($A 
  \cdot A^{-1}=\mathbbold{1}$). We only need to consider Lie 
  groups of invertible (hence square) $N \times N$ 
  matrices with the ordinary product as group operation
  and the commutator as the Lie bracket.
  In particular, a Lie group is a group of elements with a main group operation 
  mapping its elements smoothly to form a 
  differentiable manifold $\mathcal{M}$ (topological 
  condition), but also fulfilling an algebraic condition:
  the result of combining the elements of $\mathcal{M}$ 
  together with the operation of commutation stays in $\mathcal{M}$. The interest in Lie
  groups for solving differential equations is that being differentiable, hence analytical, there exists tangents to them at any point $p\in \mathcal{M}$. 
  A tangent vector at $p$ can be defined by differentiating a smooth parametric curve $\gamma(s)$ such that $\gamma(0)=p$ 
\citep[e.g.,][]{Bonfiglioli:2011aa}:
\begin{equation}
  \boldsymbol{v}=\left. \frac{d\gamma(s-s_0)}{ds}\right|_{s=s_0}
\label{eq:tangent}
\end{equation}
As illustrated in Fig. \ref{fig:fig1}, the set of all possible tangent vectors at $p$ forms the \textit{linear} vector
space $\mathcal{T}$ (the tangent space of the group). If locally (in the neighourghood of
each $p$) we associate $\gamma(s)$
with the evolution of the solution to a linear ODE
 on $\mathcal{M}$, then the infinitesimal advance of the solution at $p$
occurs in a direction contained in its tangent
space. Hence, instead of working in the nonlinear manifold $\mathcal{M}$,
Lie groups provide the mathematical foundation for solving the ODE in
a linear vector space while preserving the local structure of the group \citep[e.g.,][]{Hall_Lie_groups2015}. 
%\citep[e.g.,][]{Costa:2012aa}

The key to do this is that the elements in $\mathcal{M}$ can also be obtained by mapping (e.g., exponentiating) those of the so-called Lie algebra $\mathfrak{g}$ of the group. The algebra is defined as the combination of the tangent vector space around the group identity element with the operation of commutation. In that way, the algebraic structure of a Lie group is captured by its Lie algebra, a simpler object (since it is a vector space).

\section{Calculation of real and imaginary parts of the propagation module}\label{sec:acalc}

If $\vv{a} = \vv{\eta}+ i\vv{\rho}=(\eta_1+ i\rho_1,\eta_2+ i\rho_2,\eta_3+ i\rho_3)$, the modules of $\vv{a}(\equiv\vv{a}_{+})$ and $\vv{a}^* (\equiv\vv
{a}_{-})$ are complex numbers that we call $a_+$ and $a_-$, 
respectively. To calculate them we state:
  \begin{subequations}
    \begin{align}
      a_{\pm} &=\hat{a}\pm i\tilde{a} \quad \quad(\hat{a}, \tilde{a} \in \mathbb{R}), \nonumber\\
      a^2_{\pm}&=\vv{\eta}^2-\vv{\rho}^2 \pm i 2\vv{\eta}\cdot\vv
      {\rho}=\hat{a}^2 -\tilde{a}^2 \pm i2\hat{a}\tilde{a} \nonumber
      \end{align}
\end{subequations}
where we identify the real numbers  
\begin{subequations}
  \begin{align}
    r &= \vv{\eta}^2-\vv{\rho}^2=\hat{a}^2 -\tilde{a}^2\nonumber\\ 
    q & = 2\vv{\eta}\cdot\vv{\rho}= 2\hat{a}\tilde{a}, \nonumber\\
    h &=[r^2+q^2]^{1/2}= a_- \cdot a_+ = \hat{a}^2 +\tilde{a}^2 
    \nonumber
  \end{align}
\end{subequations}
Substituting $\hat{a}=q/(2\tilde{a})$ in $r=\hat{a}^2 -\tilde{a}^2$ we obtain $4\tilde{a}^4+4r\tilde{a}^2-q^2=0$ and find the solutions:
\begin{subequations}
  \begin{align}
    \hat{a}
    =\pm\left(\frac{\pm h + r}{2} \right) ^{1/2},\quad \tilde{a}&=\pm\left(\frac{\pm h-r}{2}\right)^{1/2}.\nonumber
    \end{align}
\end{subequations}
The signs acompanying $h$ must be chosen positive for $\hat{a}$, $\tilde{a}\in \mathbb{R}$. Regarding the outer signs of the roots, one is tempted to choose the sign of $\hat{a}$ positive because $\hat{a}$ works as an attenuation factor. \citetalias{Landi-Deglinnocenti:1985a} solved a similar problem to this before us, imposing positive sign for a proportional quantity equivalent to our $\hat{a}$ (which they called $\alpha$). However this 
imposition is unnecesary to solve the above 
equations and is not suitable for a general case in which stimulated emission could dominate. Note that the total absorption coefficient $\eta_0\pm |\hat{a}|>0$, appearing e.g. in the eigenvalues of 
the propagation matrix or in the expressions of 
the evolution operator, is always positive with $\eta_0>0$ and $|\hat{a}|<\eta_0$ when stimulated 
emission does not dominate. For this reason, we  
avoid to impose the outer signs, labelling them as $\hat{s}$ and $\tilde{s}$:
\begin{subequations}
  \begin{align}\label{eq:app1-3b}
    \hat{a}=\hat{s}\left(\frac{h + r}{2} \right) ^{1/2},\quad\quad 
    \tilde{a}&=\tilde{s}\left(\frac{h-r}{2}\right)^{1/2}.\nonumber
    \end{align}
\end{subequations}
The signs are however constrained by the angle $\theta$ between $\vv{\eta}$ and $\vv{\rho}$ in the QUV space, because $\hat{s}\cdot \tilde{s} =\text{sign}(\hat{a}\cdot\tilde{a})=\text{sign}(\vv{\eta}\cdot\vv{\rho})=\text{sign}(\cos\theta)$. Thus, when $\vv{\eta} \perp \vv{\rho}$ the signs are undetermined, but 
then the expressions involucrating them become 
independent on them.

\section{Exact exponential of the integral of a matrix}\label{sec:oursolution}
Assume a matrix $\mathbf{N}=f\,\vv{n}\cdot\vv{\sigma}$ that can be decomposed in an arbitrary 
constant $f$, a vector of components $\vv{n}=(n_1,\ldots,n_d)$, and a vector $\vv{\mathbf{\sigma}}=(\mathbf{\sigma}_1, \ldots,\mathbf{\sigma}_d)$ of basis matrices fulfilling:
\begin{equation}\label{eq:condition_theorem_app}
  \mathbf{\sigma}_k\cdot\sigma_{\ell} = \delta_{k{\ell}}\mathbbold{1} + h\cdot\epsilon_{k{\ell}m}\cdot\sigma_m, \quad\quad(h= ct.)
\end{equation}
This condition implies both $\sigma^2_k=\mathbbold{1}$ and vanishing anticommutators $[\sigma_k,\sigma_{\ell}]_+=\sigma_k\sigma_{\ell}+\sigma_{\ell}\sigma_k=0$ for all $k\neq {\ell}$. Then, the integral of $N$ is:
\begin{subequations}\label{eq:demo_theorem1}
  \begin{align}
    \mathcal{I}=\int ds \, \mathbf{N}(s)=\int ds \,f\,\vv{n}(s)\cdot\vv{\sigma}= f \sum^d_{k=1}b_k\sigma_k=f\,\vv{b}\cdot\vv{\sigma}
  \end{align}
\end{subequations}
defining the integrated vector $\vv{b}=(b_1,\ldots,b_d)$, with module $b$ and components:
\begin{equation}\label{eq:demo_theorem2}
  b_k=\int n_k(s) ds.
\end{equation}
Condition (\ref{eq:condition_theorem_app}) simplify calculation of $\mathcal{I}^2$ as:
\begin{subequations}\label{eq:demo_theorem3}
  \begin{align}
    \frac{\mathcal{I}^2}{f^2}= \left(\sum^d_{k=1}b_k\sigma_k \right)^2=\sum_{k={\ell}}b^2_k\sigma^2_k +\sum_{k\neq{\ell}}b_kb_{\ell}[\sigma_k,\sigma_{\ell}]_+ = b^2\mathbbold{1} 
  \end{align}
\end{subequations}
 Hence, the even and odd powers of the integral are $\mathcal{I}^{2k}=(fb)^{2k}\mathbbold{1}$ and $\mathcal{I}^{2k+1}=\mathcal{I}^{2k}\cdot\mathcal{I}=(fb)^{2k+1} \vv{u}\cdot\vv{\sigma}$, in terms of the unitary vector
 \begin{equation}\label{eq:demo_theorem4}
  \vv{u}=\frac{\vv{b}}{b}.
\end{equation}
Therefore, the exponential of $\mathcal{I}$ can be finally calculated as:
\begin{subequations}\label{demo_theorem_5}
  \begin{align}
    e^{\pm\int ds \,f\,\vv{n}\cdot\vv{\sigma}}=& \sum^{\infty}_{k=0}
    \frac{(\pm\mathcal{I})^k}{k!}=\sum^{\infty}_{k=0}\frac{\mathcal{I}^
    {2k}}{(2k)!} \pm \sum^{\infty}_{k=0}\frac{\mathcal{I}^{2k+1}}{(2k+1)!}=\nonumber\\
  =&\sum^{\infty}_{k=0}\frac{(fb)^{2k}}{(2k)!}
  \mathbbold{1}
  \pm \sum^{\infty}_{k=0}\frac{(fb)^{2k+1}}{(2k+1)!}\vv{u}
  \cdot\vv{\sigma}=\nonumber\\
  =&\, \mathrm{ch} (fb) \mathbbold{1} \pm \mathrm{sh} (fb) \mathbf{U},
  \end{align}
\end{subequations}
 where the matrix $\mathbf{U}=\vv{u}\cdot\vv{\sigma}$, and $b$ and $\vv{u}$ are given by Eqs. (\ref{eq:demo_theorem2}) and 
 (\ref{eq:demo_theorem4}).
 \section{Trigonometrical expressions for the coefficients of the evolution operator}\label{sec:trigo}
 The following four coefficients with complex argument $x=\hat{x}+i\tilde{x}$ are developed to find several equivalent convenient expressions:
 \begin{subequations}\label{trigo_coeff_1}
  \begin{align}
    c_0= \cosh(x)\cosh(x^*)&=\cosh^2(\hat{x})+\sinh^2(i\tilde{x})=\nonumber\\
    &=\cosh^2(\hat{x})-\sin^2(\tilde{x})=\nonumber\\
    &=\frac{\cosh(2\hat{x})+\cos(2\tilde{x})}{2}\nonumber\\
    c_1= \cosh(x)\sinh(x^*)&=\frac{\sinh(2\hat{x})-i\sin(2\tilde{x})}{2}\nonumber\\
    c_2= \cosh(x^*)\sinh(x)&=\frac{\sinh(2\hat{x})+i\sin(2\tilde{x})}{2}\nonumber\\
    c_3= \sinh(x)\sinh(x^*)&=\cosh^2(\hat{x})-\cosh^2(i\tilde{x})=\nonumber\\
    &=\cosh^2(\hat{x})-\cos^2(\tilde{x})\nonumber\\
    &=\frac{\cosh(2\hat{x})-\cos(2\tilde{x})}{2}.
  \end{align}
 \end{subequations}
  As trigonometrical expressions for double angle miminimizes numerical error for small arguments, we choose: 
  \begin{subequations}\label{trigo_coeff_2}
    \begin{align}
      c_{0,3}&=\frac{\cosh(2\hat{x})\pm\cos(2\tilde{x})}{2}\nonumber\\
      c_{1,2}&=\frac{\sinh(2\hat{x})\mp i\sin(2\tilde{x})}{2}.
    \end{align}
  \end{subequations}
 
  \newpage

 \section{Example of a type of atmosphere in HAZEL-Exp}\label{sec:atm_hazel_exp}

 Figure \ref{fig:figlast} was calculated with the 
atmosphere in cyan lines plotted below in Fig. \ref{fig:fig_app} after sampling with 7 and 97 
points. The conclusions of our tests 
are independent on the exact functional variation and values chosen for the atmosphere parameters.

\begin{figure}[h!]
  \centering
  \includegraphics[width=0.5\textwidth]{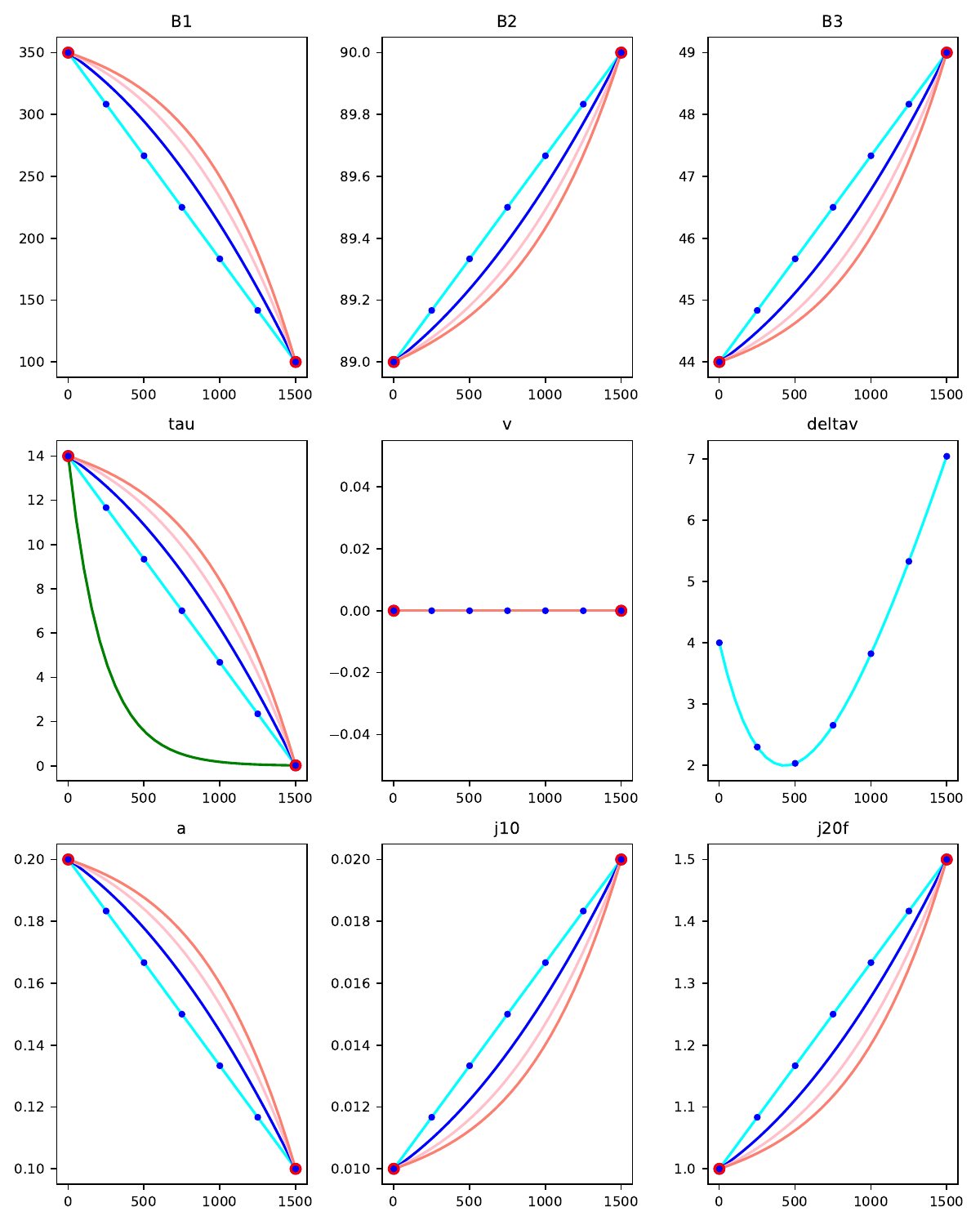} 
  \caption{Example of a parametric solar atmosphere with $7$ points as represented by our HAZEL-Exp code. All variables and units are here the standard for HAZEL-v2.0. Horizontal axes represent height in kilometers. The upper row gives the magnetic field spherical components ($B,\theta_B, \chi_B$), the middle row gives the total line-center optical depth, the velocity, and a Doppler broadening corresponding to an atmosphere with a minimum of temperature. The bottom row contains the damping coefficient of the absorption profiles, the multipolar component of the anisotropy factor $J^1_0$ in percentage, and the adimensional factor multiplying the Allen 
  anisotropy $J^2_0$ as typically done in HAZEL. None of these parameters affect the results of the present paper. 
  The curves in colors other than cyan were not used in our calculations, they are just possible functional variations that could be selected by default in HAZEL-Exp and always appear as plot references.}
  \label{fig:fig_app}
\end{figure}   

\end{appendix}

\end{document}